\pdfoutput=1
\RequirePackage{ifpdf}
\ifpdf
\documentclass[pdftex]{sigma}
\else
\documentclass{sigma}
\fi

\numberwithin{equation}{section}

\usepackage{mathtools}
\usepackage[ps,arrow,matrix]{xy} 

\DeclareMathOperator{\tens}{\otimes}
\DeclareMathOperator{\ctens}{\hat{\otimes}}

\newcommand{\sig}[1]{[#1]}
\newcommand{\fdg}[1]{|#1|}
\newcommand{\cs}[1]{\psi[#1]}

\theoremstyle{definition}
\newtheorem{dfn}{Definition}[section]

\newtheorem{rem}[dfn]{Remark}
\theoremstyle{plain}
\newtheorem{lem}[dfn]{Lemma}
\newtheorem{prop}[dfn]{Proposition}
\newtheorem{thm}[dfn]{Theorem}

\begin{document}

\allowdisplaybreaks

\renewcommand{\PaperNumber}{028}

\FirstPageHeading

\ShortArticleName{Free Fermi and Bose Fields in TQFT and GBF}

\ArticleName{Free Fermi and Bose Fields in TQFT and GBF}

\Author{Robert OECKL}

\AuthorNameForHeading{R.~Oeckl}

\Address{Centro de Ciencias Matem\'aticas, Universidad Nacional Aut\'onoma de M\'exico,
\\
Campus Morelia, C.P.~58190, Morelia, Michoac\'an, Mexico}

\Email{\href{mailto:robert@matmor.unam.mx}{robert@matmor.unam.mx}} \URLaddress{\url{http://www.matmor.unam.mx/~robert/}}

\ArticleDates{Received August 31, 2012, in f\/inal form April 02, 2013; Published online April 05, 2013}

\Abstract{We present a~rigorous and functorial quantization scheme for linear fermionic and bosonic f\/ield theory
targeting the topological quantum f\/ield theory (TQFT) that is part of the general boundary formulation (GBF).
Motivated by geometric quantization, we generalize a~previous axiomatic characterization of classical linear bosonic
f\/ield theory to include the fermionic case.
We proceed to describe the quantization scheme, combining a~Fock space quantization for state spaces with the Feynman
path integral for amplitudes.
We show rigorously that the resulting quantum theory satisf\/ies the axioms of the TQFT, in a~version generalized to
include fermionic theories.
In the bosonic case we show the equivalence to a~previously developed holomorphic quantization scheme.
Remarkably, it turns out that consistency in the fermionic case requires state spaces to be Krein spaces rather than
Hilbert spaces.
This is also supported by arguments from geometric quantization and by the explicit example of the Dirac f\/ield theory.
Contrary to intuition from the standard formulation of quantum theory, we show that this is compatible with
a~consistent probability interpretation in the GBF.
Another surprise in the fermionic case is the emergence of an algebraic notion of time, already in the classical
theory, but inherited by the quantum theory.
As in earlier work we need to impose an integrability condition in the bosonic case for all TQFT axioms to hold, due to
the gluing anomaly.
In contrast, we are able to renormalize this gluing anomaly in the fermionic case.}

\Keywords{general boundary formulation; topological quantum f\/ield theory; fermions; free f\/ield theory; functorial
quantization; foundations of quantum theory; quantum f\/ield theory}

\Classification{57R56; 81T70; 81P16; 81T20}

\tableofcontents

\section{Introduction}

\looseness=1
Free f\/ield theories and their quantization are usually the f\/irst examples treated in text books of quantum f\/ield
theory.
Apart from being simple toy examples they also serve to illustrate foundations of the formalism, its elementary objects
and elucidate the physical interpretation of quantum f\/ield theory at its most basic level.
Moreover, they serve to exhibit quantization schemes which are then ref\/ined for the treatment of more complicated
theories.
They also play a~crucial role as the basis for perturbation theory and $S$-matrix theory.
Finally, they are part of the small class of quantum f\/ield theories that have been def\/ined mathematically
rigorously.

It is natural to expect an equally important role for free f\/ield theories in approaches to quantum theory that aim to
go beyond the context of spacetime with a~f\/ixed background metric.
Here we shall be interested in the general boundary formulation (GBF), an axiomatic approach to quantum theory with
precisely that aim~\cite{Oe:gbqft}.
In the GBF, spacetime is modeled through manifolds that need not carry any structure in addition to a~topological one.
Its basic objects are that of a~specif\/ic version of topological quantum f\/ield theory (TQFT)~\cite{Ati:tqft}.
The other fundamental ingredient of the GBF consists of rules that allow to extract predictions of measurement outcomes
from these basic objects.

In quantum f\/ield theory in Minkowski spacetime a~comprehensive and universal description of free f\/ield theory can
be given, see e.g.~\cite{BaSeZh:algconstqft}.
By universal we mean here that it applies not just to specif\/ic examples, but to free f\/ield theories as a~class.
For the GBF such a~universal description of free, i.e., linear bosonic f\/ield theory has been presented
in~\cite{Oe:holomorphic}.
More specif\/ically, motivated by properties of Lagrangian f\/ield theory, an axiomatic def\/inition of classical
linear bosonic f\/ield theories as a~class was given.
Then a~functorial quantization scheme was exhibited that produces a~corresponding quantum theory in the sense of the
GBF.
In particular, it was proven rigorously that the quantum theory so produced satisf\/ies the axioms of the GBF.
It was also shown that the employed quantization is ``correct'' for certain examples of quantum f\/ield theories.
Later, even the formal equivalence of the quantization to the Feynman path integral (where applicable) was
established~\cite{Oe:affine,Oe:feynobs}.

Our main interest in the present work shall be a~functorial quantization scheme for linear fermionic f\/ield theory in
the GBF, paralleling that of the linear bosonic theory in~\cite{Oe:holomorphic}.
That is, we start with an axiomatization of classical f\/ield theory, exhibit a~quantization prescription, and show
that the quantum theory obtained satisf\/ies the axioms of the GBF.
We supplement this with an explicit example of a~realistic quantum f\/ield theory, in this case the Dirac f\/ield in
Minkowski spacetime.
This serves the dual purpose of demonstrating on the one hand that standard quantum f\/ield theories f\/it into the
framework, while on the other hand exhibiting features of the latter not accessible via the standard formulation.
Concretely, we provide here the consistent implementation of states on certain timelike hypersurfaces.

We recall that the quantization scheme in~\cite{Oe:holomorphic} used the holomorphic representation and relied heavily
on coherent states.
While there exist notions of coherent states also in the fermionic case, these have rather dif\/ferent properties and
are not amenable to an analogous treatment.
We thus opt for a~Fock space quantization.
While this seems simple enough, it turns out to complicate considerably the demonstration that the quantized theory
satisf\/ies the GBF axioms.
On the other hand, it makes possible a~treatment of the bosonic theory alongside the fermionic one in a~unif\/ied
fashion and with little additional ef\/fort.
In this way we also gain new insight into the bosonic theory from a~perspective rather dif\/ferent than the one
employed in~\cite{Oe:holomorphic}.

Even though we work within the specif\/ic framework of the GBF, much of our treatment should be applicable to a~large
class of TQFTs.
In particular, it is designed to be universal also in the sense of being independent of, but compatible with various
types of additional structure on the manifolds modeling spacetime, such as a~metric or conformal structure.
We also note that for other specif\/ic types of TQFT frameworks there is a~well developed understanding of the notion
of free f\/ield theory.
This is notably so in the TQFT approach to conformal f\/ield theory due to Segal~\cite{Seg:cftproc,Seg:cftdef}.
It is beyond the scope of the present work, however, to explore the potential connections to this approach.

\looseness=-1
An early and intriguing result of the present work is the necessity, in the fermionic case, to abandon Hilbert spaces
as (generalized) state spaces of the quantum theory in favor of \emph{Krein spaces}.
Apart from the justif\/ications for this laid out in the main body of this work, certain fermionic examples with Krein
spaces in~\cite{Seg:cftdef} served as an additional motivation.
The appearance of indef\/inite inner product spaces in quantization prescriptions in quantum f\/ield theory is not at
all new, a~famous example is the Gupta--Bleuler quantization of the electromagnetic
f\/ield~\cite{Ble:longphot, Gup:longphot}.
What is new here, however, is that such spaces not only appear at intermediate stages of the quantization process, but
as the f\/inal physical state spaces.
The appearance of Krein spaces might lead to concerns about the feasibility of a~consistent probability interpretation.
We remind the reader, however, that the state spaces of the GBF can not in general be identif\/ied with state spaces of
the standard formulation of quantum theory, where this would indeed be a~problem.
What is more, we show in this work explicitly (in Section~\ref{sec:prob}) that a~consistent probability interpretation
is possible.

Another intriguing aspect of linear fermionic f\/ield theories, present already at the classical level, is something
that may by called an \emph{emergent notion of time} in the dynamics.
We refer the reader to Section~\ref{sec:cevol} for details.

Finally, we remark on the integrability condition that was necessary to impose in~\cite{Oe:holomorphic} for a~classical
linear bosonic f\/ield theory to possess a~quantization satisfying all the axioms of the GBF.
This integrability condition arises due the presence of the \emph{gluing anomaly factor}.
Roughly speaking the integrability condition amounts to requiring that this anomaly factor be f\/inite.
In the present work we exhibit the same gluing anomaly factor.
However, in the fermionic case we are able to carry out a~renormalization procedure so that even if the gluing anomaly
factor is ``inf\/inite'', a~renormalized version of the relevant gluing axiom still holds.
Thus the quantization of any linear fermionic f\/ield theory satisfying the axioms of the classical theory satisf\/ies
the renormalized version of the GBF axioms, without exception.

In Section~\ref{sec:motivate} we provide motivation for the axiomatization of the classical theory and aspects of the
quantization scheme in the fermionic case.
In Section~\ref{sec:ingreds} we give a~short review of some necessary mathematical ingredients, notably Krein spaces
and the associated Fock spaces.
The axiomatization of classical linear f\/ield theory is discussed in Section~\ref{sec:classdata}.
The example of the free Dirac f\/ield is treated in Section~\ref{sec:dirac}, both with spacelike and timelike
hypersurfaces.
The axioms of the GBF are reviewed and generalized to include the fermionic case in Section~\ref{sec:gbfax}.
The basics of the quantization scheme are described in Section~\ref{sec:quantization}.
In Section~\ref{sec:amplip} relations between amplitude and inner product are examined.
Section~\ref{sec:composition} treats the gluing axioms and exhibits their proofs, while concentrating the most
dif\/f\/icult part into a~lemma.
Due to its length, the proof of this lemma is delegated to Appendix~\ref{sec:gproof}.
The gluing anomaly and its renormalization are the subjects of Section~\ref{sec:glanom}.
The probability interpretation in the context of Krein spaces is elaborated on in Section~\ref{sec:prob}.
In Section~\ref{sec:comphol} a~more detailed comparison to the holomorphic approach of~\cite{Oe:holomorphic} is given
in the bosonic case.
Finally, a~brief outlook is presented in Section~\ref{sec:outlook}.

\section{Motivation}
\label{sec:motivate}

In this section we shall provide some motivation for concrete aspects of the way we axiomatize classical f\/ield
theory, and also for certain aspects of the quantization scheme we describe in the following.

\subsection{Schematic view of quantization}
\label{sec:quants}

In the following we recall in a~very abbreviated and schematic fashion the f\/irst steps in the quantization of free
f\/ields from the point of view of geometric quantization~\cite{Woo:geomquant}.

In both the bosonic and the fermionic case the complex Hilbert space $\mathcal{H}$ of states for a~free f\/ield is most
conveniently constructed starting from another complex Hilbert space $L$.
This space $L$ may be thought of as the space of solutions to the equations of motion of the underlying classical
system to be quantized.
Most commonly (and also in the present work), $\mathcal{H}$ is then constructed as the Fock space over~$L$.
However, there are other, but equivalent ways to construct~$\mathcal{H}$ from $L$ which may more convenient depending
on the context.
We shall remark on this in Section~\ref{sec:comphol}.

Taking the point of view of geometric quantization, the space $L$ is obtained in the bosonic case as follows: At
f\/irst we take~$L$ to be the real vector space of solutions of the equations of motion with suitable regularity
properties.
This space becomes a~symplectic vector space by equipping it with the symplectic form $\omega:L\times L\to\mathbb{R}$
that is derived from the Lagrangian of the theory in the standard way.
The next step is then to introduce a~complex structure $J:L\to L$ that leaves~$\omega$ invariant and is compatible with
the dynamics.
The latter usually means something along the lines that~$J$ leads to a~useful notion of ``positive energy'' versus
``negative energy'' solutions and respects relevant symmetries, such as Lorentz transformations.
The details of this are not important here.
Combining the complex structure with the symplectic structure leads to a~real inner product $g:L\times L\to\mathbb{R}$,
which is usually required to be positive def\/inite (possibly requiring taking a~quotient),
\begin{gather*}
g(\phi',\phi)\coloneqq2\omega(\phi',J\phi).
\end{gather*}
This in turn is combined with the symplectic structure to yield a~complex inner product
\begin{gather}
\{\phi',\phi\}\coloneqq g(\phi',\phi)+2\mathrm{i}\omega(\phi',\phi).
\label{eq:cipripsympl}
\end{gather}
This makes $L$ (possibly upon completion) into a~complex Hilbert space.

In the fermionic case the construction of $L$ is slightly more complicated.
The symplectic structure $\tilde{\omega}:L\times L\to\mathbb{R}$ obtained from the Lagrangian does not in itself serve
to def\/ine the f\/inal complex inner product.
Rather, it has to be combined with a~complex structure $\tilde{J}:L\to L$ which is usually a~natural complex structure
on the spinor bundle that would serve to def\/ine the classical system.
This yields the real inner product $g:L\times L\to\mathbb{R}$,
\begin{gather*}
g(\phi',\phi)\coloneqq2\tilde{\omega}(\phi',\tilde{J}\phi).
\end{gather*}
Another complex structure $J:L\to L$, related to the dynamics as in the bosonic case, then yields the relevant
symplectic structure
\begin{gather}
\omega(\phi',\phi)\coloneqq-\tfrac{1}{2}g(\phi',J\phi).
\label{eq:riptosympl}
\end{gather}
The real inner product and the symplectic structure are then combined as in the bosonic case via
equation~\eqref{eq:cipripsympl}.

\subsection{Quantization in the GBF}
\label{sec:quantgbf}

For the convenience of the reader we have described quantization so far as if we were interested in a~single Hilbert
space describing our quantum system.
This is not the case.
Rather, in the GBF we need one such Hilbert space for each (admissible) hypersurface in spacetime.
It turns out that the schematic procedure described above is already naturally adapted to this~\cite{Oe:holomorphic}.
Given a~hypersurface $\Sigma$ we think of $L_{\Sigma}$ as the space of solutions in a~neighborhood of $\Sigma$ (or more
precisely as the space of germs of solutions).
Consider the bosonic case f\/irst.
The symplectic structure $\omega$ derived from the Lagrangian is actually not necessarily a~structure on a~global space
of solutions.
Rather it arises as the integral of a~$(d-1)$-form (where $d$ is the spacetime dimension) over a~hypersurface.
Moreover, the dependence of the $(d-1)$-form on the f\/ield is completely local so that the symplectic structure depends
only on the f\/ield and its derivatives on the hypersurface.
Taking the hypersurface to be~$\Sigma$, it is thus naturally a~structure on~$L_\Sigma$ and we denote it by
$\omega_{\Sigma}$.
The complex structure by contrast is generally somewhat non-local, but can nevertheless be specif\/ied as a~complex
structure on $L_{\Sigma}$.
We denote it thus by $J_{\Sigma}$.
We write the resulting complex inner product as~$\{\cdot,\cdot\}_{\Sigma}$.

Looking more closely at the above description of the space~$L_\Sigma$ and associated structures~$\omega_\Sigma$,~$g_\Sigma$ and $\{\cdot,\cdot\}_\Sigma$ it turns out that the latter not only depend on the choice of hypersurface
$\Sigma$ itself, but also on its \emph{orientation}.
We shall write~$\overline{\Sigma}$ to denote the same hypersurface $\Sigma$, but with opposite orientation.
Now, recall that in the bosonic case the symplectic structure $\omega_\Sigma$ arises as the integral of a~$(d-1)$-form
over the hypersurface~$\Sigma$.
But reversing the orientation changes the sign of this integral.
The complex structure~$J_{\Sigma}$, encoding in some sense the distinction between ``positive energy'' and ``negative
energy'' solutions, also changes sign under change of orientation of~$\Sigma$.
As a~consequence, the real inner product is invariant under orientation change while the complex inner product is
complex conjugated.
In summary,
\begin{gather}
\omega_{\overline{\Sigma}}=-\omega_{\Sigma},
\qquad
J_{\overline{\Sigma}}=-J_{\Sigma},
\qquad
g_{\overline{\Sigma}}=g_{\Sigma},
\qquad
\{\phi',\phi\}_{\overline{\Sigma}}=\overline{\{\phi',\phi\}_{\Sigma}}.
\label{eq:kinsbo}
\end{gather}

Precisely these ingredients (the spaces $L_\Sigma$ with the given structures) have been taken as the ``kinematical''
part of the data of an axiomatic formalization of classical linear bosonic f\/ield theory in~\cite{Oe:holomorphic}.
Moreover, they have been shown there to integrate nicely into a~functorial quantization scheme which moreover
reproduces and extends basic examples from quantum f\/ield theory.
We consider that this conf\/irms the ``correctness'' of the above schematic description.

In the fermionic case it is the symplectic form $\tilde{\omega}$ that arises as the integral of a~f\/ield-local
$(d-1)$-form on a~hypersurface.
The complex structure $\tilde{J}$ is also completely local in its f\/ield dependence.
So, the real inner product $g$ is the integral of a~f\/ield-local $(d-1)$-form on the hypersurface.
For a~hypersurface $\Sigma$ we denote it by~$g_\Sigma$.
As in the bosonic case the complex structure~$J_{\Sigma}$ introduces a~certain non-locality which is thus inherited by
the symplectic structure~$\omega_{\Sigma}$.
$\{\cdot,\cdot\}_{\Sigma}$ denotes the resulting complex inner product.
The symplectic form $\tilde{\omega}_{\Sigma}$ depends on the orientation of $\Sigma$ as in the bosonic case.
$\tilde{J}_{\Sigma}$ on the other hand does not depend on the orientation while $g_\Sigma$ does, coming from
a~f\/ield-local $(d-1)$-form.
$J_{\Sigma}$ does depend on the orientation as in the bosonic case and for the same reason.
So $\omega_{\Sigma}$ does not depend on the orientation.
In summary,
\begin{gather}
\omega_{\overline{\Sigma}}=\omega_{\Sigma},
\qquad
J_{\overline{\Sigma}}=-J_{\Sigma},
\qquad
g_{\overline{\Sigma}}=-g_{\Sigma},
\qquad
\{\phi',\phi\}_{\overline{\Sigma}}=-\overline{\{\phi',\phi\}_{\Sigma}}.
\label{eq:kinsfer}
\end{gather}
This is in precise analogy to the bosonic case~\eqref{eq:kinsbo} if we think of the dif\/ference between bosons and
fermions as due to the interchange of symmetric with anti-symmetric structures.

A closer look at the relations~\eqref{eq:kinsfer} reveals a~profound and surprising implication.
In contrast to the bosonic case, it cannot be consistently required that all the spaces $L_{\Sigma}$ have
positive-def\/inite inner product.
Indeed, suppose $g_{\Sigma}$ was positive-def\/inite for some hypersurface $\Sigma$.
Then $g_{\overline{\Sigma}}$ would necessarily be negative-def\/inite.
Moreover, if $L_{\Sigma}$ carries an indef\/inite inner product then so does the Fock space $\mathcal{H}_{\Sigma}$
based on it.
We are thus forced to give up the insistence that state spaces should be Hilbert spaces.

While this narrative of the fermionic case appears compelling in view of its tight analogy to the more established
bosonic case, the abandonment of Hilbert spaces calls for the presentation of strong evidence in its favor.
The purpose of the present work is in part to provide just that, by embedding it into a~functorial quantization scheme
that successfully reproduces known quantum f\/ield theories.
We shall also alleviate fears (in Section~\ref{sec:prob}) that this may destroy a~consistent probability interpretation.
Moreover, we will explain why the usual formulation of quantum theory gets along nicely without ``seeing'' these
indef\/inite inner product spaces in fermionic quantum f\/ield theories (see Section~\ref{sec:evol}).

\subsection{Dynamical aspects}

We have so far exclusively concentrated on the ``kinematical'' aspects of the classical theory and its quantization.
We proceed to discuss aspects of the dynamics.
For linear bosonic f\/ield theory this was formalized in~\cite{Oe:holomorphic} as follows.
Given a~spacetime region $M$, the ``dynamics'' in $M$ is given by a~real vector space $L_M$ of solutions of the
equations of motion.
This is a~subspace of the space $L_{\partial M}$ of (germs of) solutions on the boundary $\partial M$ of $M$.
Indeed, the key is that it is a~\emph{Lagrangian} subspace with respect to the symplectic form $\omega_{\partial M}$ on
$L_{\partial M}$.
This was motivated extensively in~\cite{Oe:holomorphic} from Lagrangian f\/ield theory and conf\/irmed in the sense
that it f\/its precisely into the quantization scheme given there.
Moreover, it also f\/its with the examples of known quantum f\/ield theories considered there.

In view of the analogy between fermions and bosons we should thus expect the dynamics in a~``classical'' fermionic
f\/ield theory to be encoded in a~subspace $L_{M}\subseteq L_{\partial M}$ that is ``Lagrangian'' with respect to the
symmetric bilinear form $g_{\partial M}$.
That is the subspace $L_{M}$ should be a~neutral subspace which is maximal in a~certain sense.
It turns out that the right notion is that of a~\emph{hypermaximal neutral subspace}.

The quantum dynamics for a~region $M$ is encoded in the amplitude map $\rho_M:\mathcal{H}_{\partial M}\to\mathbb{C}$.
This was given for the bosonic theory in~\cite{Oe:holomorphic} in terms of a~certain integral of the holomorphic wave
function on the boundary over the space of solutions in the interior, $L_M$.
It was later shown~\cite{Oe:affine,Oe:feynobs} that this is precisely equivalent to the usual Feynman path integral and
thus ``correct'' from a~quantum f\/ield theory point of view.
This same quantization rule, translated to the Fock space setting is the basis for the quantization of the dynamics in
the bosonic case also here.
(We shall have much more to say about this in Section~\ref{sec:comphol}.) The quantization rule in the fermionic case
is deduced by analogy.
Again, our overall results may be seen to ``conf\/irm'' it.

\section{Ingredients}
\label{sec:ingreds}

\subsection{Krein space}
\label{sec:krein}

Let $V$ be an indef\/inite inner product space with an orthogonal direct sum decomposition $V=V^+\oplus V^-$ such that
$V^+$ is positive def\/inite and $V^-$ is negative def\/inite.
Let $\overline{V^-}$ denote the space $V^-$ with the sign of its inner product inverted.
Then $\overline{V^-}$ is positive def\/inite and $V$ is canonically isomorphic as a~vector space to the positive inner
product space $V^+\oplus\overline{V^-}$.
If $V^+\oplus\overline{V^-}$ is a~Hilbert space then we say that $V$ is a~\emph{Krein space} and equip it with the
Hilbert space topology.

In general there are many ways to decompose a~Krein space into an orthogonal direct sum $V=V^+\oplus V^-$ with the
described properties.
We shall be interested, however, exclusively in Krein spaces that come with a~canonical such decomposition.
In order not to complicate notation, \emph{Krein space} in the following always refers to this strict version of Krein
space.

We call $V^+$ the \emph{positive part} and $V^-$ the \emph{negative part} of $V$.
We also view the decomposition as a~$\mathbb{Z}_2$-grading with the notation
\begin{gather*}
\sig{v}\coloneqq
\begin{cases}
0&\text{if} \ v\in V^+,
\nonumber\\
1&\text{if} \ v\in V^-.
\end{cases}
\end{gather*}
Let $W$ be a~subset of the Krein space $V$.
Then the subspace
\begin{gather*}
W^\perp\coloneqq\{v\in V:\langle v,w\rangle=0 \ \forall\, w\in W\}
\end{gather*}
is called the \emph{orthogonal companion} of~$W$.
A subspace $W$ of a~Krein space~$V$ is called \emph{neutral} if\/f the inner product of each element of~$W$ with itself
vanishes.
This is equivalent to the property that the real part of the inner product vanishes on~$W$.
$W$ is called \emph{maximal neutral} if\/f moreover, $W$ is not a~proper subspace of any neutral subspace of~$V$.
Since the closure of a~neutral subspace is neutral, a~maximal neutral subspace is necessarily closed.
A~subspace~$W$ is called \emph{hypermaximal neutral} if\/f $W=W^\perp$.
This implies in particular that $W$ is maximal neutral.
In case of a~complex Krein space we use in addition the adjective \emph{real} if the subspace in question is a~real
subspace only and its property refers to the real part of the inner product only.

Let $V_1$, $V_2$ be Krein spaces with decompositions $V_1=V_1^+\oplus V_1^-$ and $V_2=V_2^+\oplus V_2^-$.
A subspace $W\subset V_1$ is called \emph{adapted} if\/f $W$ admits a~decomposition as a~direct sum $W=W^+\oplus W^-$
such that $W^+\subseteq V_1^+$ and $W^-\subseteq V_1^-$.
A linear map $a:V_1\to V_2$ is called an \emph{isometry} if\/f $a$ preserves the inner product.
It is called an \emph{adapted isometry} if in addition $a$ respects the decompositions of $V_1$ and $V_2$ by mapping
$V_1^+$ to $V_2^+$ and $V_1^-$ to $V_2^-$.
A linear map $a:V_1\to V_2$ is called a~\emph{anti-isometry} if\/f $a$ reverses the sign of the inner product.
It is called an \emph{adapted anti-isometry} if in addition $a$ respects the decompositions of $V_1$ and $V_2$ by
mapping $V_1^+$ to $V_2^-$ and $V_1^-$ to $V_2^+$.
Note that an adapted isometry and an adapted anti-isometry are both isometries for the canonical Hilbert space
structures of $V_1$ and $V_2$.
In particular, they are continuous.
In case of a~complex Krein space we use in addition the adjective \emph{real} if the map in question is only real
linear and is an isometry or anti-isometry only with respect to the real part of the inner product.
\begin{lem}
\label{lem:hnsunit}
Let $V=V^+\oplus V^-$ be a~real Krein space.
Then there is a~natural one-to-one correspondence between involutive adapted anti-isometries $u:V\to V$ and
hypermaximal neutral subspaces $W\subseteq V$.
Moreover, this correspondence is such that $W$ is the fixed point set of $u$.
\end{lem}

\begin{proof}
Let $u$ be an involutive adapted anti-isometry $u:V\to V$.
Def\/ine $W$ to be the f\/ixed point set of $u$.
Then $W$ is obviously a~closed subspace of $V$.
Moreover, we must have
\begin{gather*}
\langle v,v\rangle=\langle u(v),u(v)\rangle=-\langle v,v\rangle
\qquad
\forall\, v\in W.
\end{gather*}
That is, $W$ is neutral.
Since we are in the real case this implies $W\subseteq W^\perp$.
Now let $v\in V\setminus W$.
This implies $v-u(v)\neq0$.
By the non-degeneracy of $V$ there is $w\in V$ such that
\begin{gather*}
0\neq\langle w,v-u(v)\rangle=\langle w,v\rangle-\langle w,u(v)\rangle=\langle w,v\rangle+\langle u(w),v\rangle=
\langle w+u(w),v\rangle.
\end{gather*}
Since $w+u(w)\in W$ this implies $v\notin W^\perp$.
Hence $W^\perp\subseteq W$.
Combining this with the above result yields $W=W^\perp$, i.e., $W$ is hypermaximal neutral.

Conversely, suppose that $W$ is a~hypermaximal neutral subspace of $V$.
Let $M\subseteq V^+$ be the subset of vectors $v\in V^+$ such that there exists $w\in V^-$ with $v+w\in W$.
It is easy to see that $M$ is a~subspace of $V^+$ and that it is closed (due to $W$ being closed).
In particular, $M$~has an orthogonal complement $N$ in the Hilbert space $V^+$.
Let $n\in N$ and $v\in W$.
Decompose $v=v^++v^-$ with $v^+\in V^+$ and $v^-\in V^-$.
Then $v^+\in M$, so $n$ is orthogonal both to $v^+$ and to $v^-$.
In particular, $n$ is orthogonal to $v$.
We see that any element of $N$ is orthogonal to any element of $W$.
Since $W$ is hypermaximal neutral this implies $N\subseteq W$.
On the other hand $V^+\cap W=\{0\}$ so we must have $N=\{0\}$.
This implies in turn $M=V^+$.

On the other hand given $v\in V^+$ let $w,w'\in V^-$ be such that $v+w$ and $v+w'$ are elements of $W$.
Thus we have both $w-w'\in W$ and $w-w'\in V^-$.
But $W\cap V^-=\{0\}$, so $w=w'$.
Suppose now there was an involutive adapted anti-isometry $u:V\to V$ with f\/ixed point set $W$.
Then by adaptedness we would need to have $u(v)\in V^-$ while also having $v+u(v)\in W$ by involutiveness.
Thus, we would necessarily have $u(v)=w$.
We take this here as the def\/inition of a~map $u:V^+\to V^-$ and def\/ine analogously $u:V^-\to V^+$.
It is easy to see that the so def\/ined map $u:V\to V$ is indeed adapted and involutive.
To see that it is an anti-isomorphism take $v,w\in V$ with canonical decompositions $v=v^++v^-$ and $w=w^++w^-$.
Then
\begin{gather*}
\langle u\big(v^++v^-\big),u(w^++w^-)\rangle=\langle u(v^+),u(w^+)\rangle+\langle u(v^-),u(w^-)\rangle
\\
\qquad{}=\langle v^++u(v^+),w^++u(w^+)\rangle-\langle v^+,w^+\rangle+\langle v^-+u(v^-),w^-+u(w^-)\rangle-\langle v^-,w^-\rangle
\\
\qquad{}=-\langle v^+,w^+\rangle-\langle v^-,w^-\rangle=-\langle v^++v^-,w^++w^-\rangle.
\end{gather*}
This completes the proof.
\end{proof}
\begin{rem}
The lemma shows in particular, that not every real Krein space admits a~hypermaximal neutral subspace.
Indeed, the existence of such a~subspace is equivalent to the existence of an involutive adapted anti-isometry.
The latter is equivalent to an anti-isometry between the positive and negative parts of the Krein space.
Taking the Hilbert space inner product this is just an ordinary isometry between Hilbert spaces.
Such an isometry thus exists precisely when both parts have the same cardinality.
\end{rem}

Let $V$ be a~complex Krein space.
Then $V$ is also a~real Krein space by forgetting the imaginary part of the inner product.
Suppose $V$ has a~real hypermaximal neutral subspace $W$.
Denote the complex structure of $V$ by $J:V\to V$.
We say that $J$ and $W$ are \emph{compatible} if\/f the induced involutive adapted real anti-isometry $u:V\to V$ is
conjugate linear, i.e., $u\circ J=-J\circ u$.
\begin{lem}
\label{lem:hnsdec}
Let $V$ be a~complex Krein space with a~compatible real hypermaximal neutral subspace $W$.
Then $V$ as a~real vector space decomposes as a~direct sum $V=W\oplus JW$.
Moreover, $JW$~is the eigenspace of the induced adapted real anti-isometry $u:V\to V$ with eigenvalue~$-1$.
Also, $JW$~is a~real hypermaximal neutral subspace of $V$.
Furthermore, $W$~and~$JW$ are symplectic complements with respect to the symplectic form given by the imaginary part of
the inner product of~$V$.
\end{lem}
We leave the straightforward proof to the reader.
\begin{lem}
\label{lem:lagdec}
Let $V$ be a~complex Krein space and $W$ be a~closed Lagrangian subspace with respect to the imaginary part of the
inner product.
Then $JW$~is also a~closed Lagrangian subspace and~$V$ as a~real Krein space has the orthogonal direct sum
decomposition $V=W\oplus JW$.
Moreover, the real linear map $u:V\to V$ defined as the identity on~$W$ and minus the identity on~$JW$ is a~conjugate
linear involutive real isometry of~$V$.
Also, if $W$ is adapted then so are~$JW$ and~$u$.
\end{lem}

This lemma is essentially a~straightforward generalization of Lemma~4.1 in~\cite{Oe:holomorphic} to the Krein space
case.
We leave the proof to the reader.
In the following it will be useful to view the structures appearing in Lemmas~\ref{lem:hnsunit} and~\ref{lem:hnsdec} on
the one hand and in Lemma~\ref{lem:lagdec} on the other hand from a~common perspective.
To this end let $\kappa=-1$ in the former and $\kappa=1$ in the latter case.
The map $u$ is then a~conjugate linear involution with the property
\begin{gather}
\langle u(v),u(v')\rangle=\kappa\,\overline{\langle v,v'\rangle}.
\label{eq:ipu}
\end{gather}
Moreover the map $u$ has precisely the structure of a~\emph{complex conjugation} or \emph{real structure} on the Krein
space $V$.

Let $V$ be a~Krein space with decomposition $V=V^+\oplus V^-$.
We say that a~subset $B\subseteq V$ is an \emph{orthonormal basis} (short: ON-basis) of $V$ if\/f $B$ is the union of
subsets $B^+\subseteq V^+$ and $B^-\subseteq V^-$ such that $B^+$ is an ON-basis of $V^+$ as a~Hilbert space and $B^-$
is an ON-basis of $\overline{V^-}$.

\subsection{Fock space}
\label{sec:fock}

We recall in the following some elementary facts about Fock space and introduce our notation.
It will be convenient to treat both the bosonic and the fermionic case at once.
To this end we introduce the constant $\kappa$, def\/ined as 
\begin{gather*}
\kappa\coloneqq1
\quad
\text{in the bosonic case},
\qquad
\kappa\coloneqq-1
\quad
\text{in the fermionic case}.
\end{gather*}
The fermionic or bosonic Fock space $\mathcal{F}$ is a~Krein space which arises as the completion of an
$\mathbb{N}_0$-graded Krein space,
\begin{gather*}
\mathcal{F}=\widehat{\bigoplus_{n=0}^\infty}\,\mathcal{F}_n.
\end{gather*}
To denote the degree we write
\begin{gather*}
\fdg{\psi}\coloneqq n
\qquad
\text{if}
\quad
\psi\in\mathcal{F}_n.
\end{gather*}
We refer to this grading as the \emph{Fock grading}.
Another important grading is the \emph{fermionic grading} or \emph{f-grading}.
This is a~$\mathbb{Z}_2$-grading def\/ined as follows.
All elements of a~bosonic Fock space are assigned degree zero.
Elements of a~fermionic Fock space have degree zero if their Fock degree is even and degree one if their Fock degree is
odd.

The space $\mathcal{F}_0$ is the canonical one-dimensional Hilbert space over $\mathbb{C}$.
We write it also as $\mathcal{F}_0=\mathbf{1} \mathbb{C}$ with $\langle\mathbf{1},\mathbf{1}\rangle=1$.
If $L$ is the Krein space that \emph{generates} $\mathcal{F}$, then $\mathcal{F}_n$ is def\/ined to be the space of
continuous $n$-linear maps from the $n$-fold direct product of $L$ with itself to $\mathbb{C}$ that are either
symmetric (bosonic case) or anti-symmetric (fermionic case).
Thus
\begin{gather*}
\mathcal{F}_n\coloneqq
\big\{\psi:L\times\cdots\times L\to\mathbb{C},\ n\text{-linear continuous}:\psi\circ\sigma=
\kappa^{|\sigma|}\psi, \ \forall\, \sigma\in S^n\big\}.
\end{gather*}
Here $S^n$ denotes the symmetric group acting on $L\times\cdots\times L$, while $|\sigma|$ equals $0$ or $1$ depending
on whether $\sigma\in S^n$ is even or odd.

Let $\xi_1,\dots,\xi_n\in L$.
We def\/ine a~corresponding element $\cs{\xi_1,\dots,\xi_n}\in\mathcal{F}_n$ as follows
\begin{gather*}
\cs{\xi_1,\dots,\xi_n}(\eta_1,\dots,\eta_n)\coloneqq\frac{1}{n!}\sum_{\sigma\in S^n}\kappa^{|\sigma|}\prod_{i=1}^n\{\xi_i,\eta_{\sigma(i)}\}.
\end{gather*}
Here, $\{\cdot,\cdot\}$ denotes the inner product in $L$.
Note that this expression is conjugate linear in the variables $\xi_1,\dots,\xi_n$.
We def\/ine the inner product between these elements in $\mathcal{F}_n$ as
\begin{gather}
\langle\cs{\eta_1,\dots,\eta_n},\cs{\xi_1,\dots,\xi_n}\rangle\coloneqq2^n\sum_{\sigma\in S^n}\kappa^{|\sigma|}\prod_{i=
1}^n\{\xi_i,\eta_{\sigma(i)}\}.
\label{eq:ipfs}
\end{gather}
This makes $\mathcal{F}_n$ into a~Krein space with the subspace generated by elements of the form
$\psi[\xi_1,\dots$, $\xi_n]$ dense.
We shall refer to these special states as \emph{generating states}.

Let $L=L^+\oplus L^-$ be the canonical decomposition of $L$ as a~Krein space.
We def\/ine subspaces of $\mathcal{F}_n$,
\begin{gather*}
\mathcal{F}_n^+\coloneqq\{\psi\in\mathcal{F}_n:\psi(\xi_1,\dots,\xi_n)=
0~\text{if}~\sig{\xi_1}+\cdots+\sig{\xi_n}~\text{odd}\},
\\
\mathcal{F}_n^-\coloneqq\{\psi\in\mathcal{F}_n:\psi(\xi_1,\dots,\xi_n)=
0~\text{if}~\sig{\xi_1}+\cdots+\sig{\xi_n}~\text{even}\}.
\end{gather*}
Then $\mathcal{F}_n^+$ is a~complete positive-def\/inite inner product space (a~Hilbert space) while $\mathcal{F}_n^-$
is a~complete negative-def\/inite inner product space.
Moreover, $\mathcal{F}_n$ is the orthogonal direct sum $\mathcal{F}_n=\mathcal{F}_n^+\oplus\mathcal{F}_n^-$.
This yields the Krein space structure of $\mathcal{F}_n$.
Note also that $\mathcal{F}_0^+=\mathcal{F}_0$ while $\mathcal{F}_0^-=\{0\}$.
The induced canonical decomposition of $\mathcal{F}$ as a~Krein space we denote by $\mathcal{F}^+\oplus\mathcal{F}^-$.

\begin{lem}
\label{lem:fockaisom}
Let $L$ be a~Krein space.
Suppose that $u:L\to L$ is a~conjugate linear involutive adapted real isometry in the bosonic case or anti-isometry in
the fermionic case.
Define $U_n:\mathcal{F}_n\to\mathcal{F}_n$ by
\begin{gather}
(U_n\psi)(\xi_1,\dots,\xi_n)\coloneqq\overline{\psi(u(\xi_n),\dots,u(\xi_1))}.
\label{eq:deffaisom}
\end{gather}
Then $U_n$ is a~conjugate linear involutive adapted real anti-isometry in the fermionic case if $n$ is odd and an
isometry otherwise.
Combining the maps $U_n$ to a~map $U:\mathcal{F}\to\mathcal{F}$ yields a~conjugate linear involutive adapted real
$f$-graded isometry in the sense,
\begin{gather*}
\langle U\psi',U\psi\rangle=\kappa^{\fdg{\psi}}\overline{\langle\psi',\psi\rangle}.
\end{gather*}
\end{lem}
We leave the straightforward proof to the reader.
It also straightforward to check the action of $U$ on generating states,
\begin{gather}
U\cs{\xi_1,\dots,\xi_n}=\kappa^n\cs{u(\xi_n),\dots,u(\xi_1)}
\qquad
\forall\, \xi_1,\dots,\xi_n\in L.
\label{eq:faisomg}
\end{gather}
Note also that $U$ has the structure of a~\emph{complex conjugation} or \emph{real structure} for the Fock space~$\mathcal{F}$.

Let $L$ be a~Krein space that arises as the orthogonal direct sum of Krein spaces $L=L_1\oplus L_2$.
Let $m,n\in\mathbb{N}_0$.
We write elements in $L$ as pairs $(\eta,\xi)$ with $\eta\in L_1$ and $\xi\in L_2$.
Then we have an isometric morphism of Krein spaces
$\mathcal{F}_{m}(L_1)\ctens\mathcal{F}_{n}(L_2)\to\mathcal{F}_{m+n}(L)$ given by $\psi_1\tens\psi_2\mapsto\psi$, where
\begin{gather}
\psi((\eta_1,\xi_1),\dots,(\eta_{m+n},\xi_{m+n}))
\nonumber\\
\qquad
\coloneqq\frac{1}{(m+n)!}\sum_{\sigma\in S^{m+n}}\kappa^{|\sigma|}\psi_1(\eta_{\sigma(1)},\dots,\eta_{\sigma(m)})\psi_2(\xi_{\sigma(m+1)},\dots,\xi_{\sigma(m+n)}).
\label{eq:isomtpf}
\end{gather}
For generating states this takes the form
\begin{gather}
\cs{\eta_1,\dots,\eta_m}\tens\cs{\xi_1,\dots,\xi_n}\mapsto\cs{(\eta_1,0),\dots,(\eta_m,0),(0,\xi_1),\dots,(0,\xi_n)}.
\label{eq:isomtpfg}
\end{gather}
Extending over all degrees we get an isometric isomorphism of Krein spaces
$\mathcal{F}(L_1)\ctens\mathcal{F}(L_2)\to\mathcal{F}(L)$, which is additive in the degree.
Moreover, this isometric isomorphism induces the natural isometric isomorphism
$\mathcal{F}(L_1)\ctens\mathcal{F}(L_2)\to\mathcal{F}(L_2)\ctens\mathcal{F}(L_1)$ given by
\begin{gather}
\psi_1\tens\psi_2\mapsto\kappa^{\fdg{\psi_1}\cdot\fdg{\psi_2}}\psi_2\tens\psi_1.
\label{eq:tpst}
\end{gather}

Let $L$ be a~non-trivial separable Krein space.
Let $N=\{1,\dots,\dim L\}$ if $L$ is f\/inite-dimensional and $N=\mathbb{N}$ otherwise.
Let $\{\xi_i\}_{i\in N}$ be an ON-basis of $L$.
Then an ON-basis of $L$ is given in the fermionic case by
\begin{gather}
\left\{\frac{1}{\sqrt{2^m}}\cs{\xi_{a_1},\dots,\xi_{a_m}}\right\}_{m\in\mathbb{N}_0,\,a_1<\cdots<a_m\in N},
\label{eq:fbfermi}
\end{gather}
and in the bosonic case by
\begin{gather}
\left\{\frac{1}{\sqrt{2^m K_{a_1,\dots,a_m}}}\cs{\xi_{a_1},\dots,\xi_{a_m}}\right\}_{m\in\mathbb{N}_0,\, a_1\le\cdots\le a_m\in N}.
\label{eq:fbbose}
\end{gather}
Here $K_{a_1,\dots,a_m}$ denotes the number of ways the indices $a_1,\dots,a_m$ can be permuted without changing the
value of the $m$-tuple $(a_1,\dots,a_m)$.
In particular, $K_{a_1,\dots,a_m}$ dif\/fers from~$1$ only if there are coincidences between some of the indices
$a_1,\dots,a_m$.
We also observe that the bosonic Fock space over~$L$ is always countably inf\/inite-dimensional while the fermionic
Fock space over~$L$ is countably inf\/inite-dimensional only if~$L$ is.
If~$L$ has dimension~$d$, then the fermionic Fock space over~$L$ has dimension~$2^d$.

\section{Classical data}
\label{sec:classdata}

In this section we provide an axiomatic description of a~linear classical f\/ield theory, either fermionic or bosonic.
Here the attribute ``classical'' has to be taken with a~grain of salt as we already include certain data that belongs
more properly into the quantum realm, notably complex structures.

\subsection{Geometric data}
\label{sec:geomax}

We recall brief\/ly the formalization of the notion of spacetime in terms of a~\emph{spacetime system} in the GBF.
The presentation here is a~ref\/ined version of previous presentations such as the one in~\cite{Oe:holomorphic}.

There is a~f\/ixed positive integer $d\in\mathbb{N}$, the \emph{dimension} of spacetime.
We are given a~collection~$\mathcal{M}_0^{\textrm{c}}$ of connected oriented topological manifolds of dimension $d$,
possibly with boundary, that we call connected \emph{regular regions}.
Furthermore, there is a~collection $\mathcal{M}_1^{\mathrm{c}}$ of connected orien\-ted topological manifolds without
boundary of dimension~$d-1$ that we call \emph{hypersurfaces}.
The manifolds are either abstract manifolds or they are all concrete submanifolds of a~given f\/ixed \emph{spacetime
manifold}.
In the former case we call the spacetime system \emph{local}, in the latter we call it \emph{global}.

There is an operation of union both for regular regions and for hypersurfaces.
This leads to the collection $\mathcal{M}_0$, of all formal f\/inite unions of elements of
$\mathcal{M}_0^{\mathrm{c}}$, and to the collection $\mathcal{M}_1$, of all formal f\/inite unions of elements of
$\mathcal{M}_1^{\mathrm{c}}$.
In case the spacetime system is global, only unions with members who are disjoint are allowed in $\mathcal{M}_1$ and
only unions with members whose interiors are disjoint are allowed in $\mathcal{M}_0$.
For simplicity, and to capture the intuitive meaning of these unions we use the term \emph{disjoint unions} uniformly.
Note that in the global case the elements of $\mathcal{M}_1$ are actual submanifolds of the spacetime manifold.
This is not the case for all elements of $\mathcal{M}_0$ as overlaps on boundaries may occur.

The collection $\mathcal{M}_1$ has to be closed under orientation reversal.
That is, given $\Sigma\in\mathcal{M}_1$, the hypersurface $\overline{\Sigma}$ which is identical to $\Sigma$, but with
reversed orientation is also in $\mathcal{M}_1$.
Also, any boundary of a~regular region in $\mathcal{M}_0$ is a~hypersurface in $\mathcal{M}_1$.
That is, taking the boundary def\/ines a~map $\partial:\mathcal{M}_0\to\mathcal{M}_1$.
When we want to emphasize explicitly that a~given manifold is in one of those collections we also use the attribute
\emph{admissible}.

It is convenient to also introduce the concept of \emph{slice regions}\footnote{In previous papers slice regions were
called ``empty regions''.
We modify our terminology here in order to make it more descriptive and at the same time avoid possible confusion with
the empty set.}. A slice region is topologically simply a~hypersurface, but thought of as an inf\/initesimally thin
region.
Concretely, the slice region associated to a~hypersurface $\Sigma$ will be denoted by $\hat{\Sigma}$ and its boundary
is def\/ined to be the disjoint union $\partial\hat{\Sigma}=\Sigma\cup\overline{\Sigma}$.
There is one slice region for each hypersurface (forgetting its orientation).
The reason for the terminology is that slice regions can be treated in certain respects in the same way as regular
regions.
Thus, we refer to regular regions and slice regions collectively as \emph{regions}.

There is also a~notion of \emph{gluing} of regions.
Suppose we are given a~region $M$ with its boundary a~disjoint union $\partial
M=\Sigma_1\cup\Sigma\cup\overline{\Sigma'}$, where $\Sigma'$ is a~copy of $\Sigma$
($\Sigma_1$ may be empty). Then we may obtain a~new region $M_1$ by gluing $M$ to itself along
$\Sigma$, $\overline{\Sigma'}$.
That is, we identify the points of $\Sigma$ with corresponding points of~$\Sigma'$ to obtain~$M_1$.
The resulting region~$M_1$ might be inadmissible, in which case the gluing is not allowed.

Depending on the theory one wants to model, the manifolds may carry additional structure such as for example
a~dif\/ferentiable structure or a~metric.
This has to be taken into account in the gluing and will modify the procedure as well as its admissibility in the
f\/irst place.
Our description above is merely meant as a~minimal one.
Moreover, there might be important information present in dif\/ferent ways of identifying the boundary hypersurfaces
that are glued.
Such a~case can be incorporated into our present setting by encoding this information explicitly through suitable
additional structure on the manifolds.

\subsection{Axioms of the classical theory}
\label{sec:classax}

Given a~spacetime system, we axiomatize a~linear classical theory on the spacetime system as follows.
As previously, we set $\kappa=-1$ in the fermionic case and $\kappa=1$ in the bosonic case.
In the bosonic case these axioms coincide with those previously proposed in~\cite{Oe:holomorphic} up to the small
dif\/ference that we allow for Krein spaces here rather than only for Hilbert spaces.
\begin{description}\itemsep=0pt
\item[\rm (C1)] Associated to each hypersurface $\Sigma$ is a~complex separable Krein space $L_\Sigma$ with indef\/inite
inner product denoted by $\{\cdot,\cdot\}_\Sigma$.
We also def\/ine $g_\Sigma(\cdot,\cdot)\coloneqq\Re\{\cdot,\cdot\}_\Sigma$ and
$\omega_\Sigma(\cdot,\cdot)\coloneqq\frac{1}{2}\Im\{\cdot,\cdot\}_\Sigma$ and denote by $J_\Sigma:L_\Sigma\to L_\Sigma$
the scalar multiplication with $\mathrm{i}$ in $L_\Sigma$.
\item[\rm (C2)] Associated to each hypersurface $\Sigma$ there is a~conjugate linear involution $L_\Sigma\to
L_{\overline\Sigma}$, written as an identif\/ication, under which the inner product is transformed as follows
\begin{gather}
\{\phi',\phi\}_{\overline{\Sigma}}=\kappa\,\overline{\{\phi',\phi\}_{\Sigma}}
\qquad
\forall\, \phi,\phi'\in L_{\Sigma}.
\label{eq:ocip}
\end{gather}
\item[\rm (C3)] Suppose the hypersurface $\Sigma$ decomposes into a~disjoint union of hypersurfaces
$\Sigma=\Sigma_1\cup$ $\cdots\cup\Sigma_n$.
Then there is an isometric isomorphism of complex Krein spaces $L_{\Sigma_1}\oplus\cdots\oplus L_{\Sigma_n} $ $\to
L_\Sigma$.
We will not write this map explicitly, but rather think of it as an identif\/ication.
\item[\rm (C4)] Associated to each region $M$ is a~real vector space $L_M$.
\item[\rm (C5)] Associated to each region $M$ there is a~linear map of real vector spaces $r_M:L_M\to L_{\partial M}$.
In the fermionic case the image $L_{\tilde{M}}$ of $r_M$ is a~real hypermaximal neutral subspace of $L_{\partial M}$,
compatible with the complex structure $J_{\partial M}$.
In the bosonic case the image $L_{\tilde{M}}$ of $r_M$ is a~closed adapted Lagrangian subspace of $L_{\partial M}$.
\item[\rm (C6)] Let $M_1$ and $M_2$ be regions and $M\coloneqq M_1\cup M_2$ be their disjoint union.
Then $L_M$ is the orthogonal direct sum $L_{M}=L_{M_1}\oplus L_{M_2}$.
Moreover, $r_M=r_{M_1}+r_{M_2}$.
\item[\rm (C7)] Let $M$ be a~region with its boundary decomposing as a~disjoint union $\partial
M=\Sigma_1\cup\Sigma\cup\overline{\Sigma'}$, where $\Sigma'$ is a~copy of $\Sigma$.
Let $M_1$ denote the gluing of $M$ to itself along $\Sigma,\overline{\Sigma'}$ and suppose that $M_1$ is a~region.
Note $\partial M_1=\Sigma_1$.
Then there is an injective linear map $r_{M;\Sigma,\overline{\Sigma'}}:L_{M_1}\hookrightarrow L_{M}$ such that
\begin{gather}
L_{M_1}\hookrightarrow L_{M}\rightrightarrows L_\Sigma
\label{eq:xsbdy}
\end{gather}
is an exact sequence.
Here the arrows on the right hand side are compositions of the map $r_M$ with the orthogonal projections of
$L_{\partial M}$ to $L_\Sigma$ and $L_{\overline{\Sigma'}}$ respectively (the latter identif\/ied with $L_\Sigma$).
Moreover, the following diagram commutes, where the bottom arrow is the orthogonal projection,
\begin{gather*}
\xymatrix{
  L_{M_1} \ar[rr]^{r_{M;\Sigma,\overline{\Sigma'}}} \ar[d]_{r_{M_1}} & & L_{M} \ar[d]^{r_{M}}\\
  L_{\partial M_1}  & & L_{\partial M} \ar[ll]}
\end{gather*}
\end{description}

We add the following observations: $g_\Sigma$ is a~real symmetric non-degenerate bilinear form making $L_\Sigma$ into
a~real Krein space.
$\omega_\Sigma$ is a~real anti-symmetric non-degenerate bilinear form making $L_\Sigma$ into a~symplectic vector space.
Also
\begin{gather*}
g_{\overline{\Sigma}}=\kappa\,g_\Sigma,
\qquad
J_{\overline{\Sigma}}=-J_{\Sigma},
\qquad
\omega_{\overline{\Sigma}}=-\kappa\,\omega_{\Sigma}.
\end{gather*}
Moreover, for all $\phi,\phi'\in L_\Sigma$,
\begin{alignat*}{3}
& g_\Sigma(\phi,\phi')=2\omega_\Sigma(\phi,J_\Sigma\phi'),
\qquad &&
\{\phi,\phi'\}_\Sigma=
g_\Sigma(\phi,\phi')+2\mathrm{i}\omega_\Sigma(\phi,\phi'),&
\\
& g_\Sigma(\phi,\phi')=g_\Sigma(J_\Sigma\phi,J_\Sigma\phi'),
\qquad &&
\omega_\Sigma(\phi,\phi')=
\omega_\Sigma(J_\Sigma\phi,J_\Sigma\phi'). &
\end{alignat*}

The structures coming from Lemma~\ref{lem:hnsdec} in the fermionic case or Lemma~\ref{lem:lagdec} in the bosonic case
for a~region $M$ will play an important role in the following.
In the fermionic case the real subspace $L_{\tilde{M}}\subseteq L_{\partial M}$ is a~compatible real hypermaximal
neutral subspace and we have the real direct sum decomposition $L_{\partial M}=L_{\tilde{M}}\oplus J L_{\tilde{M}}$.
Moreover, $L_{\tilde{M}}$ and $J L_{\tilde{M}}$ are both hypermaximal neutral subspaces with respect to~$g_{\partial
M}$ and they are symplectic complements with respect to~$\omega_{\partial M}$.
In the bosonic case the real subspace $L_{\tilde{M}}\subseteq L_{\partial M}$ is a~closed adapted Lagrangian subspace
and we have the real orthogonal direct sum decomposition $L_{\partial M}=L_{\tilde{M}}\oplus J L_{\tilde{M}}$.
Moreover, $L_{\tilde{M}}$~and~$J L_{\tilde{M}}$ are both closed adapted Lagrangian subspaces with respect to~$\omega_{\partial M}$.
In both cases, we denote the associated complex conjugation by $u_M:L_{\partial M}\to L_{\partial M}$.

\subsection{A notion of evolution}
\label{sec:cevol}

For a~region $M$ the physical interpretation of the space~$L_M$ is that of the space of solutions of the equations of
motions inside~$M$.
It is as such naturally a~subspace of the space~$L_{\partial M}$ of germs of solutions on the boundary of~$M$.
It is in this sense that we may think of the spaces~$L_M$ as encoding the ``dynamics'' of a~theory and of the spaces~$L_{\partial M}$ the ``kinematics''.
In order to connect this to a~more traditional view of ``dynamics'' versus ``kinematics'' we make use of the map~$u_M$.
\begin{lem}
\label{lem:uclasseva}
Suppose that $L_2\oplus L_1=L_{\partial M}$ is a~decomposition as an orthogonal direct sum of complex Krein spaces such
that $u_M(L_1)=L_2$.
Then we have the identity
\begin{gather*}
L_{\tilde{M}}=\{(u_M(\phi),\phi):\phi\in L_1\}.
\end{gather*}
\end{lem}
The proof is immediate from the properties of $u_M$ exhibited in Section~\ref{sec:krein}.
This statement admits the following interpretation: $u_M$ (restricted to $L_1$) describes the ``evolution'' of the
``initial data'' encoded in $L_1$ to the ``f\/inal data'' encoded in $L_2$.
Of course, the role of $L_1$ and $L_2$ may be interchanged at will, since $u_M$ is involutive.
Moreover, there may be many such decompositions of $L_{\partial M}$, leading to dif\/ferent such notions of
``evolution''.

Such an algebraic notion of ``evolution'' of classical data may be seen to have a~geometric underpinning when the
underlying decomposition of the space $L_{\partial M}$ arises from a~corresponding decomposition of the boundary
hypersurface $\partial M$.
\begin{prop}
\label{prop:uclassevg}
Suppose that $\partial M$ decomposes as a~disjoint union $\partial M=\overline{\Sigma_2}\cup\Sigma_1$ such that
$u_M(L_{\Sigma_1})=L_{\overline{\Sigma_2}}$.
Then the restriction of $u_M$ to $L_{\Sigma_1}$ viewed as a~map $\tilde{u}_M:L_{\Sigma_1}\to L_{\Sigma_2}$ is
a~complex linear isometric isomorphism of Krein spaces.
\end{prop}
\begin{proof}
Preservation of the inner product follows from the combination of property~\eqref{eq:ipu} of~$u_M$ with
property~\eqref{eq:ocip} associated to hypersurface orientation change according to axiom~(C2).
Complex linearity comes from the fact that both~$u_M$ and the map implementing orientation change are complex conjugate
linear.
\end{proof}
In this context the ``initial data'' $L_{\Sigma_1}$ and the ``f\/inal data'' $L_{\Sigma_2}$ are really localized on
separate hypersurfaces, giving stronger justif\/ication to the attributes ``initial'' and ``f\/inal''.

To make the link with a~traditional notion of dynamics even more pertinent consider the following setting.
Suppose there is a~f\/ixed spacetime manifold with a~metric.
Moreover, suppose the metric is Lorentzian and the spacetime is globally hyperbolic.
Now consider only hypersurfaces that are spacelike (and possibly have some additional nice properties).
Consider regions that are submanifolds of spacetime, bounded by pairs of such hypersurfaces.
Adding f\/inite disjoint unions yields a~spacetime system.
Suppose that a~classical f\/ield theory is given by hyperbolic partial dif\/ferential equations of motion so that all
the admissible connected hypersurfaces are Cauchy.
Supposing the theory is Lagrangian, proceeding roughly along the lines sketched in Sections~\ref{sec:quants}
and~\ref{sec:quantgbf}, we should obtain a~model satisfying the axioms of Section~\ref{sec:classax}.
Crucially, we are assuming here that in addition to the purely classical information, sensible complex structures exist
on the hypersurfaces.
Suppose moreover that the complex structures for the dif\/ferent hypersurfaces are chosen in a~compatible way (e.g., if
they come from a~global complex structure).
We would then have the assumption $u_M(L_{\Sigma_1})=L_{\overline{\Sigma_2}}$ of Proposition~\ref{prop:uclassevg}
satisf\/ied for any connected admissible region $M$ with initial boundary component $\Sigma_1$ and f\/inal boundary
component $\Sigma_2$.
(Here we take all admissible connected hypersurfaces to be oriented as past boundaries of an admissible region lying in
the future.) Then $\tilde{u}_M$ literally is the time-evolution map from the initial data $L_{\Sigma_1}$ on $\Sigma_1$
to the f\/inal data $L_{\Sigma_2}$ on $\Sigma_2$.
For future reference we shall refer to this setting as the \emph{standard globally hyperbolic setting}.

So far we have treated here the bosonic and fermionic case on the same footing.
However, the axioms of Section~\ref{sec:classax} imply a~remarkable asymmetry between the two cases with respect to the
(algebraic) notion of ``evolution''.
In contrast to the bosonic case the fermionic case exhibits a~\emph{preferred} decomposition of $L_{\partial M}$ of the
required type, inducing thus a~\emph{preferred} (algebraic) notion of ``evolution''.
This is the decomposition into positive and negative parts $L_{\partial M}=L_{\partial M}^+\oplus L_{\partial M}^-$,
which satisf\/ies naturally $u_M(L_{\partial M}^+)=L_{\partial M}^-$.
Moreover, for standard examples of free fermionic f\/ield theories in globally hyperbolic spacetimes formalized along
the lines sketched in the previous paragraph, this decomposition is precisely the one induced by the geometric
decomposition of $\partial M$ into initial and f\/inal hypersurface.
More formally we have $L_{\partial M}^+=L_{\Sigma_1}$ and $L_{\partial M}^-=L_{\overline{\Sigma_2}}$ if $\partial
M=\Sigma_1\cup\overline{\Sigma_2}$ as above\footnote{We have f\/ixed here an overall choice of sign.
The other choice would interchange $L_{\partial M}^+$ and $L_{\partial M}^-$.}. In Section~\ref{sec:dirac} we shall
exhibit this concretely for the example of the Dirac f\/ield in Minkowski spacetime.
On the other hand, the preferred algebraic notion of ``evolution'' is present even if the spacetime system does not
come equipped with a~notion of time.
Moreover, it behaves very much like time-evolution in that it is directed and compatible with composition of regions
etc.
This suggests to speak of an \emph{emergent notion of time} inherent in fermionic theories satisfying the axioms of
Section~\ref{sec:classax}.

\section{The free Dirac f\/ield in Minkowski spacetime}
\label{sec:dirac}

In this section we review the free Dirac f\/ield in Minkowski spacetime and show explicitly how it f\/its into the
axiomatic system of Section~\ref{sec:classax}.

\subsection{The real inner product}

We consider the free Dirac f\/ield in Minkowski spacetime.
We use the $\gamma$-matrices and other notation of high energy physics such as in~\cite{PeSc:qft}.
The f\/ield $X$ representing the Dirac spinor is a~section of a~trivial vector bundle over Minkowski spacetime with
f\/iber isomorphic to $\mathbb{C}^4$.
The Lagrangian is
\begin{gather*}
L(X)=-\Im\big(X^{\dagger}\gamma^0\gamma^{\mu}\partial_{\mu}X\big)-m X^{\dagger}\gamma^0X.
\end{gather*}
Given a~hypersurface $\Sigma$ the space $L_{\Sigma}$ is essentially the space of spinors restricted to $\Sigma$.
The Lagrangian yields on it the symplectic structure\footnote{We use the sign conventions for the symplectic structure
as in~\cite{Oe:affine}.}
\begin{gather*}
\tilde{\omega}_{\Sigma}(X,Y)=\int_{\Sigma}\Im\big(X^\dagger\gamma^0\gamma^{\mu}Y\big)n_{\mu}\mathrm{d}^3x.
\end{gather*}
Here $\mathrm{d}^3x$ denotes the $3$-form induced from the metric in the hypersurface and $n$ denotes the normal vector
to the hypersurface.
This is to be understood in the sense
\begin{gather*}
n_{\mu}\mathrm{d}^3x\coloneqq\partial_{\mu}\lrcorner\mathrm{d}^4x,
\end{gather*}
where the right-hand side denotes contraction of the partial derivative $\partial_{\mu}$ as a~vector f\/ield with the
volume $4$-form $\mathrm{d}^4x$ of Minkowski spacetime.
To f\/ix signs on the left-hand side we introduce the additional convention that the integral over $\mathrm{d}^3x$ is
locally always positive so that the orientation information of the hypersurface resides exclusively in the sign of~$n$.

Using the complex structure of the f\/iber, $\tilde{J}$ is simply the natural multiplication with $\mathrm{i}$ of the
spinor $X$ understood as a~complex vector f\/ield.
This yields the symmetric real bilinear form on~$L_{\Sigma}$,
\begin{gather*}
g_{\Sigma}(X,Y)=2\tilde{\omega}_{\Sigma}(X,\mathrm{i}Y)=
2\int_{\Sigma}\Re\big(X^\dagger\gamma^0\gamma^{\mu}Y\big)n_{\mu}\mathrm{d}^3x.
\end{gather*}
In order to analyze its properties it will be convenient to rewrite it as follows
\begin{gather}
g_{\Sigma}(X,Y)=2\int_{\Sigma}\Re\big(X^\dagger P Y\big)\mathrm{d}^3x,
\label{eq:ripp}
\end{gather}
where
\begin{gather*}
P(x)\coloneqq\gamma^0\gamma^\mu n_{\mu}(x)
\end{gather*}
is a~complex $4\times4$-matrix valued function on $\Sigma$.
We note that $P(x)$ is self-adjoint, since $\gamma^0\gamma^\mu$ is self-adjoint and $n_{\mu}(x)$ is real.
In particular, we may decompose $P(x)$ as $P(x)=P^+(x)+P^-(x)$, where $P^+(x)$ is positive and $P^-(x)$ is negative.
If $P(x)$ is non-degenerate the decomposition is unique with $P^+(x)$ strictly positive and $P^-(x)$ strictly negative.
As we will see this is the case if $n$ is not light-like.

For simplicity we shall only consider the cases that $\Sigma$ is either completely spacelike or completely timelike.
We start with the spacelike case.
If $\Sigma$ is in particular an equal-time hypersurface we have
\begin{gather}
n(x)=(1,0,0,0)
\qquad
\text{or}
\qquad
n(x)=(-1,0,0,0)
\label{eq:nvecta}
\end{gather}
depending on the orientation of $\Sigma$.
Since $\gamma^0\gamma^0=\mathbf{1}$ this implies $P(x)=\mathbf{1}$ in the f\/irst case and $P(x)=-\mathbf{1}$ in the
second\footnote{We use the sign convention $(1,-1,-1,-1)$ for the metric.}. With the positivity of the integral this
means that the real inner product $g_{\Sigma}$ given by~\eqref{eq:ripp} is either positive-def\/inite or
negative-def\/inite.

Suppose now that $\Sigma$ is an arbitrary spacelike hypersurface in Minkowski spacetime.
Picking a~point $x\in\Sigma$ we can always use a~Lorentz transformation locally to have $n(x)$ aligned with the time
axis (i.e.\  of the form~\eqref{eq:nvecta}).
Since the rank of $P(x)$ must change continuously under Lorentz transformations, but is a~discrete quantity, it remains
constant.
So $P(x)$ is strictly positive or strictly negative.
Moreover, $P(x)$ must have the same sign for all $x\in\Sigma$.
Thus~$g_{\Sigma}$ is positive-def\/inite or negative def\/inite depending on the orientation of $\Sigma$.
For spacelike hypersurfaces there is a~global notion of orientation since we may talk about the normal vector $n(x)$
being future or past pointing.
In particular, if we orient them all in the same way the inner products $g_{\Sigma}$ will be all positive-def\/inite or
all negative-def\/inite.
Since the inner product associated to spacelike hypersurfaces with a~f\/ixed orientation is the starting point for
conventional quantization prescriptions, it is natural there to consider only positive-def\/inite inner products.

We proceed to consider the case that the hypersurface $\Sigma$ is timelike.
Proceeding in a~manner similar to that of the spacelike case it will be suf\/f\/icient for our purposes to consider
just one special normal vector.
All others are related to this by a~Lorentz transformation, due to their spacelike nature.
We choose
\begin{gather*}
n(x)=(0,0,0,1).
\end{gather*}
With either the standard or the chiral representation of the $\gamma$-matrices~\cite{PeSc:qft} this yields
\begin{gather}
P(x)=-\gamma^0\gamma^3=
\begin{pmatrix}
1&0&0&0
\nonumber\\
0&-1&0&0
\nonumber\\
0&0&-1&0
\nonumber\\
0&0&0&1
\end{pmatrix}
.
\label{eq:ptl}
\end{gather}
We read of\/f immediately that $P(x)$ has one positive eigenvalue,~$1$, and one negative eigenvalue,~$-1$, each with an
associated two-dimensional eigenspace.
By the same argument as in the spacelike case we may conclude that a~decomposition of~$P(x)$ into a~positive and
a~negative part $P(x)=P^+(x)+P^-(x)$, each of rank~$2$, holds everywhere on the hypersurface.
Moreover, it holds for any timelike hypersurface.
(Note the symmetry of the statement under orientation reversal.)

The decomposition of $P$ induces a~corresponding decomposition of the space $L_{\Sigma}$ of sections of the spinor
bundle on $\Sigma$ induced by the eigenspace decomposition of each f\/iber.
That is, we have the orthogonal decomposition $L_{\Sigma}=L_{\Sigma}^+\oplus L_{\Sigma}^-$, where $L_{\Sigma}^+$ is
positive-def\/inite and $L_{\Sigma}^-$ is negative-def\/inite.

\subsection{Plane waves}

As already mentioned, the complex structure is in general not local in the f\/ield, making its discussion in the
abstract setting we have used so far dif\/f\/icult.
It is convenient, and for our motivational purposes here suf\/f\/icient, to consider a~global space of solutions of the
Dirac equation in Minkowski spacetime in terms of plane wave spinors.
Moreover, we shall consider only two special choices of hypersurface, one spacelike and one timelike.
In the treatment of spinors we follow mostly text book style conventions, in particular those of~\cite{PeSc:qft}.

Recall the Dirac equation
\begin{gather*}
\big(\mathrm{i}\gamma^{\mu}\partial_{\mu}-m\big)X=0.
\end{gather*}
We expand its solutions in Minkowski spacetime in terms of plane waves in the standard way as
\begin{gather}
X(t,x)=\int\frac{\mathrm{d}^3k}{(2\pi)^32E}\sum_{s=
1,2}\left(X_a^s(k)u^s(k)e^{-\mathrm{i}(Et-kx)}+\overline{X_b^s(k)}v^s(k)e^{\mathrm{i}(Et-kx)}\right).
\label{eq:dpwparam}
\end{gather}
Here $u^s$ and $v^s$ with $s\in\{1,2\}$ are certain spinors in momentum space.
That is, they are complex 4-dimensional vector valued functions on momentum space $\mathbb{R}^3$, forming a~convenient
basis of the solutions of the equations
\begin{gather}
\big(\gamma^\mu k_{\mu}-m\big)u^s(k)=0
\qquad
\text{and}
\qquad
\big(\gamma^\mu k_{\mu}+m\big)v^s(k)=0.
\label{eq:demoms}
\end{gather}
They satisfy moreover the following properties
\begin{alignat}{3}
& {u^r}^{\dagger}(k)\gamma^0\gamma^\mu u^{s}(k)=2k^\mu\delta^{r,s},
\qquad &&
{v^r}^{\dagger}(k)\gamma^0\gamma^\mu v^{s}(k)=2k^\mu\delta^{r,s},&
\label{eq:momsorth1}
\\
& {u^r}^{\dagger}(k)v^s(-k)=0,
\qquad &&
{v^r}^{\dagger}(k)u^s(-k)=0.&
\label{eq:momsorth2}
\end{alignat}

Let $\Sigma$ be an equal-time hypersurface located at time $t$.
We shall denote the corresponding quantities with a~subscript $t$ rather than a~subscript $\Sigma$.
For the space $L_t$ we simply take the global solution space parametrized by~\eqref{eq:dpwparam}.
The inner product $g_{t}$ is given by~\eqref{eq:ripp} with $P(x)=1$.
This yields
\begin{gather*}
g_t(X,Y)=2\int\frac{\mathrm{d}^3k}{(2\pi)^32E}\sum_{s=
1,2}\Re\Bigl(\overline{X_a^s(k)}Y_a^s(k)+X_b^s(k)\overline{Y_b^s(k)}\Bigr).
\end{gather*}

The example of a~timelike hypersurface we shall consider is the hypersurface $\Sigma$ located at a~constant value of
the coordinate $x^3$, which we shall denote by $z$ for simplicity.
It is then convenient to parametrize the plane wave solutions in a~sightly dif\/ferent (but strictly equivalent) way
compared to~\eqref{eq:dpwparam}.
We base the parametrization on Fourier modes on the hypersurface $\Sigma$ which we shall denote $(E,\tilde{k})$ where
the ``energy'' $E$ can be positive or negative while $\tilde{k}=(k_1,k_2)$.
We restrict the integral over modes to $|E|\ge m$, which means that we only consider the propagating wave modes
contained in~\eqref{eq:dpwparam} and exclude evanescent waves,
\begin{gather}
X(t,\tilde{x},z)=\int_{|E|\ge m}\frac{\mathrm{d}E\,\mathrm{d}^2\tilde{k}}{(2\pi)^32k_3}
\label{eq:dpwparamtl}\\
\hphantom{X(t,\tilde{x},z)=}{}\times
\sum_{s=1,2}
\left(\tilde{X}_a^s(E,\tilde{k})\tilde{u}^s(E,\tilde{k})e^{-\mathrm{i}(Et-\tilde{k}\tilde{x}-k_3z)}
+\overline{\tilde{X}_b^s(E,\tilde{k})}\tilde{v}^s(E,\tilde{k})e^{\mathrm{i}(Et-\tilde{k}\tilde{x}-k_3z)}\right).
\nonumber
\end{gather}
Here $\tilde{x}$ is a~shorthand for $\tilde{x}=(x^1,x^2)$ and $k_3=\sqrt{E^2-\tilde{k}^2-m^2}$ is chosen positive.
The momentum space spinors $\tilde{u}$, $\tilde{v}$ are def\/ined in terms of the spinors $u$, $v$ as follows
\begin{gather*}
\tilde{u}^s(E,\tilde{k})\coloneqq
\begin{cases}
u^s(\tilde{k},k_3)&\text{if}~E>0,
\\
v^s(-\tilde{k},-k_3)&\text{if}~E<0,
\end{cases}
\qquad
\tilde{v}^s(E,\tilde{k})\coloneqq
\begin{cases}
v^s(\tilde{k},k_3)&\text{if}~E>0,
\\
u^s(-\tilde{k},-k_3)&\text{if}~E<0.
\end{cases}
\end{gather*}
They satisfy the equations
\begin{gather*}
\big(\gamma^0E-\tilde{\gamma}\tilde{k}-\gamma^3k_3-m\big)\tilde{u}^s(E,\tilde{k})=0,
\qquad
\big(\gamma^0E-\tilde{\gamma}\tilde{k}-\gamma^3k_3+m\big)\tilde{v}^s(E,\tilde{k})=0,
\end{gather*}
analogous to~\eqref{eq:demoms}.
They have the following important properties analogous to~\eqref{eq:momsorth1},~\eqref{eq:momsorth2}:
\begin{gather}
{\tilde{u}^r}{}^{\dagger}(E,\tilde{k})\gamma^0\gamma^3\tilde{u}^{s}(E,\tilde{k})=-2\frac{E}{|E|}k_3\delta^{r,s},
\label{eq:tlsid1}
\\
{\tilde{v}^r}{}^{\dagger}(E,\tilde{k})\gamma^0\gamma^3\tilde{v}^{s}(E,\tilde{k})=-2\frac{E}{|E|}k_3\delta^{r,s},
\label{eq:tlsid2}
\\
{\tilde{u}^r}{}^{\dagger}(E,\tilde{k})\gamma^0\gamma^3\tilde{v}^s(-E,-\tilde{k})=0,
\label{eq:tlsid3}
\\
{\tilde{v}^r}{}^{\dagger}(E,\tilde{k})\gamma^0\gamma^3\tilde{u}^s(-E,-\tilde{k})=0.
\label{eq:tlsid4}
\end{gather}

We denote quantities related to the hypersurface $\Sigma$ of constant $z$ with a~subscript $z$.
The inner product $g_{z}$ is given by~\eqref{eq:ripp} with $P$ as in~\eqref{eq:ptl}.
Using the identities~\eqref{eq:tlsid1}--\eqref{eq:tlsid4}, we obtain
\begin{gather*}
g_{z}(X,Y)=2\int_{|E|\ge m}\frac{\mathrm{d}E\,\mathrm{d}^2\tilde{k}}{(2\pi)^32k_3}\frac{E}{|E|}\sum_{s=
1,2}\Re\left(\overline{\tilde{X}_a^s(E,\tilde{k})}\tilde{Y}_a^s(E,\tilde{k})+\tilde{X}_b^s(E,\tilde{k})\overline{\tilde{Y}_b^s(E,\tilde{k})}\right).
\end{gather*}
We see that the space $L_{z}$ of spinors on the hypersurface splits as a~direct sum $L_{z}=L_{z}^+\oplus L_{z}^-$ into
a~part $L_{z}^+$ with $E>0$ where $g_{z}$ is positive-def\/inite and a~part $L_{z}^-$ where $g_{z}$ is
negative-def\/inite.

It is also instructive to express this inner product in terms of the original parametrization~\eqref{eq:dpwparam} of
the global solution space.
The relation between the two parametrizations is the following
\begin{gather*}
\tilde{X}_a^s(E,\tilde{k})=
\begin{cases}
X_a^s(\tilde{k},k_3)&\text{if}~E>0,
\\
\overline{X_b^s(-\tilde{k},-k_3)}&\text{if}~E<0,
\end{cases}
\qquad
\tilde{X}_b^s(E,\tilde{k})=
\begin{cases}
X_b^s(\tilde{k},k_3)&\text{if}~E>0,
\\
\overline{X_a^s(-\tilde{k},-k_3)}&\text{if}~E<0.
\end{cases}
\end{gather*}
We obtain
\begin{gather*}
g_{z}(X,Y)=2\int\frac{\mathrm{d}^3k}{(2\pi)^32E}\frac{k_3}{|k_3|}\sum_{s=
1,2}\Re\left(\overline{X_a^s(k)}Y_a^s(k)+X_b^s(k)\overline{Y_b^s(k)}\right).
\end{gather*}
In this parametrization the subspaces $L_{z}^+$ and $L_{z}^-$ are distinguished by the direction of the momentum
component~$k_3$ that is perpendicular to the hypersurface.

\subsection{The complex structure}
\label{sec:diraccompl}

We proceed to consider the complex structure $J_{\Sigma}$ which, together with the real inner product, gives rise to
the complex inner product on $L_{\Sigma}$.
As already mentioned this encodes the distinction between ``positive energy'' and ``negative energy'' solutions and is
not a~f\/ield-local object.

In the case where the hypersurface $\Sigma$ is a~hypersurface of constant time $t$ the standard complex structure $J_t$
is given by
\begin{gather*}
(J_t X)_a^s(k)=\mathrm{i}X_a^s(k),
\qquad
(J_t X)_b^s(k)=\mathrm{i}X_b^s(k).
\end{gather*}
Looking at the parametrization~\eqref{eq:dpwparam} this means that solutions with a~temporal dependence of the form
$e^{-\mathrm{i}Et}$ are multiplied by $\mathrm{i}$, while solutions with a~temporal dependence of the form
$e^{\mathrm{i}Et}$ are multiplied by $-\mathrm{i}$.
(Here $E>0$ by convention.) The symplectic form according to~\eqref{eq:riptosympl} results to be
\begin{gather}
\omega_t(X,Y)=\int\frac{\mathrm{d}^3k}{(2\pi)^32E}\sum_{s=
1,2}\Im\left(\overline{X_a^s(k)}Y_a^s(k)+\overline{X_b^s(k)}Y_b^s(k)\right).
\label{eq:symplt}
\end{gather}
Combining real inner product and symplectic form according to~\eqref{eq:cipripsympl} yields
\begin{gather}
\{X,Y\}_t=2\int\frac{\mathrm{d}^3k}{(2\pi)^32E}\sum_{s=
1,2}\left(\overline{X_a^s(k)}Y_a^s(k)+\overline{X_b^s(k)}Y_b^s(k)\right).
\label{eq:ctip}
\end{gather}
This is indeed the usual complex inner product for the 1-particle space of the Dirac f\/ield.

We proceed to the timelike case with $\Sigma$ being the hypersurface of constant coordinate value $z=x^3$.
The analogue of ``positive energy'' versus ``negative energy'' solutions is now given by the distinction between
solutions with a~dependence in $z$-direction of the form $e^{\mathrm{i}k_3}$ versus $e^{-\mathrm{i}k_3}$ in the
parametrization~\eqref{eq:dpwparamtl}.
(Here $k_3>0$ by convention.) That is, the complex structure $J_{z}$ is given by
\begin{gather*}
(J_{z}\tilde{X})_a^s(E,\tilde{k})=\mathrm{i}\tilde{X}_a^s(E,\tilde{k}),
\qquad
(J_{z}\tilde{X})_b^s(E,\tilde{k})=\mathrm{i}\tilde{X}_b^s(E,\tilde{k}).
\end{gather*}
With~\eqref{eq:riptosympl} we obtain the symplectic form
\begin{gather*}
\omega_{z}(X,Y)=\int_{|E|\ge m}\frac{\mathrm{d}E\,\mathrm{d}^2\tilde{k}}{(2\pi)^32k_3}\frac{E}{|E|}\sum_{s=
1,2}\Im\left(\overline{\tilde{X}_a^s(E,\tilde{k})}\tilde{Y}_a^s(E,\tilde{k})+\overline{\tilde{X}_b^s(E,\tilde{k})}\tilde{Y}_b^s(E,\tilde{k})\right).
\end{gather*}
Combining real inner product and symplectic form according to~\eqref{eq:cipripsympl} yields the complex inner product
\begin{gather}
\{X,Y\}_{z}=2\int_{|E|\ge m}\frac{\mathrm{d}E\,\mathrm{d}^2\tilde{k}}{(2\pi)^32k_3}\frac{E}{|E|}\sum_{s=
1,2}\left(\overline{\tilde{X}_a^s(E,\tilde{k})}\tilde{Y}_a^s(E,\tilde{k})+\overline{\tilde{X}_b^s(E,\tilde{k})}\tilde{Y}_b^s(E,\tilde{k})\right).
\label{eq:czip}
\end{gather}

Remarkably, expressing the symplectic form $\omega_{z}$ in terms of the standard f\/ield
parametrization~\eqref{eq:dpwparam} yields
\begin{gather*}
\omega_{z}(X,Y)=\int\frac{\mathrm{d}^3k}{(2\pi)^32E}\sum_{s=
1,2}\Im\left(\overline{X_a^s(k)}Y_a^s(k)+\overline{X_b^s(k)}Y_b^s(k)\right),
\end{gather*}
identical to~\eqref{eq:symplt} for $\omega_t$.
This suggests that the symplectic structure here is the same not only for all reasonable spacelike hypersurfaces (which
is known), but also for a~class of timelike hypersurfaces.

\subsection{Dynamics and axiomatic form}

We proceed to integrate the structures derived so far in terms of the axiomatic system of Section~\ref{sec:classax},
including the already implicit dynamics.
We start with the case of constant-time hypersurfaces.
Thus, admissible hypersurfaces are constant-time hypersurfaces and their f\/inite disjoint unions.
Admissible regions are regions bounded by pairs of such hypersurfaces and their f\/inite disjoint
unions\footnote{Recall from Section~\ref{sec:geomax}, that ``disjoint'' only refers to the interior.
An admissible region may well be the formal union of regions that share boundary components.}. For a~time
$t\in\mathbb{R}$ the space $L_{t}$ associated to the hypersurface $\Sigma_t$ of constant time $t$ is a~copy of the
Hilbert space $L$.
This in turn is the $\mathcal{L}^2$-space of equivalence classes of square-integrable functions $X_{a,b}^s(k)$ with the
inner product~\eqref{eq:ctip}.
Here, we take $\Sigma_t$ to have the standard orientation by which we mean that it is oriented as the past boundary of
an admissible region.
The admissible regions all have the same orientation inherited from some global standard orientation of Minkowski space.

This yields the data of axiom (C1) for connected admissible hypersurfaces with standard orientation.
For a~connected admissible hypersurface $\Sigma_t$ with the opposite orientation we def\/ine the associated space
$L_{\overline{t}}$ simply as identical to $L_{t}$ as a~real vector space, but with opposite complex structure and inner
product modif\/ied according to relation~\eqref{eq:ocip}.
Note in particular that the space $L_{\overline{t}}$ is thus negative-def\/inite.
This yields the data of axioms (C1) and (C2) for all connected admissible hypersurfaces.
Next, we def\/ine the space associated to a~f\/inite disjoint union of connected hypersurfaces to be the direct sum of
the spaces associated to the individual hypersurfaces.
This yields the complete data for axioms (C1), (C2) and (C3).

Consider now a~connected admissible region $M$.
$M$ has the structure of a~time-interval $[t_1,t_2]$ times all of space $\mathbb{R}^3$.
We index associated structures with $[t_1,t_2]$.
The space $L_{[t_1,t_2]}$ of solutions associated to this region is a~copy of the global space of solutions $L$,
understood merely as a~real vector space without additional structure.
For admissible regions that are disjoint unions of connected regions the associated real vector space is taken to be
the direct sum of the vector spaces associated to the components.
This yields the data for axiom (C4).

For axiom (C5) it will be suf\/f\/icient to consider a~connected admissible region.
Say the region is determined by the time interval $[t_1,t_2]$.
We def\/ine the map $r_{[t_1,t_2]}:L_{[t_1,t_2]}\to L_{t_1}\oplus L_{\overline{t_2}}$ to be given by
\begin{gather*}
\phi\mapsto(\phi,\phi),
\end{gather*}
recalling that all the involved spaces are copies of $L$.
It is also easy to verify that the image $L_{\widetilde{[t_1,t_2]}}$ of this map is a~real hypermaximal subspace of
$L_{t_1}\oplus L_{\overline{t_2}}$ since the real inner product on the latter is given by
\begin{gather*}
g_{\partial[t_1,t_2]}((\phi_1,\phi_2),(\eta_1,\eta_2))=g_t(\phi_1,\eta_1)-g_t(\phi_2,\eta_2).
\end{gather*}
So if $\phi_1=\phi_2$ and $\eta_1=\eta_2$ the result vanishes while for $\phi_1\neq\phi_2$ we can always f\/ind some
$\eta_1=\eta_2$ giving a~non-vanishing result.
The according to Lemma~\ref{lem:hnsunit} induced involutive anti-isometry~$u_{[t_1,t_2]}$ takes the form
\begin{gather*}
u_{[t_1,t_2]}((\phi,\eta))=(\eta,\phi).
\end{gather*}
In particular, it anti-commutes with the complex structure since the latter takes opposite signs on the hypersurfaces
at $t_1$ and at $t_2$, due to opposite orientation.
Thus, the compatibility condition with the complex structure is satisf\/ied and so is axiom (C5) and also (C6).

With respect to the discussion in Section~\ref{sec:cevol} note that the map $\tilde{u}_{[t_1,t_2]}:L_{t_1}\to L_{t_2}$
as def\/ined in Proposition~\ref{prop:uclassevg} and encoding time-evolution is here simply the identity between
dif\/ferent copies of the same Hilbert space $L$.
In particular, the preferred algebraic notion of evolution discussed in Section~\ref{sec:cevol} coincides with the
time-evolution.
Moreover, given the f\/ixed standard orientation all the connected admissible hypersurfaces $L_t$ are Hilbert spaces.

For axiom (C7) it is suf\/f\/icient to consider the case of a~region $M$ that is the disjoint union of regions
$[t_1,t_2]$ and $[t_2,t_3]$ for $t_1<t_2<t_3$.
Gluing amounts to producing the region $[t_1,t_3]$.
It is then straightforward to check the details of the axiom, which we leave to the reader.
This completes the construction of our model of the axioms encoding the Dirac f\/ield theory in Minkowski space with
constant-time hypersurfaces.

We proceed to consider the setting based on hypersurfaces of constant $z$.
The construction is very much analogous to the one just performed, with the role of the time coordinate $t$ now played
by the space coordinate $z$.
We will therefore limit ourselves to highlighting the dif\/ferences.
We equip the global solution space $L$ to this end with the inner product~\eqref{eq:czip}.
More precisely, we take $L^+$ to be the $\mathcal{L}^2$-space of equivalence classes of square-integrable functions
$\tilde{X}_{a,b}^s(E,\tilde{k})$ with $E>0$ and the inner product~\eqref{eq:czip}.
We take $\overline{L^-}$ to be the $\mathcal{L}^2$-space of equivalence classes of square-integrable functions
$\tilde{X}_{a,b}^s(E,\tilde{k})$ with $E<0$ and the inner product the negative of~\eqref{eq:czip}.
The two Hilbert spaces $L^+$ and $\overline{L^-}$ are then combined to the Krein space $L=L^+\oplus L^-$ with positive
part $L^+$ and negative part $L^-$.
Apart from this dif\/ference the implementation of the axioms (C1)--(C7) is analogous to the setting with constant-time
hypersurfaces.

Concerning the discussion in Section~\ref{sec:cevol} the induced evolution map $\tilde{u}_{[z_1,z_2]}:L_{z_1}\to
L_{z_2}$ from the hypersurface $\Sigma_{z_1}$ to the hypersurface $\Sigma_{z_2}$ is again the identity, this time
between copies of the same Krein space (which is not a~Hilbert space).
However, this notion of spatial evolution is distinct from the preferred algebraic notion of evolution since the
decomposition of the boundary solution space $L_{\partial[z_1,z_2]}$ into positive part $L_{\partial[z_1,z_2]}^+$ and
negative part $L_{\partial[z_1,z_2]}^-$ now does not coincide with its decomposition into the hypersurface solution
spaces $L_{z_1}$ and $L_{\overline{z_2}}$.
Indeed, the preferred algebraic notion of evolution continues to agree with time-evolution.

\section{Axioms of the quantum theory}
\label{sec:gbfax}

\subsection{Core axioms of the GBF}
\label{sec:coreaxioms}

The core axioms of the GBF are generalized here to include fermionic theories in addition to bosonic ones.
Firstly, this is ref\/lected in the appearance of a~$\mathbb{Z}_2$-grading on state spaces.
Secondly, this is manifest in the fact that state spaces may be Krein spaces rather than merely Hilbert spaces.
In the special case that all state spaces are purely of degree $0$ and are moreover Hilbert spaces we recover the
purely bosonic core axioms in the version put forward in~\cite{Oe:holomorphic}.
\begin{description}\itemsep=0pt
\item[\rm (T1)] Associated to each hypersurface $\Sigma$ is a~complex separable $\mathbb{Z}_2$-graded Krein space
$\mathcal{H}_\Sigma$, called the \emph{state space} of $\Sigma$.
We denote its indef\/inite inner product by $\langle\cdot,\cdot\rangle_\Sigma$.
\item[\rm (T1b)] Associated to each hypersurface $\Sigma$ is a~conjugate linear adapted $\mathbb{Z}_2$-graded isometry
$\iota_\Sigma:\mathcal{H}_\Sigma\to\mathcal{H}_{\overline{\Sigma}}$.
This map is an involution in the sense that $\iota_{\overline{\Sigma}}\circ\iota_\Sigma$ is the identity on
$\mathcal{H}_\Sigma$.
\item[\rm (T2)] Suppose the hypersurface $\Sigma$ decomposes into a~disjoint union of hypersurfaces
$\Sigma=\Sigma_1\cup\cdots$ $\cup\Sigma_n$.
Then there is an isometric isomorphism of Krein spaces
$\tau_{\Sigma_1,\dots,\Sigma_n;\Sigma}:\mathcal{H}_{\Sigma_1}\ctens\cdots\ctens\mathcal{H}_{\Sigma_n}$ $\to\mathcal{H}_\Sigma$.
The maps $\tau$ satisfy obvious associativity conditions.
Moreover, in the case $n=2$ the map
$\tau_{\Sigma_2,\Sigma_1;\Sigma}^{-1}\circ\tau_{\Sigma_1,\Sigma_2;\Sigma}:\mathcal{H}_{\Sigma_1}\ctens\mathcal{H}_{\Sigma_2}\to\mathcal{H}_{\Sigma_2}\ctens\mathcal{H}_{\Sigma_1}$ is the $\mathbb{Z}_2$-graded transposition
\begin{gather*}
\psi_1\tens\psi_2\mapsto(-1)^{\fdg{\psi_1}\cdot\fdg{\psi_2}}\psi_2\tens\psi_1.
\end{gather*}
\item[\rm (T2b)] Orientation change and decomposition are compatible in a~$\mathbb{Z}_2$-graded sense.
That is, for a~disjoint decomposition of hypersurfaces $\Sigma=\Sigma_1\cup\Sigma_2$ we have
\begin{gather*}
\tau_{\overline{\Sigma}_1,\overline{\Sigma}_2;\overline{\Sigma}}\left((\iota_{\Sigma_1}\tens\iota_{\Sigma_2})(\psi_1\tens\psi_2)\right)=(-1)^{\fdg{\psi_1}\cdot\fdg{\psi_2}}\iota_\Sigma\left(\tau_{\Sigma_1,\Sigma_2;\Sigma}(\psi_1\tens\psi_2)\right).
\end{gather*}
\item[\rm (T4)] Associated with each region $M$ is a~$\mathbb{Z}_2$-graded linear map from a~dense subspace
$\mathcal{H}_{\partial M}^\circ$ of the state space $\mathcal{H}_{\partial M}$ of its boundary $\partial M$ (which
carries the induced orientation) to the complex numbers, $\rho_M:\mathcal{H}_{\partial M}^\circ\to\mathbb{C}$.
This is called the \emph{amplitude} map.
\item[\rm (T3x)] Let $\Sigma$ be a~hypersurface.
The boundary $\partial\hat{\Sigma}$ of the associated slice region $\hat{\Sigma}$ decomposes into the disjoint union
$\partial\hat{\Sigma}=\overline{\Sigma}\cup\Sigma'$, where $\Sigma'$ denotes a~second copy of $\Sigma$.
Then $\rho_{\hat{\Sigma}}$ is well def\/ined on
$\tau_{\overline{\Sigma},\Sigma';\partial\hat{\Sigma}}(\mathcal{H}_{\overline{\Sigma}}\tens\mathcal{H}_{\Sigma'})\subseteq\mathcal{H}_{\partial\hat{\Sigma}}$.
Moreover, $\rho_{\hat{\Sigma}}\circ\tau_{\overline{\Sigma},\Sigma';\partial\hat{\Sigma}}$ restricts to a~bilinear
pairing $(\cdot,\cdot)_\Sigma:\mathcal{H}_{\overline{\Sigma}}\times\mathcal{H}_{\Sigma'}\to\mathbb{C}$ such that
$\langle\cdot,\cdot\rangle_\Sigma=(\iota_\Sigma(\cdot),\cdot)_\Sigma$.
\item[\rm (T5a)] Let $M_1$ and $M_2$ be regions and $M\coloneqq M_1\cup M_2$ be their disjoint union.
Then $\partial M=\partial M_1\cup\partial M_2$ is also a~disjoint union and $\tau_{\partial M_1,\partial M_2;\partial
M}(\mathcal{H}_{\partial M_1}^\circ\tens\mathcal{H}_{\partial M_2}^\circ)\subseteq\mathcal{H}_{\partial M}^\circ$.
Moreover, for all $\psi_1\in\mathcal{H}_{\partial M_1}^\circ$ and $\psi_2\in\mathcal{H}_{\partial M_2}^\circ$,
\begin{gather*}
\rho_{M}\left(\tau_{\partial M_1,\partial M_2;\partial M}(\psi_1\tens\psi_2)\right)=
\rho_{M_1}(\psi_1)\rho_{M_2}(\psi_2).
\end{gather*}
\item[\rm (T5b)] Let $M$ be a~region with its boundary decomposing as a~disjoint union $\partial
M=\Sigma_1\cup\Sigma\cup\overline{\Sigma'}$, where $\Sigma'$ is a~copy of $\Sigma$.
Let $M_1$ denote the gluing of $M$ with itself along $\Sigma,\overline{\Sigma'}$ and suppose that $M_1$ is a~region.
Note $\partial M_1=\Sigma_1$.
Then $\tau_{\Sigma_1,\Sigma,\overline{\Sigma'};\partial M}(\psi\tens\xi\tens\iota_\Sigma(\xi))\in\mathcal{H}_{\partial
M}^\circ$ for all $\psi\in\mathcal{H}_{\partial M_1}^\circ$ and $\xi\in\mathcal{H}_\Sigma$.
Moreover, for any ON-basis $\{\zeta_i\}_{i\in I}$ of $\mathcal{H}_\Sigma$, we have for all
$\psi\in\mathcal{H}_{\partial M_1}^\circ$,
\begin{gather}
\rho_{M_1}(\psi)\cdot c(M;\Sigma,\overline{\Sigma'})=
\sum_{i\in I}(-1)^{\sig{\zeta_i}}\rho_M\left(\tau_{\Sigma_1,\Sigma,\overline{\Sigma'};\partial M}(\psi\tens\zeta_i\tens\iota_\Sigma(\zeta_i))\right),
\label{eq:glueax1}
\end{gather}
where $c(M;\Sigma,\overline{\Sigma'})\in\mathbb{C}\setminus\{0\}$ is called the \emph{gluing anomaly factor} and
depends only on the geometric data.
\end{description}

\subsection{Vacuum axioms}

In addition to the core axioms we shall impose the vacuum axioms, which we recall in the following.
These formalize the notion of a~vacuum state, suitably adapted to the GBF.
Their present form is a~slight adaption from their form in~\cite[Section~5.1]{Oe:2dqym}\footnote{We recall that the
strange numbering comes from the fact that in the original proposal~\cite{Oe:gbqft} there was an additional axiom (V4)
which was later made redundant due to a~modif\/ication of the core axioms.}.
\begin{description}\itemsep=0pt
\item[\rm (V1)] For each hypersurface $\Sigma$ there is a~distinguished state $\psi_{\Sigma,0}\in\mathcal{H}_\Sigma$ of
degree $0$, called the \emph{vacuum state}.
\item[\rm (V2)] The vacuum state is compatible with the involution.
That is, for any hypersurface $\Sigma$, $\psi_{\bar{\Sigma},0}=\iota_\Sigma(\psi_{\Sigma,0})$.
\item[\rm (V3)] The vacuum state is compatible with decompositions.
Suppose the hypersurface $\Sigma$ decomposes into components $\Sigma_1\cup\dots\cup\Sigma_n$.
Then $\psi_{\Sigma,0}=\tau_{\Sigma_1,\dots,\Sigma_n;\Sigma}(\psi_{\Sigma_1,0}\tens\cdots\tens\psi_{\Sigma_n,0})$.
\item[\rm (V5)] The amplitude of the vacuum state is unity.
That is, for any region $M$, $\rho_M(\psi_{\partial M,0})=1$.
\end{description}

\section{Quantization}
\label{sec:quantization}

We proceed to specify in this section the quantization scheme that yields, starting from classical data satisfying the
axioms of Section~\ref{sec:classdata}, the ingredients of a~corresponding quantum theory.
We shall also show here the validity of some of the core axioms of Section~\ref{sec:coreaxioms} for these data, namely
(T1), (T1b), (T2), (T2b) and (T4).
Moreover, we show that all of the vacuum axioms hold.

\subsection{State spaces}

The state space $\mathcal{H}_{\Sigma}$ associated to a~hypersurface $\Sigma$ is the bosonic or fermionic Fock space
$\mathcal{F}(L_{\Sigma})$ in the sense of Section~\ref{sec:fock}.
The $\mathbb{Z}_2$-grading of $\mathcal{H}_{\Sigma}$ is the f-grading of $\mathcal{F}(L_{\Sigma})$.
This yields (T1).
Moreover, we def\/ine the vacuum state $\psi_{\Sigma,0}$ to be the element $\mathbf{1}$ of the Fock space.
Thus, (V1) is satisf\/ied.

For $n\in\mathbb{N}$ def\/ine the map $\iota_n:\mathcal{F}_n(L_{\Sigma})\to\mathcal{F}_n(L_{\overline{\Sigma}})$ as
follows
\begin{gather*}
(\iota_n(\psi))(\xi_1,\dots,\xi_n)\coloneqq\overline{\psi(\xi_n,\dots,\xi_1)}.
\end{gather*}
$\iota_n$ is a~conjugate-linear adapted real isometry of Krein spaces in the bosonic case and in the fermionic case if
$n$ is even and a~conjugate-linear real anti-isometry of Krein spaces in the fermionic case if $n$ is odd.
Also, $\iota_n$ is involutive.
Combining theses maps for all $n\in\mathbb{N}_0$ (for $n=0$ take the identity map) and completing yields
a~conjugate-linear adapted real $f$-graded isometry $\mathcal{F}(L_{\Sigma})\to\mathcal{F}(L_{\overline{\Sigma}})$ which
we denote as $\iota_{\Sigma}$.
This satisf\/ies (T1b) and (V2).
Given $\xi_1,\dots,\xi_n\in L_{\Sigma}$ we also note for later use
\begin{gather}
\iota_{\Sigma}(\cs{\xi_1,\dots,\xi_n})=\kappa^n\cs{\xi_n,\dots,\xi_1}.
\label{eq:csiota}
\end{gather}

Let $\Sigma_1,\Sigma_2,\Sigma$ be hypersurfaces such that $\Sigma=\Sigma_1\cup\Sigma_2$ is a~disjoint union.
We def\/ine the map
$\tau_{\Sigma_1,\Sigma_2;\Sigma}:\mathcal{F}(L_{\Sigma_1})\ctens\mathcal{F}(L_{\Sigma_2})\to\mathcal{F}(L_\Sigma)$ as
in Section~\ref{sec:fock} by equation~\eqref{eq:isomtpf}.
Concretely
\begin{gather*}
\left(\tau_{\Sigma_1,\Sigma_2;\Sigma}(\psi_1\tens\psi_2)\right)\left((\eta_1,\xi_1),\dots,(\eta_{m+n},\xi_{m+n})\right)
\\
\qquad
\coloneqq\frac{1}{(m+n)!}\sum_{\sigma\in S^{m+n}}\kappa^{|\sigma|}\psi_1(\eta_{\sigma(1)},\dots,\eta_{\sigma(m)})\psi_2(\xi_{\sigma(m+1)},\dots,\xi_{\sigma(m+n)}).
\end{gather*}
As seen there this is an isometric isomorphism of Krein spaces.
Due to associativity, which is easily verif\/ied, this extends to a~prescription for decompositions of hypersurfaces
with an arbitrary f\/inite number of components.
We thus satisfy axioms (T2) and (V3).
In particular, the graded transposition property comes from equation~\eqref{eq:tpst}.
It is also straightforward to verify axiom (T2b).
Given $\eta_1,\dots,\eta_m\in L_{\Sigma_1}$ and $\xi_1,\dots,\xi_m\in L_{\Sigma_2}$ we also note,
using~\eqref{eq:isomtpfg},
\begin{gather}
\tau_{\Sigma_1,\Sigma_2;\Sigma}\left(\cs{\eta_1,\dots,\eta_m}\tens\cs{\xi_1,\dots,\xi_n}\right)=
\cs{(\eta_1,0),\dots,(\eta_m,0),(0,\xi_1),\dots,(0,\xi_n)}.
\label{eq:taucs}
\end{gather}

\subsection{Amplitudes}
\label{sec:ampl}

Let $M$ be a~region.
Given $\xi\in L_{\partial M}$ we decompose it as $\xi=\xi^{\textrm{R}}+J_{\partial M}\xi^{\textrm{I}}$, where
$\xi^{\textrm{R}},\xi^{\textrm{I}}\in L_{\tilde{M}}$.
We def\/ine $\widehat{\xi}\in L_{\tilde{M}}^\mathbb{C}\subseteq L_{\partial M}^{\mathbb{C}}$ as
\begin{gather*}
\widehat{\xi}\coloneqq\xi^{\textrm{R}}-\mathrm{i}\xi^{\textrm{I}}.
\end{gather*}
Here $L_{\tilde{M}}^\mathbb{C}$ and $L_{\partial M}^\mathbb{C}$ are complexif\/ications of the real vector spaces
$L_{\tilde{M}}$ and $L_{\partial M}$, where we denote the new complex structure by $\mathrm{i}$.
The real bilinear forms $\{\cdot,\cdot\}_{\partial M}$, $g_{\partial M}$, $\omega_{\partial M}$ all extend to
$L_{\partial M}^\mathbb{C}$ as complex bilinear forms.
This is understood in the following.

We def\/ine the subspace $\mathcal{H}_{\partial M}^\circ\subseteq\mathcal{H}_{\partial M}$ on which $\rho_M$ will be
def\/ined to be the space of linear combinations of vectors $\cs{\xi_1,\dots,\xi_{n}}$ for $n\in\mathbb{N}$ and
$\xi_1,\dots,\xi_{n}\in L_{\partial M}$ together with the vector $\mathbf{1}$.

If $n\in\mathbb{N}$ is odd and $\xi_1,\dots,\xi_{n}\in L_{\partial M}$ are arbitrary we def\/ine the amplitude map to
vanish,
\begin{gather}
\rho_M(\cs{\xi_1,\dots,\xi_{n}})\coloneqq0.
\label{eq:defamplodd}
\end{gather}
Since we view the target space $\mathbb{C}$ of $\rho_M$ as of even f-degree, this is equivalent to saying that $\rho_M$
is $f$-graded in the fermionic case.
In the bosonic case the fact that $\rho_M$ is $f$-graded is trivial.
For the vacuum we def\/ine, in accordance with axiom (V5),
\begin{gather}
\rho_M(\mathbf{1})\coloneqq1.
\label{eq:defamplvac}
\end{gather}
Now let $n\in\mathbb{N}$ and $\xi_1,\dots,\xi_{2n}\in L_{\partial M}$.
We def\/ine then
\begin{gather}
\rho_M(\cs{\xi_1,\dots,\xi_{2n}})\coloneqq\frac{1}{n!}\sum_{\sigma\in S^{2n}}\kappa^{|\sigma|}\prod_{j=
1}^n\big\{\widehat{\xi_{\sigma(j)}},\widehat{\xi_{\sigma(2n+1-j)}}\big\}_{\partial M}.
\label{eq:defampleven}
\end{gather}
Note that the right-hand side is symmetric in the bosonic and anti-symmetric in the fermionic case under interchange of
any two elements $\xi_i$, $\xi_j$ with $i\neq j$.
In particular, in the fermionic case we note that the real parts of the inner products $\{\cdot,\cdot\}_{\partial M}$
appearing on the right-hand side vanish since the real subspace $L_{\tilde{M}}$ and its complexif\/ication
$L_{\tilde{M}}^\mathbb{C}$ is neutral with respect to $g_{\partial M}$.
Thus, we could equally well replace $\{\cdot,\cdot\}_{\partial M}$ with the symplectic form
$2\mathrm{i}\omega_{\partial M}(\cdot,\cdot)$ in~\eqref{eq:defampleven}, making the anti-symmetry manifest.
In the bosonic case the imaginary parts of the inner products $\{\cdot,\cdot\}_{\partial M}$ vanish due to the
isotropic nature of the subspace $L_{\tilde{M}}^{\mathbb{C}}\subseteq L_{\partial M}^{\mathbb{C}}$.
Thus we could replace these inner products with their real parts $g_{\partial M}$, making the symmetry manifest.
This makes $\rho_M$ well def\/ined and completes its def\/inition on $\mathcal{H}_{\partial M}^\circ$, satisfying (T4).

\section{Amplitude and inner product}
\label{sec:amplip}

\subsection{The inner product as an amplitude}

We turn to axiom (T3x).
Let $\Sigma$ be a~hypersurface.
Then $\Sigma$ def\/ines a~slice region $\hat{\Sigma}$ with boundary
$\partial\hat{\Sigma}=\overline{\Sigma}\cup\Sigma'$.
Here, $\Sigma'$ denotes a~second copy of $\Sigma$.
We have then, $L_{\partial\hat{\Sigma}}=L_{\overline{\Sigma}}\times L_{\Sigma'}$ and
$\mathcal{H}_{\partial\hat{\Sigma}}=\mathcal{H}_{\overline{\Sigma}}\ctens\mathcal{H}_{\Sigma'}$.
Moreover,
\begin{gather*}
\omega_{\partial\hat{\Sigma}}=\omega_{\overline{\Sigma}}+\omega_{\Sigma'}=\omega_{\Sigma'}-\kappa\omega_{\Sigma},
\qquad
J_{\partial\hat{\Sigma}}=J_{\overline{\Sigma}}+J_{\Sigma'}=J_{\Sigma'}-J_{\Sigma},
\\
g_{\partial\hat{\Sigma}}=g_{\overline{\Sigma}}+g_{\Sigma'}=g_{\Sigma'}+\kappa g_{\Sigma},
\qquad
\{\cdot,\cdot\}_{\partial\hat{\Sigma}}=\{\cdot,\cdot\}_{\overline{\Sigma}}+\{\cdot,\cdot\}_{\Sigma'}=
\kappa\overline{\{\cdot,\cdot\}_{\Sigma}}+\{\cdot,\cdot\}_{\Sigma'}.
\end{gather*}
It follows from axioms (C6) and (C7) applied to two copies of $\hat{\Sigma}$ and their gluing that the subspace
$L_{\tilde{\hat{\Sigma}}}\subseteq L_{\partial\hat{\Sigma}}$ is precisely the space of pairs $(\phi,\phi)$ for $\phi\in
L_{\Sigma}$.
Correspondingly, the subspace $J_{\partial\hat{\Sigma}}L_{\tilde{\hat{\Sigma}}}\subseteq L_{\partial\hat{\Sigma}}$ is
the space of pairs $(\phi,-\phi)$ for $\phi\in L_{\Sigma}$.
Now note that given $\eta,\xi\in L_{\Sigma}$ the expressions $\widehat{(\eta,0)},\widehat{(0,\xi)}\in
L_{\tilde{\hat{\Sigma}}}^\mathbb{C}$ may be expanded as follows
\begin{gather*}
\widehat{(\eta,0)}=\left(\tfrac{1}{2}\eta-\mathrm{i}J_{\Sigma}\eta,\tfrac{1}{2}\eta-\mathrm{i}J_{\Sigma}\eta\right),
\qquad
\widehat{(0,\xi)}=\left(\tfrac{1}{2}\xi+\mathrm{i}J_{\Sigma}\xi,\tfrac{1}{2}\xi+\mathrm{i}J_{\Sigma}\xi\right).
\end{gather*}
Thus we have for $\eta,\eta',\xi,\xi'\in L_{\Sigma}$,
\begin{gather}
\big\{\widehat{(\eta,0)},\widehat{(\eta',0)}\big\}_{\partial\hat{\Sigma}}=0
\qquad
\big\{\widehat{(0,\xi)},\widehat{(0,\xi')}\big\}_{\partial\hat{\Sigma}}=0
\label{eq:ipssv}
\\
\big\{\widehat{(\eta,0)},\widehat{(0,\xi)}\big\}_{\partial\hat{\Sigma}}=\kappa\{\xi,\eta\}_{\Sigma}
\label{eq:ipos}
.
\end{gather}
The following proposition shows that axiom (T3x) is satisf\/ied.
\begin{prop}
\label{prop:amplip}
$\rho_{\hat{\Sigma}}$ is well defined on
$\tau_{\overline{\Sigma},\Sigma';\partial\hat{\Sigma}}(\mathcal{H}_{\overline{\Sigma}}\tens\mathcal{H}_{\Sigma'})\subseteq\mathcal{H}_{\partial\hat{\Sigma}}$.
Moreover,
\begin{gather}
\rho_{\hat{\Sigma}}\big(\tau_{\overline{\Sigma},\Sigma';\partial\hat{\Sigma}}\big(\iota_\Sigma(\psi')\tens\psi\big)\big)=
\langle\psi',\psi\rangle_\Sigma
\qquad
\forall\,\psi,\psi'\in\mathcal{H}_{\Sigma}.
\label{eq:amplip}
\end{gather}
\end{prop}
\begin{proof}
Let $\eta_1,\dots,\eta_m\in L_{\Sigma}$ and $\xi_1,\dots,\xi_n\in L_{\Sigma}$.
We have
\begin{gather}
\rho_{\hat{\Sigma}}\big(\tau_{\overline{\Sigma},\Sigma';\partial\hat{\Sigma}}\big(\iota_{\Sigma}(\cs{\eta_1,\dots,\eta_m})\tens\cs{\xi_1,\dots,\xi_n}\big)\big)
\label{eq:erip1}
\\
\qquad
=\kappa^m\rho_{\hat{\Sigma}}\big(\tau_{\overline{\Sigma},\Sigma';\partial\hat{\Sigma}}\big(\cs{\eta_m,\dots,\eta_1}\tens\cs{\xi_1,\dots,\xi_n}\big)\big)
\label{eq:erip2}
\\
\qquad
=\kappa^m\rho_{\hat{\Sigma}}\big(\cs{(\eta_m,0),\dots,(\eta_1,0),(0,\xi_1),\dots,(0,\xi_n)}\big)
\label{eq:erip3}
\\
\qquad
=\delta_{n,m}\kappa^n\;2^n\sum_{\sigma\in S^n}\kappa^{|\sigma|}\prod_{j=1}^n\big\{\widehat{(\eta_j,0)},\widehat{(0,\xi_{\sigma(j)})}\big\}_{\partial\hat{\Sigma}}
\label{eq:erip4}
\\
\qquad
=\delta_{n,m}2^n\sum_{\sigma\in S^n}\kappa^{|\sigma|}\prod_{j=1}^n\{\xi_{j},\eta_{\sigma(j)}\}_{\Sigma}
\label{eq:erip5}
\\
\qquad
=\langle\cs{\eta_1,\dots,\eta_m},\cs{\xi_1,\dots,\xi_n}\rangle_{\Sigma}
\nonumber
\end{gather}
The step from~\eqref{eq:erip1} to~\eqref{eq:erip2} consists in the application of~\eqref{eq:csiota}, while the step
to~\eqref{eq:erip3} arises from~\eqref{eq:taucs}.
In applying the amplitude map~\eqref{eq:defampleven}, many permutations do not contribute due to~\eqref{eq:ipssv} and
the remaining ones can be reorganized leading to~\eqref{eq:erip4}.
Applying~\eqref{eq:ipos} and moving the permutations from the $\xi$-variables to the $\eta$-variables
yields~\eqref{eq:erip5}.
This expression coincides with the inner product on $\mathcal{H}_{\Sigma}$, see~\eqref{eq:ipfs}.
This shows the equality~\eqref{eq:amplip} on a~dense subspace.
Since the right hand side extends continuously to the whole Krein space this def\/ines the left hand side there also.
This implies the claimed well def\/inedness of $\rho_{\hat{\Sigma}}$ on
$\tau_{\overline{\Sigma},\Sigma';\partial\hat{\Sigma}}(\mathcal{H}_{\overline{\Sigma}}\tens\mathcal{H}_{\Sigma'})$,
completing the proof.
\end{proof}

\subsection{Complex conjugation}

Let $M$ be a~region.
Then the subspace $L_{\tilde{M}}\subseteq L_{\partial M}$ gives rise to a~conjugate linear involutive adapted real
(anti-)isometry $u_M:L_{\partial M}\to L_{\partial M}$ due to axiom (C5) as explained in Section~\ref{sec:classdata}.
Recall also that $u_M$ is a~complex conjugation on $L_{\partial M}$.
With the conventions of Section~\ref{sec:ampl} we have the identity
\begin{gather*}
\big\{\widehat{\xi},\widehat{\eta}\big\}_{\partial M}=\{\xi,u_M(\eta)\}_{\partial M}
\qquad
\forall\,\xi,\eta\in L_{\partial M}.
\end{gather*}
Using this we may rewrite formula~\eqref{eq:defampleven} for the amplitude map as
\begin{gather}
\rho_M(\cs{\xi_1,\dots,\xi_{2n}})\coloneqq\frac{1}{n!}\sum_{\sigma\in S^{2n}}\kappa^{|\sigma|}\prod_{j=
1}^n\{\xi_{\sigma(j)},u_M(\xi_{\sigma(2n+1-j)})\}_{\partial M}.
\label{eq:amplevenu}
\end{gather}

Given $u_M$, Lemma~\ref{lem:fockaisom} gives rise to a~conjugate linear involutive adapted real $f$-graded isometry
$U_M:\mathcal{H}_{\partial M}\to\mathcal{H}_{\partial M}$ def\/ined by expression~\eqref{eq:deffaisom}.
Recall from Section~\ref{sec:fock} that the map $U_M$ may also be seen as a~complex conjugation.
It turns out that the amplitude map commutes with this complex conjugation in the following sense.
\begin{prop}
\begin{gather*}
\rho_M(U_M(\psi))=\overline{\rho_M(\psi)}
\qquad
\forall\, \psi\in\mathcal{H}_{\partial M}^{\circ}.
\end{gather*}
\end{prop}
\begin{proof}
It is suf\/f\/icient to carry out the proof for generating states.
Moreover, it is trivial for states of odd Fock degree so we consider states of even Fock degree only.
Let $\xi_1,\dots,\xi_{2n}\in L_{\partial M}$.
Then
\begin{gather*}
\rho_M(U_M(\cs{\xi_1,\dots,\xi_{2n}}))=\rho_M(\cs{u_M(\xi_{2n}),\dots,u_M(\xi_{1})})
\\
\hphantom{\rho_M(U_M(\cs{\xi_1,\dots,\xi_{2n}}))}{}
=\frac{1}{n!}\sum_{\sigma\in S^{2n}}\kappa^{|\sigma|}\prod_{j=
1}^n\{u_M(\xi_{\sigma(2n+1-j)}),\xi_{\sigma(j)}\}_{\partial M}
\\
\hphantom{\rho_M(U_M(\cs{\xi_1,\dots,\xi_{2n}}))}
=\frac{1}{n!}\sum_{\sigma\in S^{2n}}\kappa^{|\sigma|}\prod_{j=
1}^n\overline{\{\xi_{\sigma(j)},u_M(\xi_{\sigma(2n+1-j)})\}_{\partial M}}
\\
\hphantom{\rho_M(U_M(\cs{\xi_1,\dots,\xi_{2n}}))}{}
=\overline{\rho_M(\cs{\xi_1,\dots,\xi_{2n}})}.
\end{gather*}
Here we have used the def\/inition of $U_M$ given by expression~\eqref{eq:deffaisom}, the amplitude map in the
form~\eqref{eq:amplevenu} and the involutiveness of $u_M$.
\end{proof}

\subsection{Quantum evolution}
\label{sec:evol}

A quite dif\/ferent perspective on the roles of the maps $u_M$ and $U_M$ is af\/forded by comparing
formula~\eqref{eq:amplevenu} for the amplitude to formula~\eqref{eq:ipfs} for the inner product.
As we shall see this suggest to think of (a restricted version of) $U_M$ as an ``evolution'' operator, providing
a~quantum analogue of the role of $u_M$ in the classical theory as discussed in Section~\ref{sec:cevol}.

The following is in a~sense the quantum analogue of Lemma~\ref{lem:uclasseva}.
\begin{lem}
\label{lem:amplipu}
Suppose that $L_2\oplus L_1=L_{\partial M}$ is a~decomposition as an orthogonal direct sum of complex Krein spaces such
that $u_M(L_1)=L_2$.
Then $U_M(\mathcal{F}(L_1))=\mathcal{F}(L_2)$ with $\mathcal{F}(L_1)$ and $\mathcal{F}(L_2)$ understood as subspaces
of $\mathcal{H}_{\partial M}=\mathcal{F}(L_{\partial M})$.
With the isometric isomorphism of Fock spaces $\mathcal{F}(L_2)\ctens\mathcal{F}(L_1)\to\mathcal{F}(L_{\partial M})$
given by expression~\eqref{eq:isomtpf} we have the following identity,
\begin{gather}
\rho_M\left(\psi'\tens\psi\right)=\kappa^{\fdg{\psi}}\langle U_M(\psi),\psi'\rangle_2,
\qquad
\forall\, \psi\in\mathcal{F}(L_1),\quad \forall\,\psi'\in\mathcal{F}(L_2),
\label{eq:amplu}
\end{gather}
where $\langle\cdot,\cdot\rangle_2$ denotes the inner product of $\mathcal{H}_{\partial M}=\mathcal{F}(L_{\partial M})$
restricted to $\mathcal{F}(L_2)$.
\end{lem}
\begin{proof}
The claimed property of $U_M$ follows from the def\/inition~\eqref{eq:deffaisom}.
For the proof of equation~\eqref{eq:amplu} it is enough to consider generating states.
Thus, let $\xi_1,\dots,\xi_n\!\in\! L_1$ and $\eta_1,\dots,\eta_m\!\in\! L_2$.
Then
\begin{gather*}
\rho_M\left(\cs{\eta_1,\dots,\eta_m}\tens\cs{\xi_1,\dots,\xi_n}\right)
=\rho_M\left(\cs{(\eta_1,0),\dots,(\eta_m,0),(0,\xi_1),\dots,(0,\xi_n)}\right)
\\
\qquad
=\delta_{n,m}2^n\sum_{\sigma\in S^n}\kappa^{|\sigma|}\prod_{j=
1}^n\left\{(\eta_j,0),u_M((0,\xi_{\sigma(n+1-j)}))\right\}_{\partial M}
\\
\qquad
=\delta_{n,m}2^n\sum_{\sigma\in S^n}\kappa^{|\sigma|}\prod_{j=1}^n\left\{\eta_j,u_M(\xi_{\sigma(n+1-j)})\right\}_{2}
\\
\qquad
=\langle\cs{u_M(\xi_n),\dots,u_M(\xi_1)},\cs{\eta_1,\dots,\eta_m},\rangle_2
\\
\qquad
=\kappa^n\langle\tilde{U}_M\left(\cs{\xi_1,\dots,\xi_n}\right),\cs{\eta_1,\dots,\eta_m}\rangle_2
\end{gather*}
Here we have used equation~\eqref{eq:isomtpfg}, formula~\eqref{eq:amplevenu} for the amplitude and the fact that the
inner products of the type
\begin{gather*}
\{(\eta',0),u_M((\eta,0))\}_{\partial M}
\qquad
\text{and}
\qquad
\{(0,\xi'),u_M((0,\xi))\}_{\partial M}
\end{gather*}
vanish.
Moreover, we have used the inner product formula~\eqref{eq:ipfs} and the formula~\eqref{eq:faisomg} for $U_M$ on
generating states.
The notation $\{\cdot,\cdot\}_2$ denotes the restriction of the inner product of $L_{\partial M}$ to the subspace $L_2$.
\end{proof}

Analogous to the classical case, such an algebraic notion of ``evolution'' underlying the amplitude map may be seen to
arise from a~geometric notion of ``evolution'' when the underlying decomposition of the space $L_{\partial M}$ arises
from a~corresponding decomposition of the hypersurface~$\partial M$.
This yields the following quantum analogue of Proposition~\ref{prop:uclassevg}.
\begin{prop}
\label{prop:amplevol}
Suppose that $\partial M$ decomposes as a~disjoint union $\partial M=\overline{\Sigma_2}\cup\Sigma_1$ such that
$u_M(L_{\Sigma_1})=L_{\overline{\Sigma_2}}$.
Define $\tilde{U}_M:\mathcal{H}_{\Sigma_1}\to\mathcal{H}_{\Sigma_2}$ as the composition of
$\iota_{\overline{\Sigma_2}}$ with the restriction of $U_M$ to $\mathcal{H}_{\Sigma_1}$.
Then $\tilde{U}_M$ is a~complex linear isometric isomorphism.
Moreover,
\begin{gather*}
\rho_M\big(\tau_{\overline{\Sigma_2},\Sigma_1;\partial M}\big(\iota_{\Sigma_2}(\psi')\tens\psi\big)\big)=
\langle\psi',\tilde{U}_M(\psi)\rangle_{\Sigma_2},
\qquad
\forall\,\psi\in\mathcal{H}_{\Sigma_1},\quad \forall\,\psi'\in\mathcal{H}_{\Sigma_2}.
\end{gather*}
\end{prop}
\begin{proof}
This follows rather straightforwardly from Lemma~\ref{lem:amplipu}.
Setting $L_1=L_{\Sigma_1}$ and $L_2=L_{\overline{\Sigma_2}}$ we have for $\psi\in L_{\Sigma_1}$ and $\psi'\in
L_{\Sigma_2}$,
\begin{gather*}
\rho_M\big(\tau_{\overline{\Sigma_2},\Sigma_1;\partial M}\big(\iota_{\Sigma_2}(\psi')\tens\psi\big)\big)=
\kappa^{\fdg{\psi}}\langle U_M(\psi),\iota_{\Sigma_2}(\psi')\rangle_{\overline{\Sigma_2}}
\\
\qquad
=\kappa^{\fdg{\psi}}\langle\iota_{\Sigma_2}\circ\tilde{U}_M(\psi),\iota_{\Sigma_2}(\psi')\rangle_{\overline{\Sigma_2}}=
\langle\psi',\tilde{U}_M(\psi)\rangle_{\Sigma_2}.\tag*{\qed}
\end{gather*}
\renewcommand{\qed}{}
\end{proof}

This suggests to view $M$ as a~cobordism from $\Sigma_1$ to $\Sigma_2$, inducing the unitary ``evolution'' map~$\tilde{U}_M$ between the associated state spaces.
This would then be in line with the often preferred categorial view of TQFT, in the sense of assigning to the morphism
$M:\Sigma_1\to\Sigma_2$ the morphism $\tilde{U}_M:\mathcal{H}_{\Sigma_1}\to\mathcal{H}_{\Sigma_2}$.
In particular, if $M$ as a~cobordism takes the role of time-evolution, $\tilde{U}_M$ is the associated unitary quantum
operator.
We also note that in the special case of a~slice region with $\Sigma_1\approx\Sigma_2$, we recover
Proposition~\ref{prop:amplip}.
The ``evolution'' operator $\tilde{U}_M$ is then simply the identity.

Recall from Section~\ref{sec:cevol} that in the fermionic case (but not the bosonic one) there is a~preferred algebraic
notion of evolution, coming from the canonical decomposition $L_{\partial M}=L_{\partial M}^+\oplus L_{\partial M}^-$
which satisf\/ies automatically $u_M(L_{\partial M}^+)=L_{\partial M}^-$.
This property is inherited by the quantum theory.
Denoting the Fock space $\mathcal{F}(L_{\partial M}^+)$ by $\mathcal{H}_{\partial M}^{(+)}$ and
$\mathcal{F}(L_{\partial M}^-)$ by $\mathcal{H}_{\partial M}^{(-)}$ we have the tensor product decomposition
$\mathcal{H}_{\partial M}=\mathcal{H}_{\partial M}^{(+)}\ctens\mathcal{H}_{\partial M}^{(-)}$.
(This is not to be confused with the direct sum decomposition of $\mathcal{H}_{\partial M}$ as a~Krein space into
positive and negative parts.) The map $U_M$ then restricts to a~conjugate linear real $f$-graded isometry
$\mathcal{H}_{\partial M}^{(+)}\to\mathcal{H}_{\partial M}^{(-)}$.
If this decomposition of~$L_{\partial M}$ into positive and negative parts is induced by a~geometric decomposition of~$\partial M$ in the sense of Proposition~\ref{prop:amplevol}, the spaces $\mathcal{H}_{\Sigma_1}=\mathcal{H}_{\partial
M}^{(+)}$ and $\mathcal{H}_{\Sigma_2}=\overline{\mathcal{H}_{\partial M}^{(-)}}$ are Hilbert spaces and
$\tilde{U}_M:\mathcal{H}_{\Sigma_1}\to\mathcal{H}_{\Sigma_2}$ is a~unitary operator between
them\footnote{Alternatively we may have $\mathcal{H}_{\Sigma_1}=\mathcal{H}_{\partial M}^{(-)}$ and
$\mathcal{H}_{\Sigma_2}=\overline{\mathcal{H}_{\partial M}^{(+)}}$ in which case we would have anti-Hilbert spaces,
i.e., negative def\/inite spaces.
For the considerations that follow we may in this case simply invert the overall sign by convention.}.

Recall now the standard globally hyperbolic setting described in Section~\ref{sec:cevol}.
As already stated in that Section and as exhibited explicitly in the example of the Dirac f\/ield theory
(Section~\ref{sec:dirac}), for standard fermionic f\/ield theories in this setting the algebraic notion of evolution
coincides with the usual notion of time-evolution.
That is, the decomposition of the boundary of a~region $M$ into initial and f\/inal component induces precisely the
decomposition of the Krein space $L_{\partial M}$ into its positive and negative part.
Applying Proposition~\ref{prop:amplevol}, we obtain a~description of the dynamics of the quantum theory in terms of the
unitary maps $\tilde{U}_M$ between Hilbert spaces associated to these hypersurfaces (all oriented in the same way).
In particular, no indef\/inite inner product space appears in this way of describing the dynamics.
This clarif\/ies the absence of indef\/inite inner product spaces in the standard description of conventional fermionic
quantum f\/ield theories.

\section{Gluing}
\label{sec:composition}

\subsection{Disjoint gluing}

We proceed to consider the gluing axioms (T5a) and (T5b).
We start with the proof of (T5a), describing the disjoint gluing of regions.
\begin{prop}
\label{prop:dglueampl}
Let $M_1$ and $M_2$ be regions and $M=M_1\cup M_2$ be their disjoint union.
Then
\begin{gather}
\rho_{M}\left(\tau_{\partial M_1,\partial M_2;\partial M}(\psi_1\tens\psi_2)\right)=\rho_{M_1}(\psi_1)\rho_{M_2}(\psi_2)
\qquad
\forall\,\psi_1\in\mathcal{H}_{\partial M_1}^{\circ}, \ \psi_2\in\mathcal{H}_{\partial M_2}^{\circ}.
\label{eq:dglueampl}
\end{gather}
\end{prop}
\begin{proof}
Let $\eta_1,\dots,\eta_m\in L_{\partial M_1}$ and $\xi_1,\dots,\xi_n\in L_{\partial M_2}$.
Then with~\eqref{eq:taucs} we get
\begin{gather*}
\rho_{M}\left(\tau_{\partial M_1,\partial M_2;\partial M}\left(\cs{\eta_1,\dots,\eta_m}\tens\cs{\xi_1,\dots,\xi_n}\right)\right)\\
\qquad{}
=\rho_{M}\left(\cs{(\eta_1,0),\dots,(\eta_m,0),(0,\xi_1),\dots,(0,\xi_n)}\right).
\end{gather*}
If $n+m$ is odd, then~\eqref{eq:defamplodd} applies here (i.e., to the left hand side of~\eqref{eq:dglueampl}) and also
to one of the two factors on the right hand side of~\eqref{eq:defamplodd}, conf\/irming the equality.
We may thus suppose that $n+m$ is even and~\eqref{eq:defampleven} applies.
Since mixed terms of the type
\begin{gather*}
\big\{\widehat{(\eta_i,0)},\widehat{(0,\xi_j)}\big\}_{\partial M}=\{(\widehat{\eta_i},0),(0,\widehat{\xi_j})\}_{\partial M}=
\{\widehat{\eta_i},0\}_{\partial M_1}+\{0,\widehat{\xi_j}\}_{\partial M_2}
\end{gather*}
vanish, only pairings of the type $\{\widehat{(\eta_i,0)},\widehat{(\eta_j,0)}\}_{\partial M}$ or
$\{\widehat{(0,\xi_i)},\widehat{(0,\xi_j)}\}_{\partial M}$ contribute.
This restricts the permutations we need to sum over considerably.
In particular, if~$m$ (and hence also~$n$) is odd we do not have any contributing term and the result vanishes,
conf\/irming~\eqref{eq:dglueampl} in this case also.
It remains to consider the case where both $m$ and $n$ are even.

Thus, let $\eta_1,\dots,\eta_{2m}\in L_{\partial M_1}$ and $\xi_1,\dots,\xi_{2n}\in L_{\partial M_2}$.
Then
\begin{gather}
\rho_{M}\left(\tau_{\partial M_1,\partial M_2;\partial M}\left(\cs{\eta_1,\dots,\eta_{2m}}\tens\cs{\xi_1,\dots,\xi_{2n}}\right)\right)
\nonumber
\\
\qquad {}
=\rho_{M}\left(\cs{(\eta_1,0),\dots,(\eta_{2m},0),(0,\xi_1),\dots,(0,\xi_{2n})}\right)
\label{eq:dga2}
\\
\qquad{}
=\frac{1}{m!n!}\sum_{\sigma\in S^{2m},\sigma'\in S^{2n}}(-1)^{|\sigma|+|\sigma'|}\left(\prod_{i=
1}^m\big\{\widehat{(\eta_{\sigma(i)},0)},\widehat{(\eta_{\sigma(2m+1-i)},0)}\big\}_{\partial M}\right)
\nonumber
\\
\qquad\quad{}
\times
\left(\prod_{j=1}^n\big\{\widehat{(0,\xi_{\sigma(j)})},\widehat{(0,\xi_{\sigma(2n+1-j)})}\big\}_{\partial M}\right)
\label{eq:dga3}
\\
\qquad{}
=\frac{1}{m!n!}\sum_{\sigma\in S^{2m},\sigma'\in S^{2n}}(-1)^{|\sigma|+|\sigma'|}\left(\prod_{i=
1}^m\big\{\widehat{\eta_{\sigma(i)}},\widehat{\eta_{\sigma(2m+1-i)}}\big\}_{\partial M_1}\right)
\nonumber
\\
\qquad\quad{}
\times
\left(\prod_{j=1}^n\big\{\widehat{\xi_{\sigma(j)}},\widehat{\xi_{\sigma(2n+1-j)}}\big\}_{\partial M_2}\right).
\label{eq:dga4}
\end{gather}
In the application of~\eqref{eq:defampleven} in the step from~\eqref{eq:dga2} to~\eqref{eq:dga3} we make use of the
already mentioned simplif\/ication due to the elimination of permutations from the sum that yield no contribution.
Finally~\eqref{eq:dga4} can be easily recognized as the product of two amplitudes of the form~\eqref{eq:defampleven}.
This completes the proof of~\eqref{eq:dglueampl}.
\end{proof}

\subsection{Gluing along a~hypersurface}
\label{sec:glueh}

Axiom (T5b) is considerably more delicate.
Consider a~region $M$ with boundary decomposing as a~disjoint union $\partial
M=\Sigma_1\cup\Sigma\cup\overline{\Sigma'}$ where $\Sigma'$ is a~copy of $\Sigma$.
We suppose that $M$ can be glued to itself along $\Sigma$/$\Sigma'$ to form the new region $M_1$ with boundary
$\Sigma_1$.
In view of the def\/inition~\eqref{eq:defamplvac} of the amplitude on the vacuum state the composition
identity~\eqref{eq:glueax1} would imply the following identity for the gluing anomaly,
\begin{gather}
c(M;\Sigma,\overline{\Sigma'})=
\sum_{i\in I}(-1)^{\sig{\zeta_i}}\rho_M\big(\tau_{\Sigma_1,\Sigma,\overline{\Sigma'};\partial M}(\mathbf{1}\tens\zeta_i\tens\iota_\Sigma(\zeta_i))\big),
\label{eq:gaid}
\end{gather}
given any ON-basis $\{\zeta_i\}_{i\in I}$ of $\mathcal{H}_{\Sigma}$.
However, if the dimension of $\mathcal{H}_{\Sigma}$ is inf\/inite, it is not even clear whether the sum converges,
given some particular ordered basis.
Indeed, it is easy to construct simple models of a~spacetime system together with data satisfying the axioms of
Section~\ref{sec:classdata} such that for a~particular gluing this sum never converges.
For the core axioms to be valid we thus need to introduce an additional assumption, not contained in the axioms of
Section~\ref{sec:classdata}.
This is precisely the assumption that the gluing anomaly $c(M;\Sigma,\overline{\Sigma'})$ is well def\/ined for all
admissible gluings.
In the bosonic case this was discussed in~\cite{Oe:holomorphic} where it was shown that this is equivalent to an
integrability condition in the holomorphic representation.
Here we need to deal with this from a~Fock space point of view.

Given the Fock space structure of $\mathcal{H}_{\Sigma}$ there is a~natural family of ON-basis arising from ON-basis of
the underlying space $L_\Sigma$.
It will be convenient to formulate the well-def\/inedness criterion for the gluing anomaly using these special basis
only.
Let $\{\xi_i\}_{i\in N}$ be an ON-basis of~$L_{\Sigma}$, where $N=\{1,\dots,\dim L_{\Sigma}\}$ if $L_{\Sigma}$ is
f\/inite-dimensional and $N=\mathbb{N}$ otherwise.
Using the ON-basis~\eqref{eq:fbbose} of Fock space in the bosonic case we can then write the sum
\begin{gather}
\sum_{i\in I}(-1)^{\sig{\zeta_i}}\zeta_i\tens\iota_{\Sigma}(\zeta_i)
\label{eq:complbasis}
\end{gather}
appearing in the identities~\eqref{eq:glueax1} and~\eqref{eq:gaid} as follows
\begin{gather*}
\sum_{m=0}^\infty\sum_{a_1\le\dots\le a_m\in N}(-1)^{\sum\limits_{i=
1}^m\sig{\xi_{a_i}}}\frac{1}{2^m K_{a_1,\dots,a_m}}\,\cs{\xi_{a_1},\dots,\xi_{a_m}}\tens\iota_{\Sigma}\left(\cs{\xi_{a_1},\dots,\xi_{a_m}}\right).
\end{gather*}
In fact, this very same formula serves also in the fermionic case since the fermionic ON-basis~\eqref{eq:fbfermi}
corresponds to a~subset of the bosonic one and the extraneous terms then simply vanish due to antisymmetry.
Moreover, we can simplify the formula by eliminating the combinatorial factors $K_{a_1,\dots,a_m}$ in exchange for an
unrestricted sum of the indices over~$N^m$ while dividing by a~factor~$m!$.
This yields
\begin{gather}
\sum_{m=0}^\infty\frac{1}{2^m\,m!}\sum_{a_1,\dots,a_m\in N}(-1)^{\sum\limits_{i=
1}^m\sig{\xi_{a_i}}}\cs{\xi_{a_1},\dots,\xi_{a_m}}\tens\iota_{\Sigma}\left(\cs{\xi_{a_1},\dots,\xi_{a_m}}\right).
\label{eq:cbfock}
\end{gather}
This continues to apply to both the bosonic and the fermionic case.

Suppose now that the Krein space $L_{\Sigma}$ is f\/inite-dimensional and $N$ thus f\/inite.
The inner sum in~\eqref{eq:cbfock} is then a~well def\/ined element of the tensor product
$\mathcal{H}_{\Sigma}\tens\mathcal{H}_{\overline{\Sigma}}$.
Moreover, it does not depend on the choice of basis of $L_{\Sigma}$.
This follows from the conjugate linearity of the states $\cs{\xi_{a_1},\dots,\xi_{a_m}}$ in terms of the variables
$\xi_{a_1},\dots,\xi_{a_m}$ combined with the conjugate linearity of $\iota_{\Sigma}$.
To see this explicitly note that any other ON-basis of $L_{\Sigma}$ is of the form $\{\Lambda\xi_i\}_{i\in N}$ for
a~unitary map $\Lambda:L_{\Sigma}\to L_{\Sigma}$.
The sum for the modif\/ied basis reduces to that of the original basis as follows
\begin{gather*}
\sum_{a_1,\dots,a_m\in N}(-1)^{\sum\limits_{i=
1}^m\sig{\Lambda\xi_{a_i}}}\cs{\Lambda\xi_{a_1},\dots,\Lambda\xi_{a_m}}\tens\iota_{\Sigma}\left(\cs{\Lambda\xi_{a_1},\dots,\Lambda\xi_{a_m}}\right)
\\
\qquad
=\sum_{a_1,\dots,a_m\in N}\sum_{b_1,\dots,b_m\in N}\sum_{d_1,\dots,d_m\in N}\left(\prod_{i=
1}^m\{\xi_{d_i},\Lambda\xi_{a_i}\}_{\Sigma}\,(-1)^{\sig{\Lambda\xi_{a_i}}}\,\{\Lambda\xi_{a_i},\xi_{b_i}\}_{\Sigma}\right)
\\
\qquad\quad{}\times
\cs{\xi_{b_1},\dots,\xi_{b_m}}\tens\iota_{\Sigma}\left(\cs{\xi_{d_1},\dots,\xi_{d_m}}\right)
\\
\qquad
=\sum_{b_1,\dots,b_m\in N}(-1)^{\sum\limits_{i=
1}^m\sig{\xi_{b_i}}}\cs{\xi_{b_1},\dots,\xi_{b_m}}\tens\iota_{\Sigma}\left(\cs{\xi_{b_1},\dots,\xi_{b_m}}\right).
\end{gather*}
In the fermionic case the outer sum in~\eqref{eq:cbfock} is then also f\/inite, since terms with $m>\dim L_{\Sigma}$
vanish.

Replacing in~\eqref{eq:gaid} expression~\eqref{eq:complbasis} with expression~\eqref{eq:cbfock} yields for the gluing
anomaly,
\begin{gather}
c(M;\Sigma,\overline{\Sigma'})=\sum_{m=0}^\infty\frac{1}{2^m\,m!}\sum_{a_1,\dots,a_m\in N}(-1)^{\sum\limits_{i=
1}^m\sig{\xi_{a_i}}}
\nonumber\\
\hphantom{c(M;\Sigma,\overline{\Sigma'})=}{}
\times
\rho_M\left(\tau_{\Sigma_1,\Sigma,\overline{\Sigma'};\partial M}\left(\mathbf{1}\tens\cs{\xi_{a_1},\dots,\xi_{a_m}}\tens\iota_{\Sigma}\left(\cs{\xi_{a_1},\dots,\xi_{a_m}}\right)\right)\right).
\label{eq:gafid}
\end{gather}
If $L_{\Sigma}$ is f\/inite-dimensional, the gluing anomaly is automatically well def\/ined in the fermionic case, as
can also be seen from the fact that $\mathcal{H}_{\Sigma}$ is then f\/inite-dimensional.
In the bosonic case, $\mathcal{H}_{\Sigma}$ is inf\/inite-dimensional even when $L_{\Sigma}$ is f\/inite-dimensional.
However, only the outer sum in~\eqref{eq:gafid} may diverge then.
We declare $c(M;\Sigma,\overline{\Sigma'})$ well def\/ined whenever this sum converges absolutely.

If $L_{\Sigma}$ is inf\/inite-dimensional the situation is more complicated.
Consider the projective system $\{L_{\Sigma,\alpha}\}_{\alpha\in A}$ of all f\/inite-dimensional complex subspaces of
$L_{\Sigma}$.
(Here $A$ denotes a~suitable index set.) The projective limit of $\{L_{\Sigma,\alpha}\}_{\alpha\in A}$ is $L_{\Sigma}$.
Associated to each subspace $L_{\Sigma,\alpha}\subseteq L_{\Sigma}$ we def\/ine the corresponding restricted anomaly
factor $c_{\alpha}(M;\Sigma,\overline{\Sigma'})$ by the evaluation of the right hand side of~\eqref{eq:gafid} if it
exists.
We then say that the anomaly factor is well def\/ined if the projective limit exists,
\begin{gather*}
c(M;\Sigma,\overline{\Sigma'})\coloneqq\varinjlim_{\alpha}c_{\alpha}(M;\Sigma,\overline{\Sigma'}).
\end{gather*}
More explicitly, this may be formulated as follows.
In the fermionic case, every $c_{\alpha}(M;\Sigma,\overline{\Sigma'})$ is well def\/ined.
We say then that the limit exists and is equal to a~quantity $c\in\mathbb{C}$, i.e.,
$\varinjlim_{\alpha}c_{\alpha}(M;\Sigma,\overline{\Sigma'})=c$ if $c$ has the following property: For any $\epsilon>0$
there exists $\beta\in A$ such that for any $\gamma\in A$ with $\gamma\ge\beta$ we have
\begin{gather*}
\left|c_{\gamma}(M;\Sigma,\overline{\Sigma'})-c\right|<\epsilon.
\end{gather*}
In the bosonic case the def\/inition is the same with the additional requirement that
$c_{\beta}(M;\Sigma,\overline{\Sigma'})$ as well as $c_{\gamma}(M;\Sigma,\overline{\Sigma'})$ for any $\gamma\ge\beta$
must be well def\/ined.

Our def\/inition of the anomaly factor really amounts to saying that equation~\eqref{eq:gaid} has to make sense, but
only for certain particular choices of ON-basis of $\mathcal{H}_{\Sigma}$ and certain particular choices of orderings
of the sum.
We apply this same def\/inition to the right-hand side of~\eqref{eq:glueax1}.
Thus, our main theorem shall consist of a~proof of equation~\eqref{eq:glueax1} and thus of axiom (T5b), but with the
slight modif\/ication of restricting the ON-basis $\{\zeta_i\}_{i\in I}$ and their orderings.
\begin{thm}
\label{thm:glueaxb}
Assume the geometric context of axiom {\rm (T5b)}.
Moreover, assume that the anomaly factor $c(M;\Sigma,\overline{\Sigma'})$ is well defined and non-zero.
Then the composition identity~\eqref{eq:glueax1} holds in the sense discussed and thus axiom {\rm (T5b)} holds $($for
a~restricted choice of ON-basis and orderings$)$.
\end{thm}

\begin{proof}
By our def\/inition of the subspace $\mathcal{H}_{\partial M_1}^\circ\subseteq\mathcal{H}_{\partial M_1}$ it is
suf\/f\/icient to prove the composition identity~\eqref{eq:glueax1} for states of the form $\cs{\phi_1,\dots,\phi_n}$
for $\phi_1,\dots,\phi_n\in L_{\partial M_1}$.
Consider a~f\/inite-dimensional complex Krein subspace $L_{\Sigma,\alpha}$ of $L_{\Sigma}$, let
$N_\alpha=\{1,\dots,\dim L_{\Sigma,\alpha}\}$ and let $\{\xi_a\}_{a\in N_{\alpha}}$ be an ON-basis of
$L_{\Sigma,\alpha}$ as a~complex Krein space.
We def\/ine
\begin{gather*}
R_{\alpha}\coloneqq\sum_{m=0}^\infty\frac{1}{2^m\,m!}\sum_{a_1,\dots,a_m\in N_{\alpha}}(-1)^{\sum\limits_{i=1}^m\sig{\xi_{a_i}}}
\\
\hphantom{R_{\alpha}\coloneqq}{}
\times
\rho_M\big(\tau_{\Sigma_1,\Sigma,\overline{\Sigma'};\partial M}\big(\cs{\phi_1,\dots,\phi_{n}}\tens\cs{\xi_{a_1},\dots,\xi_{a_m}}\tens\iota_{\Sigma} (\cs{\xi_{a_1},\dots,\xi_{a_m}})\big)\big),
\end{gather*}
if the outer sum converges absolutely.
The version of the identity~\eqref{eq:glueax1} we need to prove amounts to showing the following limit,
\begin{gather*}
\rho_{M_1}(\cs{\phi_1,\dots,\phi_{n}})\cdot c(M;\Sigma,\overline{\Sigma'})=\varinjlim_{\alpha}R_{\alpha}.
\end{gather*}
More explicitly this means that given $\epsilon>0$ we need to show the existence of $\beta\in A$ such that for all
$\gamma\ge\beta$ we have
\begin{gather*}
\left|\rho_{M_1}(\cs{\phi_1,\dots,\phi_{n}})\cdot c(M;\Sigma,\overline{\Sigma'})-R_{\gamma}\right|<\epsilon.
\end{gather*}
Instead we are going to show the stronger statement that there exists $\beta\in A$ such that for all $\gamma\ge\beta$
we have
\begin{gather}
\rho_{M_1}(\cs{\phi_1,\dots,\phi_{n}})\cdot c_{\gamma}(M;\Sigma,\overline{\Sigma'})=R_{\gamma}.
\label{eq:gaxconst}
\end{gather}

To simplify the proof it is convenient to exploit complex conjugate linearity of the states
$\cs{\phi_1,\dots,\phi_{n}}$ in terms of the variables $\phi_1,\dots,\phi_n$ in conjunction with the decomposition
$L_{\partial M_1}=L_{\tilde{M}_1}\oplus J_{\partial M_1}L_{\tilde{M}_1}$.
Decomposing $\phi_i=\phi_i^0+J_{\Sigma_1}\phi_i^1$ where $\phi_i^0,\phi_i^1\in L_{\tilde{M}_1}$ we have
\begin{gather*}
\cs{\phi_1,\dots,\phi_{n}}=
\sum_{l_1,\dots,l_n\in\{0,1\}}(-\mathrm{i})^{l_1+\dots+l_n}\psi\big[\phi_1^{l_1},\dots,\phi_{n}^{l_n}\big].
\end{gather*}
Since the amplitude map is complex linear we can apply this decomposition to both sides of~\eqref{eq:gaxconst}.
It is thus suf\/f\/icient to restrict to the special case $\phi_1,\dots,\phi_n\in L_{\tilde{M}_1}$, as we shall do in
the following.

Given $\phi\in L_{\tilde{M}_1}$ we observe that due to the exact sequence~\eqref{eq:xsbdy} of axiom (C7) there is
a~unique $\tilde{\phi}\in L_{\Sigma}$ such that $(\phi,\tilde{\phi},\tilde{\phi})\in L_{\tilde{M}}\subseteq
L_{\Sigma_1}\times L_{\Sigma}\times L_{\overline{\Sigma'}}$.
With this notation let $\beta'\in A$ be determined such that $L_{\Sigma,\beta'}$ is the complex Krein subspace of
$L_{\Sigma}$ generated by $\{\tilde{\phi}_1,\dots,\tilde{\phi}_n\}$.
On the other hand let $\beta''\in A$ be such that for any $\gamma\ge\beta''$, $c_{\gamma}(M;\Sigma,\overline{\Sigma'})$
is well def\/ined.
Such a~$\beta''$ must exist since by assumption $c(M;\Sigma,\overline{\Sigma'})$ is well def\/ined.
Now let $\beta\in A$ such that $\beta\ge\beta'$ and $\beta\ge\beta''$.
We claim that with this choice of $\beta$ we satisfy~\eqref{eq:gaxconst} for all $\gamma\ge\beta$.
This is precisely the content of the following Lemma~\ref{lem:gfid}.
\end{proof}
\begin{lem}
\label{lem:gfid}
Assume the geometric context of axiom {\rm (T5b)}.
Let $\phi_1,\dots,\phi_n\in L_{\tilde{M}_1}$.
Let $L_{\Sigma,\alpha}$ be a~complex Krein subspace of $L_{\Sigma}$ such that $\tilde{\phi}_1,\dots,\tilde{\phi}_n\in
L_{\Sigma,\alpha}$ with the notation as above.
Suppose that $c_{\alpha}(M;\Sigma,\overline{\Sigma'})$ is well defined.
Define $N_{\alpha}\coloneqq\{1,\dots,\dim L_{\Sigma,\alpha}\}$ and let $\{\xi_{a}\}_{a\in N_\alpha}$ be an ON-basis of
$L_{\Sigma,\alpha}$.
Then
\begin{gather}
\rho_{M_1}(\cs{\phi_1,\dots,\phi_{n}})\cdot c_{\alpha}(M;\Sigma,\overline{\Sigma'})=\sum_{m=
0}^\infty\frac{1}{2^m\,m!}\sum_{a_1,\dots,a_m\in N_{\alpha}}(-1)^{\sum\limits_{i=1}^m\sig{\xi_{a_i}}}
\nonumber\\
\qquad
{}\times
\rho_M\big(\tau_{\Sigma_1,\Sigma,\overline{\Sigma'};\partial M}\big(\cs{\phi_1,\dots,\phi_{n}}\tens\cs{\xi_{a_1},\dots,\xi_{a_m}}\tens\iota_{\Sigma} (\cs{\xi_{a_1},\dots,\xi_{a_m}})\big)\big).
\label{eq:gfid}
\end{gather}
\end{lem}
The rather lengthy proof of this lemma is contained in Appendix~\ref{sec:gproof}.

To summarize: Given a~spacetime system and a~model satisfying the axioms of Section~\ref{sec:classdata}, we need an
additional assumption, namely the well def\/inedness of the gluing anomaly for any admissible gluing.
If this assumption is satisf\/ied, the quantization described in this section yields a~general boundary quantum f\/ield
theory satisfying the core axioms for a~slightly modif\/ied version of axiom (T5b).

\section{The gluing anomaly}
\label{sec:glanom}

\subsection{An important special case}

Although it is not obvious in general how to ensure the well def\/inedness of the gluing anomaly and thus the validity
of all of the core axioms, there is an important special case where well def\/inedness can be demonstrated easily.
This is the case where two regions are glued together, one of which can be seen as a~cobordism with an associated
notion of evolution in the sense of Sections~\ref{sec:cevol} and~\ref{sec:evol}.
\begin{prop}
Let $P$ be a~region with boundary decomposing disjointly as $\partial P=\Sigma_1\cup\overline{\Sigma_2}$ and $Q$
a~region with boundary decomposing disjointly as $\partial Q=\Sigma_2'\cup\Sigma_3$, with $\Sigma_2'$ a~copy of
$\Sigma_2$.
Suppose furthermore that $u_{P}(L_{\Sigma_1})=L_{\overline{\Sigma_2}}$.
Let $M$ be the disjoint union of $P$ and $Q$ and consider the gluing of $M$ to itself along $\Sigma_2$.
Then the associated gluing anomaly is well defined and equal to $1$, that is
\begin{gather*}
c(M;\Sigma_2',\overline{\Sigma_2})=1.
\end{gather*}
\end{prop}
\begin{proof}
We write $\rho_M$ as a~map
$\mathcal{H}_{\Sigma_1}\ctens\mathcal{H}_{\Sigma_3}\ctens\mathcal{H}_{\Sigma_2'}\ctens\mathcal{H}_{\overline{\Sigma_2}}\to\mathbb{C}$.
Then given an ON-basis $\{\zeta_i\}_{i\in I}$ of $\mathcal{H}_{\Sigma_2}$ we obtain
\begin{gather}
c(M;\Sigma_2',\overline{\Sigma_2})=
\sum_{i\in I}(-1)^{\sig{\zeta_i}}\rho_M\big(\tau_{\Sigma_1,\Sigma_2',\overline{\Sigma_2};\partial M}(\mathbf{1}\tens\mathbf{1}\tens\zeta_i\tens\iota_{\Sigma_2}(\zeta_i))\big)
\nonumber\\
\hphantom{c(M;\Sigma_2',\overline{\Sigma_2})}{}
=\sum_{i\in I}(-1)^{\sig{\zeta_i}}\rho_P\big(\tau_{\Sigma_1,\overline{\Sigma_2};\partial P}(\mathbf{1}\tens\iota_{\Sigma_2}(\zeta_i))\big)\rho_Q\big(\tau_{\Sigma_3,\Sigma_2';\partial Q}(\mathbf{1}\tens\zeta_i)\big)
\nonumber\\
\hphantom{c(M;\Sigma_2',\overline{\Sigma_2})}{}
=\sum_{i\in I}(-1)^{\sig{\zeta_i}}\langle\zeta_i,\mathbf{1}\rangle_{\Sigma_2}\;\rho_Q\big(\tau_{\Sigma_3,\Sigma_2';\partial Q}(\mathbf{1}\tens\zeta_i)\big)
=\rho_Q\big(\tau_{\Sigma_3,\Sigma_2';\partial Q}(\mathbf{1}\tens\mathbf{1})\big)
=1.
\nonumber
\end{gather}
We have used here equation~\eqref{eq:gaid}, Proposition~\ref{prop:dglueampl}, Proposition~\ref{prop:amplevol}, and   
Def\/inition~\eqref{eq:defamplvac}.
\end{proof}

In particular, this applies to the standard globally hyperbolic setting described in Section~\ref{sec:cevol}.
The quantization of this setting thus proceeds without any obstruction from a~potential ill-def\/inedness of the gluing
anomaly.

\subsection{Renormalization}

The necessity of an additional integrability condition, here in the form of the existence of the gluing anomaly factor,
in order for the quantum theory to be well def\/ined is somewhat unsatisfactory.
This is especially so given that we know that allowing for both, non-trivial topologies and inf\/inite-dimensional
state spaces, will likely lead to a~violation of this condition.
Clearly, throwing out axiom (T5b) is not an option as this forms a~corner stone of the coherence of the whole edif\/ice
of TQFT and consequently of the GBF.
It is therefore natural to look for some kind of ``renormalization'' of the gluing anomaly.
That is, one would ``regularize'' the quantities on both sides of relation~\eqref{eq:glueax1} by introducing some
``cut-of\/f''.
Then the equation would have to hold in the ``limit'' that the ``cut-of\/f'' is taken away.

In the fermionic case, to which we restrict in the following, this informal description of renormalization can be
realized in a~precise way.
Moreover, the hard work for this has already been done in Section~\ref{sec:glueh} and Appendix~\ref{sec:gproof}.
The role of the ``cut-of\/f'' will be played by restricting the space $L_{\Sigma}$ associated to the gluing
hypersurface $\Sigma$ to the subspace $L_{\Sigma,\alpha}$.
The role of the ``limit'' is taken by the projective limit in the system $\{L_{\Sigma,\alpha}\}_{\alpha\in A}$.

Def\/ine the orthogonal projector $P_{\alpha}$ as the projector onto the Fock subspace $\mathcal{F}(L_{\Sigma,\alpha})$
of the Fock space $\mathcal{F}(L_{\Sigma})$.
The way we have treated the right hand side of relation~\eqref{eq:glueax1} in Theorem~\ref{thm:glueaxb} then amounts to
saying that it is well def\/ined and coincides with the limit in $\alpha$ of the expression
\begin{gather}
\sum_{i\in I}(-1)^{\sig{\zeta_i}}\rho_M\big(\tau_{\Sigma_1,\Sigma,\overline{\Sigma'};\partial M}(\psi\tens P_{\alpha}\zeta_i\tens\iota_\Sigma(P_{\alpha}\zeta_i))\big)
\label{eq:rhsreg}
\end{gather}
whenever this limit exists.
Here $\{\zeta_i\}_{i\in I}$ is an arbitrary ON-basis of $\mathcal{H}_{\Sigma}=\mathcal{F}(L_{\Sigma})$ whose choice is
irrelevant, as we saw already in Section~\ref{sec:glueh}.
Expression~\eqref{eq:rhsreg} will play the role of the regularized version of the right hand side of
relation~\eqref{eq:glueax1}.
Our def\/inition of the regularized gluing anomaly factor from Section~\ref{sec:glueh}, essentially
equation~\eqref{eq:gafid}, can be rewritten in a~similar way as follows
\begin{gather}
c_{\alpha}(M;\Sigma,\overline{\Sigma'})\coloneqq\sum_{i\in I}(-1)^{\sig{\zeta_i}}\rho_M\big(\tau_{\Sigma_1,\Sigma,\overline{\Sigma'};\partial M}(\mathbf{1}\tens P_{\alpha}\zeta_i\tens\iota_\Sigma(P_{\alpha}\zeta_i))\big).
\label{eq:gareg}
\end{gather}
Again, the choice of ON-basis is irrelevant.
Note that $P_{\alpha}$ projects onto a~f\/inite-dimensional subspace, so expressions~\eqref{eq:rhsreg}
and~\eqref{eq:gareg} are always well def\/ined.
We are now ready to state the renormalized version of axiom (T5b).
\begin{description}
\item[\rm (T5b*)] Let $M$ be a~region with its boundary decomposing as a~disjoint union $\partial
M=\Sigma_1\cup\Sigma\cup\overline{\Sigma'}$, where $\Sigma'$ is a~copy of $\Sigma$.
Let $M_1$ denote the gluing of $M$ with itself along $\Sigma$, $\overline{\Sigma'}$ and suppose that $M_1$ is a~region.
Note $\partial M_1=\Sigma_1$.
Then $\tau_{\Sigma_1,\Sigma,\overline{\Sigma'};\partial M}(\psi\tens\xi\tens\iota_\Sigma(\xi))\in\mathcal{H}_{\partial
M}^\circ$ for all $\psi\in\mathcal{H}_{\partial M_1}^\circ$ and $\xi\in\mathcal{H}_\Sigma$.
Moreover, there is $\{c_{\alpha}(M;\Sigma,\overline{\Sigma'})\}_{\alpha\in A}$ such that for any ON-basis
$\{\zeta_i\}_{i\in I}$ of $\mathcal{H}_\Sigma$ and any $\psi\in\mathcal{H}_{\partial M_1}^\circ$,
\begin{gather}
\varinjlim_{\alpha}\Bigg(\rho_{M_1}(\psi)\cdot c_{\alpha}(M;\Sigma,\overline{\Sigma'})\nonumber\\
\qquad{} -\sum_{i\in I}(-1)^{\sig{\zeta_i}}\rho_M\big(\tau_{\Sigma_1,\Sigma,\overline{\Sigma'};\partial M}(\psi\tens P_{\alpha}\zeta_i\tens\iota_\Sigma(P_{\alpha}\zeta_i))\big)\Bigg)=0.
\label{eq:glueaxren}
\end{gather}
\end{description}

\begin{thm}
Assume the geometric context of axiom {\rm (T5b*)}.
Then the renormalized composition identity~\eqref{eq:glueaxren} holds and thus axiom {\rm (T5b*)} holds.
Moreover, $c_{\alpha}(M;\Sigma,\overline{\Sigma'})$ can be taken to be specified by equation~\eqref{eq:gareg}.
\end{thm}

\begin{proof}
By our def\/inition of the subspace $\mathcal{H}_{\partial M_1}^\circ\subseteq\mathcal{H}_{\partial M_1}$ it is
suf\/f\/icient to prove the renormalized composition identity~\eqref{eq:glueaxren} for states of the form
$\cs{\phi_1,\dots,\phi_n}$ for $\phi_1,\dots,\phi_n\in L_{\partial M_1}$.
Moreover, as in the proof of Theorem~\ref{thm:glueaxb} it will be suf\/f\/icient to even suppose
$\phi_1,\dots,\phi_n\in L_{\tilde{M}_1}$.
Also, proceeding as in that proof there are due to axiom (C2) unique elements $\tilde{\phi}_1,\dots,\tilde{\phi}_n\in
L_{\Sigma}$ such that $(\phi_i,\tilde{\phi}_i,\tilde{\phi}_i)\in L_{\tilde{M}}$ for $i\in\{1,\dots,n\}$.
Now let $\beta\in A$ such that $L_{\Sigma,\beta}$ is the Krein subspace of~$L_{\Sigma}$ generated by
$\{\tilde{\phi}_1,\dots,\tilde{\phi}_n\}$.
Then Lemma~\ref{lem:gfid} implies for any $\gamma\ge\beta$ the equality
\begin{gather*}
\rho_{M_1}(\cs{\phi_1,\dots,\phi_n})\cdot c_{\gamma}(M;\Sigma,\overline{\Sigma'})
\\
\qquad
{} -\sum_{i\in I}(-1)^{\sig{\zeta_i}}\rho_M\big(\tau_{\Sigma_1,\Sigma,\overline{\Sigma'};\partial M}(\cs{\phi_1,\dots,\phi_n}\tens P_{\gamma}\zeta_i\tens\iota_\Sigma(P_{\gamma}\zeta_i))\big)=0.
\end{gather*}
(This is just a~rewrite of equation~\eqref{eq:gfid}.)
This implies the validity of the equation in the limit, i.e.,~\eqref{eq:glueaxren} is valid for
$\psi=\cs{\phi_1,\dots,\phi_n}$.
This completes the proof.
\end{proof}

\section{Remarks on the probability interpretation}
\label{sec:prob}

In the standard formulation of quantum theory an indef\/inite inner product on a~state space would lead to serious
problems with the probability interpretation.
The situation turns out to be more nuanced in the GBF.
A key fact to keep in mind is that the concept of state space is considerably more general in the GBF.
Only state spaces associated with certain rather special hypersurfaces may be compared in their role to the state
spaces of the standard formulation.
In particular, as explained in Section~\ref{sec:evol}, in the standard globally hyperbolic setting, restricting
ourselves to state spaces that play this role leaves us with only Hilbert spaces, even in fermionic theories.

In the following we shall consider this issue from an intrinsic GBF perspective.
Recall to this end the probability interpretation for amplitudes in the GBF in the setting of Hilbert
spaces~\cite{Oe:gbqft}.
That is, we have a~region $M$ and a~boundary Hilbert space $\mathcal{H}_{\partial M}$.
To def\/ine a~measurement we need to specify a~closed subspace $\mathcal{S}\subseteq\mathcal{H}_{\partial M}$ that
encodes ``preparation'' and another closed subspace $\mathcal{A}\subseteq\mathcal{S}$ that encodes the ``question'' we
are asking in the measurement.
The probability for an af\/f\/irmative outcome of the measurement (if def\/ined) is then given by the formula
\begin{gather}
P(\mathcal{A}|\mathcal{S})=\frac{\sum\limits_{i\in J}|\rho_M(\xi_i)|^2}{\sum\limits_{i\in I}|\rho_M(\xi_i)|^2}.
\label{eq:probampl}
\end{gather}
Here $\{\xi_i\}_{i\in I}$ is an orthonormal basis of $\mathcal{S}$ that restricts to an orthonormal basis
$\{\xi_i\}_{i\in J}$ of $\mathcal{A}$ with $J\subseteq I$.
The quantity~\eqref{eq:probampl} has then all the right properties of a~probability, reduces to the usual probabilities
in the standard formulation, admits a~notion of probability conservation in spacetime etc.~\cite{Oe:gbqft,Oe:probgbf}.

So what changes if we wish to generalize from Hilbert spaces to Krein spaces? Apparently not much as the
quantity~\eqref{eq:probampl} still has all the right properties and all the relevant arguments
in~\cite{Oe:gbqft,Oe:probgbf} still apply.
However, there is a~crucial dif\/ference.
For expression~\eqref{eq:probampl} to be def\/ined in the f\/irst place the subspaces $\mathcal{A}$ and $\mathcal{S}$
have to be Krein spaces in the sense of Section~\ref{sec:krein}.
This amounts to $\mathcal{A}$ and $\mathcal{S}$ being decomposable as direct sums in terms of positive and negative
parts.
Concretely, given the decomposition $\mathcal{H}_{\partial M}=\mathcal{H}_{\partial M}^+\oplus\mathcal{H}_{\partial
M}^-$ we must have $\mathcal{A}=\mathcal{A}^+\oplus\mathcal{A}^-$ and $\mathcal{S}=\mathcal{S}^+\oplus\mathcal{S}^-$,
where $\mathcal{A}^+\subseteq\mathcal{S}^+\subseteq\mathcal{H}_{\partial M}^+$ and
$\mathcal{A}^-\subseteq\mathcal{S}^-\subseteq\mathcal{H}_{\partial M}^-$.

The Krein space case is thus more restrictive than the Hilbert space case in the following sense.
Recall that $\mathcal{H}_{\partial M}$ is also a~Hilbert space by taking the inner product of $\mathcal{H}_{\partial
M}^+\oplus\overline{\mathcal{H}_{\partial M}^-}$.
All the Krein subspaces of $\mathcal{H}_{\partial M}$ are Hilbert subspaces of $\mathcal{H}_{\partial
M}^+\oplus\overline{\mathcal{H}_{\partial M}^-}$, but there are many more Hilbert subspaces.
One potential way to make physical sense of this limitation in the Krein space case would be to think of it as
originating from a~kind of ``superselection rule''.
This superselection rule would here amount to saying that superpositions between states in the subspace
$\mathcal{H}_{\partial M}^+$ and states in the subspace $\mathcal{H}_{\partial M}^-$ do not make physical sense.

An undesirable feature of the setting as described so far is its apparent dependence on conventions.
Consider a~general boundary quantum f\/ield theory given in terms of the axioms of Section~\ref{sec:coreaxioms}.
If we invert the global orientation of spacetime, i.e., the orientation of all regions and hypersurfaces of the
spacetime system, we obtain again a~theory satisfying the axioms.
Indeed, physically the new theory should really be equivalent to the old theory.
However, given a~region $M$, the decomposition of its boundary Krein space $\mathcal{H}_{\partial M}$ into a~direct sum
of positive and negative parts is generally \emph{different} from the corresponding decomposition of
$\mathcal{H}_{\overline{\partial M}}=\iota_{\partial M}(\mathcal{H}_{\partial M})$ (the boundary Krein space
$\mathcal{H}_{\partial M}$ in the orientation-reversed theory).
This would seem to be an indication against a~physical signif\/icance of this decomposition.

However, there is another superselection rule at play here: The amplitude of a~state with odd fermionic degree must
vanish.
This is stipulated in our formulation in axiom (T4), via the $\mathbb{Z}_2$-gradedness of the amplitude map.
It is of course a~basic and long established fact of quantum f\/ield theory, interacting or not\footnote{The
conventional way of expressing this would be to say that the transition amplitude between a~state with even and a~state
with odd fermion number must vanish.}. Let us denote by $\mathcal{H}_{\partial M,0}$ the part of $\mathcal{H}_{\partial
M}$ with $\mathbb{Z}_2$-degree $0$ (equivalent to even fermion number).
By axiom (T1b) the restriction of $\iota_{\partial M}$ to $\mathcal{H}_{\partial M,0}$ is an isometry.
In particular, the decomposition into positive and negative part of $\mathcal{H}_{\partial M,0}$ is the same as that of
$\iota_{\partial M}(\mathcal{H}_{\partial M,0})$.
So, if we also enforce this superselection rule in the choice of the subspaces $\mathcal{S}$ and $\mathcal{A}$, the new
rule behaves well.
Concretely, this enforcement consists of restricting the subspaces $\mathcal{S}$ and $\mathcal{A}$ to be subspaces of
$\mathcal{H}_{\partial M,0}$.\footnote{We could more generally let them be direct sums with another part of
$\mathbb{Z}_2$-degree $1$.
However, there is little point in considering these extra parts as they do not contribute to the amplitude and hence to
the probability formula~\eqref{eq:probampl}.}

Let us emphasize that the new superselection rule does not show in predictions that can also be made within the
standard formalism for conventional bosonic or fermionic f\/ield theories.
Indeed, for bosonic theories there does not seem to be any realistic example theory that would require genuine Krein
spaces instead of Hilbert spaces.
For fermionic theories, Krein spaces do necessarily appear in a~GBF setting as we have seen.
However, the predictions of measurement outcomes that can be made in the standard formalism are tied to transition
amplitudes.
In these, $\mathcal{S}$ encodes a~condition on the initial hypersurface and $\mathcal{A}$ an additional condition on
the f\/inal hypersurface.
However, due to the coincidence of preferred algebraic and geometric notion of time in the standard globally hyperbolic
setting (recall Sections~\ref{sec:cevol} and~\ref{sec:evol}), this implies that~$\mathcal{S}$ and~$\mathcal{A}$ then
automatically satisfy the new superselection rule, supposing that we explicitly enforce the usual fermionic
$\mathbb{Z}_2$-superselection rule.

\section{A brief comparison to holomorphic quantization}
\label{sec:comphol}

The previous treatment of the free bose f\/ield and its quantization in the GBF in~\cite{Oe:holomorphic} dif\/fers from
the present one principally in the quantization method used.
While we employ a~Fock space construction here, a~holomorphic quantization was performed in~\cite{Oe:holomorphic}.
Nevertheless, the results are equivalent.
We shall provide a~basic dictionary between the two approaches here, restricted to the Hilbert space case.

We f\/irst recall the quantization of state spaces in the holomorphic approach.
Let $\Sigma$ be a~hypersurface.
Denote by $\hat{L}_{\Sigma}$ the algebraic dual of the topological dual of $L_{\Sigma}$, equipped with the weak$^*$
topology.
Note that there is a~natural inclusion $L_{\Sigma}\subseteq\hat{L}_{\Sigma}$.
The inner product of~$L_{\Sigma}$ induces a~Gaussian Borel measure $\nu_{\Sigma}$ on $\hat{L}_{\Sigma}$.
The square integrable holomorphic functions on~$\hat{L}_{\Sigma}$ form a~separable Hilbert space
$H^2(\hat{L}_{\Sigma},\nu_{\Sigma})$ with inner product
\begin{gather*}
\langle\psi',\psi\rangle=\int_{\hat{L}_{\Sigma}}\overline{\psi'(\xi)}\psi(\xi)\,\mathrm{d}\nu_{\Sigma}(\xi).
\end{gather*}
It turns out that elements of $H^2(\hat{L}_{\Sigma},\nu_{\Sigma})$ are uniquely determined by their values on
$L_{\Sigma}$.
That is, we may think of $H^2(\hat{L}_{\Sigma},\nu_{\Sigma})$ as a~space of holomorphic functions on $L_{\Sigma}$
rather than on~$\hat{L}_{\Sigma}$.
This def\/ines the state space $\mathcal{H}_{\Sigma}$ in the holomorphic approach.
The Fock space $\mathcal{F}(L_{\Sigma})$, which is the model for $\mathcal{H}_{\Sigma}$ in the present work, is related
to it by an isometric isomorphism $T:\mathcal{F}(L_{\Sigma})\to H^2(\hat{L}_{\Sigma},\nu_{\Sigma})$ with the following
characteristics.
$T$ is the sum of isometries $T_n:\mathcal{F}_n(L_{\Sigma})\to H^2(\hat{L}_{\Sigma},\nu_{\Sigma})$ given by
\begin{gather*}
\left(T_n(\psi)\right)(\xi)=\psi(\xi,\dots,\xi)
\qquad
\forall\, \psi\in\mathcal{F}_n(L_{\Sigma}),\quad \forall\, \xi\in L_{\Sigma}.
\end{gather*}
In particular, a~generating state $\cs{\xi_1,\dots,\xi_n}$ with $\xi_1,\dots,\xi_n\in L_{\Sigma}$ is mapped to the wave
function
\begin{gather}
\left(T_n(\cs{\xi_1,\dots,\xi_n})\right)(\eta)=\prod_{i=1}^n\{\xi_i,\eta\}_{\Sigma}
\qquad
\forall\,\eta\in L_{\Sigma}.
\label{eq:gswv}
\end{gather}

Of particular importance in the holomorphic setting are the coherent states $K_{\xi}$ which are associated to elements
$\xi\in L_{\Sigma}$.
As elements of $H^2(\hat{L}_{\Sigma},\nu_{\Sigma})$ these take the form
\begin{gather*}
K_{\xi}(\eta)=\exp\left(\tfrac{1}{2}\{\xi,\eta\}_{\Sigma}\right)
\qquad
\forall\, \eta\in L_{\Sigma}.
\end{gather*}
Expanding the exponential and comparing to formula~\eqref{eq:gswv}, we see that the coherent state $K_{\xi}$ may be
written in terms of generating states as the sum
\begin{gather*}
K_{\xi}=\sum_{n=0}^{\infty}\frac{1}{n!2^n}T_n(\cs{\xi,\dots,\xi}).
\end{gather*}
Conversely, we can recover a~generating state $\cs{\xi_1,\dots,\xi_n}$ for $\xi_1,\dots,\xi_n\in L_{\Sigma}$ from
a~coherent state by using partial derivatives,
\begin{gather}
T_n(\cs{\xi_1,\dots,\xi_n})=
2^n\frac{\partial}{\partial\lambda_1}\cdots\frac{\partial}{\partial\lambda_n}K_{\lambda_1\xi_1+\cdots+\lambda_n\xi_n}\bigg|_{\lambda_1=0,\dots,\lambda_n=0}.
\label{eq:cohstogs}
\end{gather}

We proceed to the comparison of the amplitude maps $\rho_M:\mathcal{H}_{\partial M}\to\mathbb{C}$.
This was given in the holomorphic approach through an integral
\begin{gather}
\rho_M(\psi)=\int_{\hat{L}_{\tilde{M}}}\psi(\xi)\,\mathrm{d}\nu_{\tilde{M}}(\xi).
\label{eq:holomquant}
\end{gather}
The integral is over the subspace $\hat{L}_{\tilde{M}}\subseteq\hat{L}_{\partial M}$ that carries an induced Gaussian
measure $\nu_{\tilde{M}}$.
This, however, was shown to lead for coherent states to the remarkably simple expression
\begin{gather*}
\rho_M(K_{\xi})=\exp\left(\tfrac{1}{4}\{\widehat{\xi},\widehat{\xi}\}_{\partial M}\right),
\end{gather*}
where we are using the conventions of Section~\ref{sec:ampl}.
It is then straightforward, using formula~\eqref{eq:cohstogs} to extract the amplitude for a~generating state.
The result is precisely given by the def\/initions~\eqref{eq:defamplvac},~\eqref{eq:defamplodd}
and~\eqref{eq:defampleven} of Section~\ref{sec:ampl}.
Thus, not only the state spaces, but also the amplitudes are equivalent in both approaches.
In fact, this is actually the way in which we obtained the bosonic versions of
formulas~\eqref{eq:defamplvac},~\eqref{eq:defamplodd} and~\eqref{eq:defampleven}.

Its was shown in~\cite{Oe:affine,Oe:feynobs} that the quantization rule~\eqref{eq:holomquant} is precisely equivalent
to the usual Feynman path integral.
So this applies also to the quantization scheme put forward in the present work.

\section{Outlook}
\label{sec:outlook}

An obvious question concerns the generalization to interacting f\/ield theories.
Neither the core axioms (Section~\ref{sec:coreaxioms}) nor the probability interpretation (Section~\ref{sec:prob}) are
specif\/ic to the linear setting.
What needs to be generalized is the encoding of the classical theory and the quantization prescription.
In the bosonic case a~perturbative approach to interactions was brief\/ly discussed in~\cite{Oe:affine}.
This basically amounts to the standard techniques of quantum f\/ield theory motivated through the Feynman path
integral, involving generating functions and leading to Feynman diagrams.
It should not be dif\/f\/icult to adapt this to the fermionic case.
In the bosonic case the holomorphic quantization approach followed in~\cite{Oe:holomorphic} is suggestive of
non-perturbative generalizations in view of its motivation from geometric quantization.
This was exploited in~\cite{Oe:affine} in a~mild way by generalizing from linear to af\/f\/ine theories.
Due to the large body of work on geometric quantization of non-linear systems (although mostly limited to systems with
f\/initely many degrees of freedom)~\cite{Woo:geomquant} one should expect that this line of investigation could be
pushed much further.
In the purely fermionic case this looks much less promising, but one would expect interesting fermionic systems to
contain also bosonic degrees of freedom.

\looseness=-1
Apart from state spaces and amplitudes, the GBF also naturally accommodates a~notion of observable~\cite{Oe:obsgbf}.
For linear bosonic f\/ield theory this was comprehensively developed in~\cite{Oe:feynobs}.
An essentially parallel treatment should be possible for linear fermionic f\/ield theory.
A likely limitation from the axiomatic point of view would be the restriction for observables to have fermionic degree
$0$.
In the light of the remarks in Section~\ref{sec:prob} this would likely also ensure compatibility with the probability
interpretation and thus a~consistent notion of expectation value.

As shown in Section~\ref{sec:cevol}, classical linear fermionic f\/ield theory, axiomatized according to
Section~\ref{sec:classax}, turns out to exhibit an ``emergent'' (algebraic) notion of time.
As explained in Section~\ref{sec:evol}, the quantized theory inherits this notion.
Moreover, for theories with Lorentzian metric backgrounds (such as the Dirac theory considered in
Section~\ref{sec:dirac}) this notion seems to agree precisely with the usual (geometric) notion of time evolution.
This suggests to examine linear fermionic theories that do not come with a~(Lorentzian) metric background to analyze
the physical meaning of this emergent notion of time.
Going beyond the linear case, the core axioms of the quantum theory (Section~\ref{sec:coreaxioms}) alone do not seem to
induce any such notion.
This begs the question whether this phenomenon is strictly limited to the linear theory or has a~more general
counterpart.

The generalization of state spaces from Hilbert spaces to Krein spaces is a~serious step in view of the justif\/ied
reservations against giving up Hilbert spaces in quantum theory.
However, as we have shown in Section~\ref{sec:prob}, taking this step in the GBF is not at all the same thing as taking
it in the standard formulation of quantum theory.
Indeed, as we have seen there, a~consistent probability interpretation with Krein spaces in the GBF is perfectly
possible.
Even though the necessity for Krein spaces only arises in the fermionic case, this suggests to permit them also in
bosonic theories.
Indeed, throughout this work, we have allowed for this theoretical possibility.
On the other hand, the known realistic bosonic f\/ield theories work f\/ine with Hilbert spaces.
This might change though when we consider them on non-standard hypersurfaces.
(However, in the specif\/ic examples considered in~\cite{Col:desitterpaper,Oe:kgtl, Oe:timelike} there seems to be no
indication to this ef\/fect.) Also, Krein spaces might become useful when abandoning Lorentzian background metrics.

A subject we have touched upon neither in~\cite{Oe:holomorphic} nor here is that of \emph{corners}, i.e., the admission
of hypersurfaces with boundaries.
The problem is not that it would be dif\/f\/icult to write down axioms for this and have them satisf\/ied.
Indeed, we could easily include corners as follows.
First we modify the axioms for a~spacetime system to include hypersurfaces with boundaries.
The notion of a~disjoint decomposition of a~hypersurface would be replaced by a~more general notion that allows the
components to have boundaries.
(Precisely this was done in~\cite{Oe:2dqym}.) We could then proceed to modify both the classical axioms of
Section~\ref{sec:classax} and the core axioms of Section~\ref{sec:coreaxioms} by simply replacing the disjoint notion
of decomposition of hypersurfaces with the more general one.
This would not af\/fect at all the ``algebraic'' side of things.
Everything would still work, including the quantization scheme and the proofs.
The problem with this is that we would no longer be able to capture standard examples of quantum f\/ield theories.
Generically, there would not be a~suitable Hilbert or Krein space to be associated with a~hypersurface with boundary,
even in the classical linear theory.
This is related to the non-locality of the complex structure $J_{\Sigma}$ associated with the hypersurface $\Sigma$,
compare Sections~\ref{sec:quantgbf} and~\ref{sec:diraccompl}.
More generally, this problem can be seen in the light of the Reeh--Schlieder theorem~\cite{ReSc:unitequiv}.
This essentially implies that state spaces cannot be localized in parts of hypersurfaces.
We remark that, on the other hand, in the case of 2-dimensional Yang--Mills theory a~certain modif\/ication of the core
axioms was shown to correctly implement corners~\cite{Oe:2dqym}.

\appendix

\section{Proof of the main lemma}
\label{sec:gproof}

To prepare for the proof of Lemma~\ref{lem:gfid} we note some useful identities.
We assume the geometric context of axiom (T5b) or equivalently that of (T5b*).
Consider $\phi\in L_{M_1}$.
Due to the exact sequence~\eqref{eq:xsbdy} of axiom (C7) there is then $\tilde{\phi}\in L_{\Sigma}$ such that
$(\phi,\tilde{\phi},\tilde{\phi})\in L_{M}\subseteq L_{\Sigma_1}\times L_{\Sigma}\times
L_{\overline{\Sigma'}}$.\footnote{As elsewhere, our simplif\/ied notation does not explicitly distinguish between $L_M$
and $L_{\tilde{M}}$ etc.} With this notation we have the following identities.
We omit the straightforward proofs.
\begin{lem}
\label{lem:subid}
Let $\xi\in L_{\Sigma}$ and $\phi\in L_{M_1}$.
Then
\begin{gather}
\big\{\widehat{(0,\xi,0)},\widehat{(\phi,0,0)}\big\}_{\partial M}=
\{\xi,\tilde{\phi}\}_{\Sigma}-\big\{\widehat{(0,\xi,0)},\widehat{(0,\tilde{\phi},\tilde{\phi}}\big\}_{\partial M},
\label{eq:subid1}
\\
\big\{\widehat{(0,0,\xi)},\widehat{(\phi,0,0)}\big\}_{\partial M}=
\{\xi,\tilde{\phi}\}_{\overline{\Sigma'}}-\big\{\widehat{(0,0,\xi)},\widehat{(0,\tilde{\phi},\tilde{\phi})}\big\}_{\partial M}.
\label{eq:subid2}
\end{gather}
\end{lem}

\begin{lem}
\label{lem:phiid}
Let $\phi_1,\phi_2\in L_{M_1}$.
Then
\begin{gather}
\{\phi_1,\phi_2\}_{\Sigma_1}=
\big\{\widehat{(\phi_1,0,0)},\widehat{(\phi_2,0,0)}\big\}_{\partial M}
-\big\{\widehat{(0,\tilde{\phi}_1,\tilde{\phi}_1)},\widehat{(0,\tilde{\phi}_2,\tilde{\phi}_2)}\big\}_{\partial M}
\nonumber
\\
\phantom{\{\phi_1,\phi_2\}_{\Sigma_1}=}
{}+\{\tilde{\phi}_1,\tilde{\phi}_2\}_{\Sigma}
+\{\tilde{\phi}_1,\tilde{\phi}_2\}_{\overline{\Sigma'}}.
\label{eq:phiid}
\end{gather}
\end{lem}
\begin{proof}[Proof of Lemma~\ref{lem:gfid}] \looseness=-1  We f\/irst note that the left hand side and the right hand side of
expression~\eqref{eq:gfid} are both necessarily zero if $n$ is odd, due to def\/inition~\eqref{eq:defamplodd}.
We may thus assume that~$n$ is even.
To simplify notation we replace~$n$ in the following by~$2n$, where the new variable~$n$ is not necessarily even.
We start by evaluating the right hand side of expression~\eqref{eq:gfid}, which we shall denote by~$R$ (in order to avoid sub-indices we shall abbreviate~$\xi_{a_i}$ by~$\xi_i$ throughout this proof)
\begin{gather}
R=\sum_{m=0}^\infty\frac{1}{2^m\,m!}\sum_{a_1,\dots,a_m\in N_{\alpha}}(-1)^{\sum\limits_{i=1}^m\sig{\xi_{i}}}
\nonumber
\\
\hphantom{R=}{}
\times
\rho_M\left(\tau_{\Sigma_1,\Sigma,\overline{\Sigma'};\partial M}\left(\cs{\phi_1,\dots,\phi_{2n}}\tens\cs{\xi_{1},\dots,\xi_{m}}\tens\iota_{\Sigma}(\cs{\xi_{1},\dots,\xi_{m}})\right)\right)
\label{eq:ga1}
\\
\hphantom{R}{}
=\sum_{m=0}^\infty\frac{\kappa^m}{2^m\,m!}\sum_{a_1,\dots,a_m\in N_{\alpha}}(-1)^{\sum\limits_{i=
1}^m\sig{\xi_{i}}}\rho_M\big(\cs{(\phi_1,0,0),\dots,(\phi_{2n},0,0),
\nonumber
\\
\left.
\hphantom{R=}{}\qquad
(0,\xi_{1},0),\dots,(0,\xi_{m},0),(0,0,\xi_{m}),\dots,(0,0,\xi_{1})}\right)
\label{eq:ga2}
\\
\hphantom{R}{}
=\sum_{m=0}^\infty\sum_{a_1,\dots,a_m\in N_{\alpha}}(-1)^{\sum\limits_{i=1}^m\sig{\xi_{i}}}\sum_{\substack{\{l,k,a,b,c,d:2n=2l+a+b,\\
m=2c+b+k=2d+a+k\}}}\frac{2^{n-l-c-d}}{l!k!a!b!c!d!}\sum_{\substack{\sigma\in S^{2n}\\
\nu\in S^m}}\kappa^{|\sigma|+|\nu|}
\nonumber
\\
\hphantom{R=}{}
\times
\kappa^{d+a+k+\frac{l(l-1)+b(b-1)+c(c-1)+d(d-1)}{2}}
\left(\prod_{i=1}^l\{\widehat{(\phi_{\sigma(i)},0,0)},\widehat{(\phi_{\sigma(l+i)},0,0)}\}_{\partial M}\right)
\nonumber
\\
\hphantom{R=}{}
\times
\left(\prod_{i=1}^a\{\widehat{(\phi_{\sigma(2l+i)},0,0)},\widehat{(0,0,\xi_{\nu(i)})}\}_{\partial M}\right)
\left(\prod_{i=1}^b\{\widehat{(\phi_{\sigma(2l+a+i)},0,0)},\widehat{(0,\xi_i,0)}\}_{\partial M}\right)
\nonumber
\\
\hphantom{R=}{}
\times
\left(\prod_{i=1}^k\{\widehat{(0,\xi_{b+i},0)},\widehat{(0,0,\xi_{\nu(a+i)})}\}_{\partial M}\right)
\left(\prod_{i=1}^c\{\widehat{(0,\xi_{b+k+i},0)},\widehat{(0,\xi_{b+k+c+i},0)}\}_{\partial M}\right)
\nonumber
\\
\hphantom{R=}{}
\times
\left(\prod_{i=1}^d\{\widehat{(0,0,\xi_{\nu(a+k+i)})},\widehat{(0,0,\xi_{\nu(a+k+d+i)})}\}_{\partial M}\right).
\label{eq:ga3}
\end{gather}
The step from expression~\eqref{eq:ga1} to~\eqref{eq:ga2} consists of evaluating the maps $\iota_{\Sigma}$ and
$\tau_{\Sigma_1,\Sigma,\overline{\Sigma'};\partial M}$.
The next step consists of applying the amplitude formula~\eqref{eq:defampleven}.
This generates all possible pairings of 1-particle states.
In order to organize these comprehensively, we introduce the variab\-les~$l$,~$k$, $a$, $b$, $c$, $d$.
These are non-negative integers that obey certain relations as indicated.
They count the dif\/ferent types of pairing that appear.
This reorganization and the associated particular summation over certain permutations leads to combinatorial factors as
indicated in the f\/irst line of expression~\eqref{eq:ga3}.
At the same time, reorderings that occur lead in the fermionic case to a~sign factor, which is written as a~power of
$\kappa$ at the end of the f\/irst line and in the second line of expression~\eqref{eq:ga3}.
Note in particular, that implicit use was made of the fact that the summation over the indices $a_1,\dots,a_m$ is
symmetrical.
This allowed to remove a~summation over permutations of the factors $\xi_1,\dots,\xi_m$ and thus a~cancellation of the
factor $1/(m!)$ that appears in~\eqref{eq:ga2}.
We proceed to perform the substitution of the identities~\eqref{eq:subid1} and~\eqref{eq:subid2} of
Lemma~\ref{lem:subid}, yielding
\begin{gather*}
R=\sum_{m=0}^\infty\sum_{a_1,\dots,a_m\in N_{\alpha}}(-1)^{\sum\limits_{i=1}^m\sig{\xi_{i}}}\sum_{\substack{\{l,k,a,b,c,d:2n=
2l+a+b,\\
m=2c+b+k=2d+a+k\}}}\frac{2^{n-l-c-d}}{l!k!a!b!c!d!}
\\
\hphantom{R=}{}
\times
\sum_{\substack{\sigma\in S^{2n}\\
\nu\in S^m}}\kappa^{|\sigma|+|\nu|}
\kappa^{d+a+k+\frac{l(l-1)+b(b-1)+c(c-1)+d(d-1)}{2}}
\left(\prod_{i=1}^l\big\{\widehat{(\phi_{\sigma(i)},0,0)},\widehat{(\phi_{\sigma(l+i)},0,0)}\big\}_{\partial M}\right)
\\
\hphantom{R=}{}
\times
\left(\prod_{i=
1}^a\left(\kappa\{\xi_{\nu(i)},\tilde{\phi}_{\sigma(2l+i)}\}_{\overline{\Sigma'}}
-\big\{\widehat{(0,\tilde{\phi}_{\sigma(2l+i)},0)},\widehat{(0,0,\xi_{\nu(i)})}\big\}_{\partial M}\right.\right.
\\
\left.\left.
\hphantom{R=}{} -\big\{\widehat{(0,0,\tilde{\phi}_{\sigma(2l+i)})},\widehat{(0,0,\xi_{\nu(i)})}\big\}_{\partial M}\right)\right)
\\
\hphantom{R=}{}
\times
\left(\prod_{i=
1}^b\left(\kappa\{\xi_i,\tilde{\phi}_{\sigma(2l+a+i)}\}_{\Sigma}-\big\{\widehat{(0,{\tilde\phi_{\sigma(2l+a+i)}},0)},\widehat{(0,\xi_i,0)}\big\}_{\partial M}\right.\right.
\\
\left.\left.
\hphantom{R=}{}
-\big\{\widehat{(0,0,\tilde{\phi}_{\sigma(2l+a+i)})},\widehat{(0,\xi_i,0)}\big\}_{\partial M}\right)\right)
\\
\hphantom{R=}{}
\times
\left(\prod_{i=1}^k\big\{\widehat{(0,\xi_{b+i},0)},\widehat{(0,0,\xi_{\nu(a+i)})}\big\}_{\partial M}\right)
\left(\prod_{i=1}^c\big\{\widehat{(0,\xi_{b+k+i},0)},\widehat{(0,\xi_{b+k+c+i},0)}\big\}_{\partial M}\right)
\\
\hphantom{R=}{}
\times
\left(\prod_{i=1}^d\big\{\widehat{(0,0,\xi_{\nu(a+k+i)})},\widehat{(0,0,\xi_{\nu(a+k+d+i)})}\big\}_{\partial M}\right).
\end{gather*}
We proceed to expand the products of sums that have arisen from the substitutions.
In doing so we introduce new variables $e$, $f$, $g$, $h$, $j$, $o$ counting the occurrences of the summands, splitting $a=e+f+g$ and
$b=h+j+o$.
At the same time the variables $a$, $b$ are eliminated.
This yields
\begin{gather*}
R=\sum_{m=0}^\infty\sum_{a_1,\dots,a_m\in N_{\alpha}}(-1)^{\sum\limits_{i=
1}^m\sig{\xi_{i}}}\sum_{\substack{\{l,k,c,d,e,f,g,h,j,o:\\
2n=2l+e+f+g+h+j+o,\\
m=2c+h+j+o+k\\
=2d+e+f+g+k\}}}\frac{2^{n-l-c-d}}{l!c!d!e!f!g!h!j!o!}
\\
\hphantom{R=}{}
\times
\sum_{\substack{\sigma\in S^{2n}\\
\nu\in S^m}}\kappa^{|\sigma|+|\nu|+d+h+f+g+k+h j+h o+j o}(-1)^{j+o+e+f}
\\
\hphantom{R=}{}
\times
\kappa^{\frac{l(l-1)+h(h-1)+j(j-1)+o(o-1)+c(c-1)+d(d-1)}{2}}
\left(\prod_{i=1}^l\big\{\widehat{(\phi_{\sigma(i)},0,0)},\widehat{(\phi_{\sigma(l+i)},0,0)}\big\}_{\partial M}\right)
\\
\hphantom{R=}{}
\times
\left(\prod_{i=1}^{e}\{\xi_{\nu(i)},\tilde{\phi}_{\sigma(2l+i)}\}_{\overline{\Sigma'}}\right)
\left(\prod_{i=1}^{f}\big\{\widehat{(0,\tilde{\phi}_{\sigma(2l+e+i)},0)},\widehat{(0,0,\xi_{\nu(e+i)})}\big\}_{\partial M}\right)
\\
\hphantom{R=}{}
\times
\left(\prod_{i=
1}^{g}\big\{\widehat{(0,0,\tilde{\phi}_{\sigma(2l+e+f+i)})},\widehat{(0,0,\xi_{\nu(e+f+i)})}\big\}_{\partial M}\right)
\left(\prod_{i=1}^{h}\{\xi_i,\tilde{\phi}_{\sigma(2l+e+f+g+i)}\}_{\Sigma}\right)
\\
\hphantom{R=}{}
\times
\left(\prod_{i=
1}^{j}\big\{\widehat{(0,\tilde{\phi}_{\sigma(2l+e+f+g+h+i)},0)},\widehat{(0,\xi_{h+i},0)}\big\}_{\partial M}\right)
\\
\hphantom{R=}{}
\times
\left(\prod_{i=
1}^{o}\big\{\widehat{(0,0,\tilde{\phi}_{\sigma(2l+e+f+g+h+j+i)})},\widehat{(0,\xi_{h+j+i},0)}\big\}_{\partial M}\right)
\\
\hphantom{R=}{}
\times
\left(\prod_{i=1}^k\{\widehat{(0,\xi_{h+j+o+i},0)},\widehat{(0,0,\xi_{\nu(e+f+g+i)})}\}_{\partial M}\right)
\\
\hphantom{R=}{}
\times
\left(\prod_{i=1}^c\big\{\widehat{(0,\xi_{h+j+o+k+i},0)},\widehat{(0,\xi_{h+j+o+k+c+i},0)}\big\}_{\partial M}\right)
\\
\hphantom{R=}{}
\times
\left(\prod_{i=1}^d\big\{\widehat{(0,0,\xi_{\nu(e+f+g+k+i)})},\widehat{(0,0,\xi_{\nu(e+f+g+k+d+i)})}\big\}_{\partial M}\right).
\end{gather*}
Next, we contract the factors of the form
\begin{gather*}
\{\xi_{\nu(i)},\tilde{\phi}_{\sigma(2l+i)}\}_{\overline{\Sigma'}}
\end{gather*}
with corresponding pairs, by viewing the summation in the participating $\xi$ variables as sums over complete basis.
To this end we use the identity
\begin{gather*}
\sum_{a_i\in N_{\alpha}}(-1)^{\sig{\xi_i}}\xi_i\{\xi_{i},\eta\}_{\overline{\Sigma'}}=\kappa\,\eta
\end{gather*}
for $\eta\in L_{\overline{\Sigma'}}$.
Note that this leads to a~factor $\kappa^e$.
The contraction involves a~distribution of factors over the dif\/ferent types of matching pairs.
This makes necessary the introduction of additional counting variables $p$, $q$, $r$, $s$, $t$, $u$, $v$, $w$, $x$, $y$, $z$.
At the same time the variables $e$, $h$, $j$, $o$, $k$, $c$ are eliminated.
The relation between old and new variables is the following
\begin{gather*}
e=p+q+r+s+t+2u,
\qquad
h=p+v,
\qquad
j=q+w,
\\
o=r+x,
\qquad
k=s+y,
\qquad
c=z+t+u.
\end{gather*}
Since some of the summations over basis of $L_{\Sigma}$ are removed in the process, the variable $m$ is also replaced
by a~new variable $m'$ with the relation $m=m'+e$.
However, for simplicity of notation we rename the new variable $m'$ again $m$.
After some opportune reorderings we obtain the following expression
\begin{gather}
R=\sum_{m=0}^\infty\sum_{a_1,\dots,a_m\in N_{\alpha}} \!(-1)^{\sum\limits_{i=
1}^m\sig{\xi_{i}}} \!\!\! \sum_{\substack{\{l,d,f,g,p,q,r,s,t,u,v,w,x,y,z:\\
2n=2l+2p+2q+2r+2u+f+g+s+t+v+w+x,\\
m=2z+t+v+w+x+y=2d+f+g+s+y\}}}
\!\!\! \frac{2^{n-l-d-u-z}}{l!d!f!g!p!q!r!s!t!u!v!w!x!y!z!}\nonumber\\
\hphantom{R=}{}
\times
\sum_{\substack{\sigma\in S^{2n}\\
\nu\in S^m}}\kappa^{|\sigma|+|\nu|}(-1)^{f+g+q+r+w+x}
\kappa^{d+f+g+s+v+y+(t+w)(f+g+s+v+x)+(v+x)(f+g+s)+v x}
\nonumber
\\
\hphantom{R=}{}
\times
\kappa^{\frac{l(l-1)+d(d-1)+p(p-1)+q(q-1)+r(r-1)+u(u-1)+v(v-1)+x(x-1)+z(z-1)+(t+w)(t+w-1)}{2}}
\nonumber
\\
\hphantom{R=}{}
\times
\left(\prod_{i=1}^l\big\{\widehat{(\phi_{\sigma(i)},0,0)},\widehat{(\phi_{\sigma(l+i)},0,0)}\big\}_{\partial M}\right)
\left(\prod_{i=1}^{p}\{\tilde{\phi}_{\sigma(2l+i)},\tilde{\phi}_{\sigma(2l+p+i)}\}_{\Sigma}\right)
\nonumber
\\
\hphantom{R=}{}
\times
\left(\prod_{i=1}^{q}\big\{\widehat{(0,\tilde{\phi}_{\sigma(2l+2p+i)},0)},\widehat{(0,\tilde{\phi}_{\sigma(2l+2p+q+i)},0)}\big\}_{\partial M}\right)
\nonumber
\\
\hphantom{R=}{}
\times
\left(\prod_{i=1}^{u}\big\{\widehat{(0,\tilde{\phi}_{\sigma(2l+2p+2q+i)},0)},\widehat{(0,\tilde{\phi}_{\sigma(2l+2p+2q+u+i)},0)}\big\}_{\partial M}\right)
\nonumber
\\
\hphantom{R=}{}
\times
\left(\prod_{i=1}^{r}\big\{\widehat{(0,0,\tilde{\phi}_{\sigma(2l+2p+2q+2u+i)})},\widehat{(0,\tilde{\phi}_{\sigma(2l+2p+2q+2u+r+i)},0)}\big\}_{\partial M}\right)
\nonumber
\\
\hphantom{R=}{}
\times
\left(\prod_{i=1}^{v}\{\xi_i,\tilde{\phi}_{\sigma(2l+2p+2q+2u+2r+i)}\}_{\Sigma}\right)
\nonumber
\\
\hphantom{R=}{}
\times
\left(\prod_{i=1}^{w}\big\{\widehat{(0,\tilde{\phi}_{\sigma(2l+2p+2q+2u+2r+v+i)},0)},\widehat{(0,\xi_{v+i},0)}\big\}_{\partial M}\right)
\nonumber
\\
\hphantom{R=}{}
\times
\left(\prod_{i=1}^{t}\big\{\widehat{(0,\tilde{\phi}_{\sigma(2l+2p+2q+2u+2r+v+w+i)},0)},\widehat{(0,\xi_{v+w+i},0)}\big\}_{\partial M}\right)
\nonumber
\\
\hphantom{R=}{}
\times
\left(\prod_{i=1}^{x}\big\{\widehat{(0,0,\tilde{\phi}_{\sigma(2l+2p+2q+2u+2r+v+w+t+i)})},\widehat{(0,\xi_{v+w+t+i},0)}\big\}_{\partial M}\right)
\nonumber
\\
\hphantom{R=}{}
\times
\left(\prod_{i=1}^{s}\big\{\widehat{(0,\tilde{\phi}_{\sigma(2l+2p+2q+2u+2r+v+w+t+x+i)},0)},\widehat{(0,0,\xi_{\nu(i)})}\big\}_{\partial M}\right)
\nonumber
\\
\hphantom{R=}{}
\times
\left(\prod_{i=1}^{f}\big\{\widehat{(0,\tilde{\phi}_{\sigma(2l+2p+2q+2u+2r+v+w+t+x+s+i)},0)},\widehat{(0,0,\xi_{\nu(s+i)})}\big\}_{\partial M}\right)
\nonumber
\\
\hphantom{R=}{}
\times
\left(\prod_{i=1}^{g}\big\{\widehat{(0,0,\tilde{\phi}_{\sigma(2l+2p+2q+2u+2r+v+w+t+x+s+f+i)})},\widehat{(0,0,\xi_{\nu(s+f+i)})}\big\}_{\partial M}\right)
\nonumber
\\
\hphantom{R=}{}
\times
\left(\prod_{i=1}^{y}\big\{\widehat{(0,\xi_{v+w+t+x+i},0)},\widehat{(0,0,\xi_{\nu(s+f+g+i)})}\big\}_{\partial M}\right)
\nonumber
\\
\hphantom{R=}{}
\times
\left(\prod_{i=1}^{z}\big\{\widehat{(0,\xi_{v+w+t+x+y+i},0)},\widehat{(0,\xi_{v+w+t+x+y+z+i},0)}\big\}_{\partial M}\right)
\nonumber
\\
\hphantom{R=}{}
\times
\left(\prod_{i=1}^d\big\{\widehat{(0,0,\xi_{\nu(s+f+g+y+i)})},\widehat{(0,0,\xi_{\nu(s+f+g+y+d+i)})}\big\}_{\partial M}\right).
\label{eq:gd}
\end{gather}
It turns out that certain sums in the above expression can be identif\/ied as binomial sums and simplif\/ied.
In particular, f\/ix all variables except for $q$ and $u$.
Leaving out all expressions that do not depend on these two variables (but may involve their sum), the remaining sum
$S$ is the following
\begin{gather*}
S=\sum_{\{q,u:2n=2l+2p+2q+2r+2u+f+g+s+t+v+w+x\}}\frac{2^{-u}(-1)^q}{q!u!}\,\kappa^{\frac{q(q-1)+u(u-1)}{2}}
\\
\hphantom{S=}{}
\times
\left(\prod_{i=
1}^{q}\big\{\widehat{(0,\tilde{\phi}_{\sigma(2l+2p+i)},0)},\widehat{(0,\tilde{\phi}_{\sigma(2l+2p+q+i)},0)}\big\}_{\partial M}\right)
\\
\hphantom{S=}{}
\times
\left(\prod_{i=
1}^{u}\big\{\widehat{(0,\tilde{\phi}_{\sigma(2l+2p+2q+i)},0)},\widehat{(0,\tilde{\phi}_{\sigma(2l+2p+2q+u+i)},0)}\big\}_{\partial M}\right).
\end{gather*}
A reordering yields the following expression
\begin{gather*}
S=\sum_{\{q,u:2n=2l+2p+2(q+u)+2r+f+g+s+t+v+w+x\}}\frac{2^{-u}(-1)^q}{q!u!}\,\kappa^{\frac{(q+u)(q+u-1)}{2}}
\\
\hphantom{S=}{}
\times
\left(\prod_{i=1}^{q}\big\{\widehat{(0,\tilde{\phi}_{\sigma(2l+2p+i)},0)},\widehat{(0,\tilde{\phi}_{\sigma(2l+2p+q+u+i)},0)}\big\}_{\partial M}\right)
\\
\hphantom{S=}{}
\times
\left(\prod_{i=1}^{u}\big\{\widehat{(0,\tilde{\phi}_{\sigma(2l+2p+q+i)},0)},\widehat{(0,\tilde{\phi}_{\sigma(2l+2p+q+u+q+i)},0)}\big\}_{\partial M}\right).
\end{gather*}
Introducing the new variable $a=q+u$ we recognize the above as a~binomial sum, leading to the following
simplif\/ication
\begin{gather*}
S=\sum_{\{a:2n=2l+2p+2a+2r+f+g+s+t+v+w+x\}}\frac{1}{a!}\left(\sum_{q=
0}^a\binom{a}{q}\left(\frac{1}{2}\right)^{a-q}(-1)^{q}\right)\kappa^{\frac{a(a-1)}{2}}
\\
\hphantom{S=}{}
\times
\left(\prod_{i=
1}^{a}\big\{\widehat{(0,\tilde{\phi}_{\sigma(2l+2p+i)},0)},\widehat{(0,\tilde{\phi}_{\sigma(2l+2p+a+i)},0)}\big\}_{\partial M}\right)
\\
\hphantom{S}{}
=\sum_{\{a:2n=2l+2p+2a+2r+f+g+s+t+v+w+x\}}\frac{2^{-a}(-1)^{a}}{a!}\,\kappa^{\frac{a(a-1)}{2}}
\\
\hphantom{S=}{}
\times
\left(\prod_{i=1}^{a}\big\{\widehat{(0,\tilde{\phi}_{\sigma(2l+2p+i)},0)},\widehat{(0,\tilde{\phi}_{\sigma(2l+2p+a+i)},0)}\big\}_{\partial M}\right).
\end{gather*}
We may proceed similarly with the variables~$t$ and~$w$.
Introducing $b=t+w$ we have a~residual sum
\begin{gather*}
S=\sum_{\substack{\{w,t:2n=2l+2p+2a+2r+f+g+s+t+v+w+x,
\\
m=2z+t+v+w+x+y\}}}\frac{(-1)^w}{t!w!}\kappa^{(t+w)(f+g+s+v+x)+\frac{(t+w)(t+w-1)}{2}}
\\
\hphantom{S=}{}
\times
\left(\prod_{i=
1}^{w}\big\{\widehat{(0,\tilde{\phi}_{\sigma(2l+2p+2q+2u+2r+v+i)},0)},\widehat{(0,\xi_{v+i},0)}\big\}_{\partial M}\right)
\\
\hphantom{S=}{}
\times
\left(\prod_{i=
1}^{t}\big\{\widehat{(0,\tilde{\phi}_{\sigma(2l+2p+2q+2u+2r+v+w+i)},0)},\widehat{(0,\xi_{v+w+i},0)}\big\}_{\partial M}\right)
\\
\hphantom{S}{}
=\sum_{\substack{\{b:2n=2l+2p+2a+2r+f+g+s+b+v+x,
\\
m=2z+b+v+x+y\}}}\frac{1}{b!}\left(\sum_{w=0}^b\binom{b}{w}(-1)^w\right)\kappa^{b(f+g+s+v+x)+\frac{b(b-1)}{2}}
\\
\hphantom{S=}{}
\times
\left(\prod_{i=
1}^{b}\big\{\widehat{(0,\tilde{\phi}_{\sigma(2l+2p+2q+2u+2r+v+i)},0)},\widehat{(0,\xi_{v+i},0)}\big\}_{\partial M}\right).
\end{gather*}
In this case, however, the binomial expression vanishes whenever $b\neq0$.
That is, we may eliminate the variables $t$ and $w$ from expression~\eqref{eq:gd} by simply setting $t=0$ and $w=0$.
The same applies to the pair of variables $f$ and $s$, as the reader is invited to verify.
All in all we obtain the following simplif\/ied expression for $R$,
\begin{gather*}
R=\sum_{m=0}^\infty\sum_{a_1,\dots,a_m\in N_{\alpha}}(-1)^{\sum\limits_{i=
1}^m\sig{\xi_{i}}}
\sum_{\substack{\{l,a,d,g,p,r,v,x,y,z:\\
2n=2l+2a+2p+2r+g+v+x,\\
m=2z+v+x+y=2d+g+y\}}}
\frac{2^{n-l-a-d-z}}{l!a!d!g!p!r!v!x!y!z!}
\\
\hphantom{R=}{}
\times
\sum_{\substack{\sigma\in S^{2n}\\
\nu\in S^m}}\kappa^{|\sigma|+|\nu|}(-1)^{a+g+r+x}\kappa^{d+g+v+y+(v+x)g+v x}
\\
\hphantom{R=}{}
\times
\kappa^{\frac{l(l-1)+a(a-1)+d(d-1)+p(p-1)+r(r-1)+v(v-1)+x(x-1)+z(z-1)}{2}}
\\
\hphantom{R=}{}
\times
\left(\prod_{i=1}^l\big\{\widehat{(\phi_{\sigma(i)},0,0)},\widehat{(\phi_{\sigma(l+i)},0,0)}\big\}_{\partial M}\right)
\left(\prod_{i=1}^{p}\{\tilde{\phi}_{\sigma(2l+i)},\tilde{\phi}_{\sigma(2l+p+i)}\}_{\Sigma}\right)
\\
\hphantom{R=}{}
\times
\left(\prod_{i=1}^{a}\big\{\widehat{(0,\tilde{\phi}_{\sigma(2l+2p+i)},0)},\widehat{(0,\tilde{\phi}_{\sigma(2l+2p+a+i)},0)}\big\}_{\partial M}\right)
\\
\hphantom{R=}{}
\times
\left(\prod_{i=1}^{r}\big\{\widehat{(0,0,\tilde{\phi}_{\sigma(2l+2p+2a+i)})},\widehat{(0,\tilde{\phi}_{\sigma(2l+2p+2a+r+i)},0)}\big\}_{\partial M}\right)
\\
\hphantom{R=}{}
\times
\left(\prod_{i=1}^{v}\{\xi_i,\tilde{\phi}_{\sigma(2l+2p+2a+2r+i)}\}_{\Sigma}\right)
\\
\hphantom{R=}{}
\times
\left(\prod_{i=1}^{x}\big\{\widehat{(0,0,\tilde{\phi}_{\sigma(2l+2p+2a+2r+v+i)})},\widehat{(0,\xi_{v+i},0)}\big\}_{\partial M}\right)
\\
\hphantom{R=}{}
\times
\left(\prod_{i=1}^{g}\big\{\widehat{(0,0,\tilde{\phi}_{\sigma(2l+2p+2a+2r+v+x+i)})},\widehat{(0,0,\xi_{\nu(i)})}\big\}_{\partial M}\right)
\\
\hphantom{R=}{}
\times
\left(\prod_{i=1}^{y}\big\{\widehat{(0,\xi_{v+x+i},0)},\widehat{(0,0,\xi_{\nu(g+i)})}\big\}_{\partial M}\right)
\\
\hphantom{R=}{}
\times
\left(\prod_{i=1}^{z}\big\{\widehat{(0,\xi_{v+x+y+i},0)},\widehat{(0,\xi_{v+x+y+z+i},0)}\big\}_{\partial M}\right)
\\
\hphantom{R=}{}
\times
\left(\prod_{i=1}^d\big\{\widehat{(0,0,\xi_{\nu(g+y+i)})},\widehat{(0,0,\xi_{\nu(g+y+d+i)})}\big\}_{\partial M}\right).
\end{gather*}
We proceed to contract the factors of the form
\begin{gather*}
\{\xi_{i},\tilde{\phi}_{\sigma(2l+2p+2a+2r+i)}\}_{\Sigma}
\end{gather*}
with corresponding pairs, by viewing the summation in the participating $\xi$ variables as sums over complete basis.
To this end we use the identity
\begin{gather*}
\sum_{a_i\in N_{\alpha}}(-1)^{\sig{\xi_i}}\xi_i\{\xi_{i},\eta\}_{\Sigma}=\eta
\end{gather*}
for $\eta\in L_{\Sigma}$.
The contraction involves a~distribution of factors over the dif\/ferent types of matching pairs.
This makes necessary the introduction of additional counting variables $b$, $c$,~$e$,~$f$,~$h$,~$j$,~$k$.
At the same time the variables $v$, $g$, $y$, $d$ are eliminated.
The relation between old and new variables is the following
\begin{gather*}
v=b+c+e+2f,\qquad g=b+h,\qquad y=c+j,\qquad d=k+e+f.
\end{gather*}
Since some of the summations over basis of $L_{\Sigma}$ are removed in the process, the variable $m$ is also replaced
by a~new variable $m'$ with the relation $m=m'+v$.
However, for simplicity of notation we rename the new variable $m'$ again $m$.
After some reorderings we obtain the following expression
\begin{gather*}
R=\sum_{m=0}^\infty\sum_{a_1,\dots,a_m\in N_{\alpha}}(-1)^{\sum\limits_{i=
1}^m\sig{\xi_{i}}}\sum_{\substack{\{l,a,b,c,e,f,h,j,k,p,r,x,z:\\
2n=2l+2a+2b+2f+2p+2r+c+e+h+x,\\
m=2z+c+j+x=2k+e+h+j\}}}
\frac{2^{n-l-a-f-k-z}}{l!a!b!c!e!f!h!j!k!p!r!x!z!}
\\
\hphantom{R=}{}
\times
\sum_{\substack{\sigma\in S^{2n}\\
\nu\in S^m}}\kappa^{|\sigma|+|\nu|}(-1)^{a+b+h+r+x}\kappa^{e+h+j+k+(e+h)(c+x)}
\\
\hphantom{R=}{}
\times
\kappa^{\frac{l(l-1)+a(a-1)+b(b-1)+f(f-1)+k(k-1)+p(p-1)+r(r-1)+z(z-1)+(c+x)(c+x-1)}{2}}
\\
\hphantom{R=}{}
\times
\left(\prod_{i=1}^l\big\{\widehat{(\phi_{\sigma(i)},0,0)},\widehat{(\phi_{\sigma(l+i)},0,0)}\big\}_{\partial M}\right)
\left(\prod_{i=1}^{p}\{\tilde{\phi}_{\sigma(2l+i)},\tilde{\phi}_{\sigma(2l+p+i)}\}_{\Sigma}\right)
\\
\hphantom{R=}{}
\times
\left(\prod_{i=1}^{a}\big\{\widehat{(0,\tilde{\phi}_{\sigma(2l+2p+i)},0)},\widehat{(0,\tilde{\phi}_{\sigma(2l+2p+a+i)},0)}\big\}_{\partial M}\right)
\\
\hphantom{R=}{}
\times
\left(\prod_{i=1}^{r}\big\{\widehat{(0,0,\tilde{\phi}_{\sigma(2l+2p+2a+i)})},\widehat{(0,\tilde{\phi}_{\sigma(2l+2p+2a+r+i)},0)}\big\}_{\partial M}\right)
\\
\hphantom{R=}{}
\times
\left(\prod_{i=1}^{b}\big\{\widehat{(0,0,\phi_{\sigma(2l+2p+2a+2r+i)})},\widehat{(0,0,\tilde{\phi}_{\sigma(2l+2p+2a+2r+b+i)})}\big\}_{\partial M}\right)
\\
\hphantom{R=}{}
\times
\left(\prod_{i=1}^{f}\big\{\widehat{(0,0,\tilde{\phi}_{\sigma(2l+2p+2a+2r+2b+i)})},\widehat{(0,0,\tilde{\phi}_{\sigma(2l+2p+2a+2r+2b+f+i)})}\big\}_{\partial M}\right)
\\
\hphantom{R=}{}
\times
\left(\prod_{i=1}^{c}\big\{\widehat{(0,0,\tilde{\phi}_{\sigma(2l+2p+2a+2r+2b+2f+i)})},\widehat{(0,\xi_{i},0)}\big\}_{\partial M}\right)
\\
\hphantom{R=}{}
\times
\left(\prod_{i=1}^{x}\big\{\widehat{(0,0,\tilde{\phi}_{\sigma(2l+2p+2a+2r+2b+2f+c+i)})},\widehat{(0,\xi_{c+i},0)}\big\}_{\partial M}\right)
\\
\hphantom{R=}{}
\times
\left(\prod_{i=1}^{h}\big\{\widehat{(0,0,\tilde{\phi}_{\sigma(2l+2p+2a+2r+2b+2f+c+x+i)})},\widehat{(0,0,\xi_{\nu(i)})}\big\}_{\partial M}\right)
\\
\hphantom{R=}{}
\times
\left(\prod_{i=1}^{e}\big\{\widehat{(0,0,\tilde{\phi}_{\sigma(2l+2p+2a+2r+2b+2f+c+x+h+i)})},\widehat{(0,0,\xi_{\nu(h+i)})}\big\}_{\partial M}\right)
\\
\hphantom{R=}{}
\times
\left(\prod_{i=1}^{j}\big\{\widehat{(0,\xi_{c+x+i},0)},\widehat{(0,0,\xi_{\nu(h+e+i)})}\big\}_{\partial M}\right)
\\
\hphantom{R=}{}
\times
\left(\prod_{i=1}^{z}\big\{\widehat{(0,\xi_{c+x+j+i},0)},\widehat{(0,\xi_{c+x+j+z+i},0)}\big\}_{\partial M}\right)
\\
\hphantom{R=}{}
\times
\left(\prod_{i=1}^{k}\big\{\widehat{(0,0,\xi_{\nu(h+e+j+i)})},\widehat{(0,0,\xi_{\nu(h+e+j+k+i)})}\big\}_{\partial M}\right).
\end{gather*}
Again, we identify binomial sums in this expression that can be evaluated.
Leaving the details to the reader we note that the variables $c$ and $x$ form a~pair, with a~vanishing binomial if we
sum over $c$ and $x$, holding their sum f\/ixed.
This leads to the substitution $c=0$ and $x=0$.
Similarly, $h$ and $e$ form a~pair yielding a~vanishing binomial and we may set $h=0$ and $e=0$.
Also, $b$ and $f$ form a~pair.
In this case the binomial does not vanish and we introduce for the sum the new variable $d=b+f$, discarding $b$ and $f$
in the process.
This yields
\begin{gather*}
R=\sum_{m=0}^\infty\sum_{a_1,\dots,a_m\in N_{\alpha}}(-1)^{\sum\limits_{i=1}^m\sig{\xi_{i}}}\sum_{\substack{\{l,a,d,j,k,p,r,z:\\n=l+a+d+p+r,\\
m=2z+j=2k+j\}}}\frac{2^{n-l-a-d-k-z}}{l!a!d!j!k!p!r!z!}
\\
\hphantom{R=}{}
\times
\sum_{\substack{\sigma\in S^{2n}\\
\nu\in S^m}}\kappa^{|\sigma|+|\nu|}(-1)^{a+d+r}\kappa^{j+k}
\kappa^{\frac{l(l-1)+a(a-1)+d(d-1)+k(k-1)+p(p-1)+r(r-1)+z(z-1)}{2}}
\\
\hphantom{R=}{}
\times
\left(\prod_{i=1}^l\big\{\widehat{(\phi_{\sigma(i)},0,0)},\widehat{(\phi_{\sigma(l+i)},0,0)}\big\}_{\partial M}\right)
\left(\prod_{i=1}^{p}\{\tilde{\phi}_{\sigma(2l+i)},\tilde{\phi}_{\sigma(2l+p+i)}\}_{\Sigma}\right)
\\
\hphantom{R=}{}
\times
\left(\prod_{i=1}^{a}\big\{\widehat{(0,\tilde{\phi}_{\sigma(2l+2p+i)},0)},\widehat{(0,\tilde{\phi}_{\sigma(2l+2p+a+i)},0)}\big\}_{\partial M}\right)
\\
\hphantom{R=}{}
\times
\left(\prod_{i=1}^{r}\big\{\widehat{(0,0,\tilde{\phi}_{\sigma(2l+2p+2a+i)})},\widehat{(0,\tilde{\phi}_{\sigma(2l+2p+2a+r+i)},0)}\big\}_{\partial M}\right)
\\
\hphantom{R=}{}
\times
\left(\prod_{i=1}^{d}\big\{\widehat{(0,0,\phi_{\sigma(2l+2p+2a+2r+i)})},\widehat{(0,0,\tilde{\phi}_{\sigma(2l+2p+2a+2r+d+i)})}\big\}_{\partial M}\right)
\\
\hphantom{R=}{}
\times
\left(\prod_{i=1}^{j}\big\{\widehat{(0,\xi_{i},0)},\widehat{(0,0,\xi_{\nu(i)})}\big\}_{\partial M}\right)
\left(\prod_{i=1}^{z}\big\{\widehat{(0,\xi_{j+i},0)},\widehat{(0,\xi_{j+z+i},0)}\big\}_{\partial M}\right)
\\
\hphantom{R=}{}
\times
\left(\prod_{i=1}^{k}\big\{\widehat{(0,0,\xi_{\nu(j+i)})},\widehat{(0,0,\xi_{\nu(j+k+i)})}\big\}_{\partial M}\right).
\end{gather*}
This expression exhibits the desired factorization into a~term depending on the variables $\phi_1,\dots$, $\phi_{2n}$ and
a~term involving sums over basis of $L_{\Sigma}$.
Denoting the former term by~$R_1$ and the latter by~$R_2$ we have $R=R_1R_2$, with $R_1$ given as follows
\begin{gather}
R_1=\sum_{\substack{\{l,a,d,p,r:\\
n=l+a+d+p+r\}}}\frac{2^{n-l-a-d}}{l!a!d!p!r!}\sum_{\sigma\in S^{2n}}\kappa^{|\sigma|}(-1)^{a+d+r}
\kappa^{\frac{l(l-1)+a(a-1)+d(d-1)+p(p-1)+r(r-1)}{2}}
\nonumber
\\
\hphantom{R_1=}{}
\times
\left(\prod_{i=1}^l\big\{\widehat{(\phi_{\sigma(i)},0,0)},\widehat{(\phi_{\sigma(l+i)},0,0)}\big\}_{\partial M}\right)
\left(\prod_{i=1}^{p}\{\tilde{\phi}_{\sigma(2l+i)},\tilde{\phi}_{\sigma(2l+p+i)}\}_{\Sigma}\right)
\nonumber
\\
\hphantom{R_1=}{}
\times
\left(\prod_{i=1}^{a}\big\{\widehat{(0,\tilde{\phi}_{\sigma(2l+2p+i)},0)},\widehat{(0,\tilde{\phi}_{\sigma(2l+2p+a+i)},0)}\big\}_{\partial M}\right)
\nonumber
\\
\hphantom{R_1=}{}
\times
\left(\prod_{i=1}^{r}\big\{\widehat{(0,0,\tilde{\phi}_{\sigma(2l+2p+2a+i)})},\widehat{(0,\tilde{\phi}_{\sigma(2l+2p+2a+r+i)},0)}\big\}_{\partial M}\right)
\nonumber
\\
\hphantom{R_1=}{}
\times
\left(\prod_{i=1}^{d}\big\{\widehat{(0,0,\phi_{\sigma(2l+2p+2a+2r+i)})},\widehat{(0,0,\tilde{\phi}_{\sigma(2l+2p+2a+2r+d+i)})}\big\}_{\partial M}\right).
\label{eq:gar1}
\end{gather}
Taking into account that the summation constraint implies $z=k$, $R_2$ may be written as follows
\begin{gather}
R_2=\sum_{m=0}^\infty\sum_{a_1,\dots,a_m\in N_{\alpha}}(-1)^{\sum\limits_{i=1}^m\sig{\xi_{i}}}\sum_{\substack{\{j,k:\\
m=2k+j\}}}\frac{2^{-2k}}{j!(k!)^2}\sum_{\nu\in S^m}\kappa^{|\nu|}\kappa^{j+k}
\nonumber
\\
\hphantom{R_2=}{}
\times
\left(\prod_{i=1}^{j}\big\{\widehat{(0,\xi_{i},0)},\widehat{(0,0,\xi_{\nu(i)})}\big\}_{\partial M}\right)
\left(\prod_{i=1}^{k}\big\{\widehat{(0,\xi_{j+i},0)},\widehat{(0,\xi_{j+k+i},0)}\big\}_{\partial M}\right)
\nonumber
\\
\hphantom{R_2=}{}
\times
\left(\prod_{i=1}^{k}\big\{\widehat{(0,0,\xi_{\nu(j+i)})},\widehat{(0,0,\xi_{\nu(j+k+i)})}\big\}_{\partial M}\right).
\label{eq:gar2}
\end{gather}
Observation reveals that $R_2$ coincides precisely with expression~\eqref{eq:ga3} for $R$ in the special case $n=0$.
Therefore, $R_2$ coincides with expression~\eqref{eq:gafid} for the regularized gluing anomaly factor associated to~$\alpha$.
That is
\begin{gather*}
R_2=c_{\alpha}(M;\Sigma,\overline{\Sigma'}).
\end{gather*}
Returning to expression~\eqref{eq:gar1} for $R_1$, we observe that it may be simplif\/ied by identifying the various
factors as arising from expanding the left hand side of identity~\eqref{eq:phiid} of Lemma~\ref{lem:phiid} in terms of
the right hand side.
This eliminates the variables $l$, $a$, $d$, $p$, $r$, leading to the much simpler expression
\begin{gather*}
R_1=\frac{1}{n!}\sum_{\sigma\in S^{2n}}\kappa^{|\sigma|}\left(\prod_{i=
1}^n\{\phi_{\sigma(i)},\phi_{\sigma(2n+1-i)}\}_{\Sigma_1}\right).
\end{gather*}
But this coincides precisely with the def\/inition~\eqref{eq:defampleven} for an amplitude in $M_1$,
\begin{gather*}
R_1=\rho_{M_1}(\cs{\phi_1,\dots,\phi_{2n}}).
\end{gather*}
So, summarizing, we obtain
\begin{gather*}
\rho_{M_1}(\cs{\phi_1,\dots,\phi_{2n}})\cdot c_{\alpha}(M;\Sigma,\overline{\Sigma'})=R.
\end{gather*}
This is expression~\eqref{eq:gfid} of Lemma~\ref{lem:gfid} (with our redef\/inition of the variable $n$ as $2n$).
This completes the proof.
As a~f\/inal remark, observe that absolute convergence of expression~\eqref{eq:gar2} is ensured by the assumption of
well def\/inedness of $c_{\alpha}(M;\Sigma,\overline{\Sigma'})$.
This then implies absolute convergence of the sums appearing in the preceding expressions for $R$.
This is the case in spite of the fact that some reorderings are taking place (recall the substitutions of the variable
$m$), since the reorderings are bounded in range by the f\/ixed constant $2n$.
\end{proof}

\subsection*{Acknowledgements}

This work was supported in part by UNAM--DGAPA--PAPIIT through project grant IN100212.

\LastPageEnding

\end{document}